\documentclass[12pt]{article}
\usepackage[utf8]{inputenc}
\usepackage{graphicx}
\usepackage{animate}
\usepackage{amsmath,bm,amsfonts,amssymb,enumerate,epsfig,bbm,calc,color,ifthen,capt-of,multimedia,bm,multirow}
\usepackage{blindtext}
\usepackage{booktabs}
\usepackage{MnSymbol}
\usepackage{url}
\usepackage{natbib}
\usepackage{changepage}
\usepackage{fancyhdr}
\usepackage[flushleft]{threeparttable}

\usepackage{setspace}
\usepackage[flushleft]{threeparttable}

\DeclareMathOperator*{\argmax}{arg\,max}

\title{Weighted Q-learning for optimal dynamic treatment regimes with nonignorable missing covariates}
\author{Jian Sun \thanks{School of Data Science,  Fudan University, Shanghai 200433, China. Email: jsun19@fudan.edu.cn.}, Bo Fu \thanks{School of Data Science,  Fudan University, Shanghai 200433, China. Email: fu@fudan.edu.cn.}, and Li Su \thanks{MRC Biostatistics Unit, University of Cambridge, Cambridge, CB2 0SR, UK. Email: li.su@mrc-bsu.cam.ac.uk.}}

\date{}

\begin{document}

\maketitle

\begin{abstract}
Dynamic treatment regimes (DTRs)  formalize medical decision-making as a sequence of rules for different stages, mapping patient-level information to recommended treatments. In practice, estimating an optimal DTR using observational data from electronic medical  record (EMR) databases can be complicated by nonignorable missing covariates resulting from 
informative monitoring of patients. Since complete case analysis can provide consistent estimation of outcome model parameters under the assumption of outcome-independent missingness, Q-learning is a natural approach to accommodating nonignorable missing covariates. 
However, the backward induction algorithm used in Q-learning can introduce challenges, as nonignorable missing covariates at later stages can result in nonignorable missing pseudo-outcomes at earlier stages, leading to suboptimal DTRs, even if the longitudinal outcome variables are fully observed. To address this unique missing data problem in DTR settings, we propose two weighted Q-learning approaches where inverse probability weights for missingness of the pseudo-outcomes are obtained through estimating equations with valid nonresponse instrumental variables or sensitivity analysis. The asymptotic properties of the weighted Q-learning estimators are derived, and
the finite-sample performance of the proposed methods is evaluated and compared with alternative methods through extensive simulation studies. Using EMR data from the Medical Information Mart for Intensive Care database, we apply the proposed methods to investigate the optimal fluid strategy for sepsis patients in intensive care units.
\end{abstract}


\section{Introduction}

Dynamic treatment regimes (DTRs) formalize medical decision-making as a sequence of decision rules, each corresponding to a specific decision stage, that map available information at patient-level to a recommended treatment \citep{Murphy_2003}. Optimal DTRs have been estimated and evaluated in various biomedical areas, such as cancer \citep{wang2012evaluation} and sepsis \citep{Yu_Zhang_2022}. In practice, estimating optimal DTRs becomes challenging when there are missing data in covariates, treatments, or outcomes.

There is a rich literature on identifying optimal DTRs with complete data. Various methods have been developed, including  Q-learning \citep{Murphy_2003}, A-learning \citep{Schulte_et_al_2014}, and their numerous variants \citep{Kosorok_Laber_2019}. These methods rely on approximate dynamic programming methods. Direct-optimization methods, such as simultaneous outcome weighted learning \citep{Zhao_et_al_2015}, and model-based planning via g-computation \citep{Xu_et_al_2016} have also been employed. In contrast, limited research to date has been devoted to accommodating missing data when estimating optimal DTRs. For survival outcomes,  methods for optimal DTRs estimation were developed by addressing missingness from right-censoring \citep{Simoneau_et_al_2020}.  For handling missing data in covariates, treatments and outcomes in general, \cite{Shortreed_et_al_2014} proposed a time-ordered nested conditional imputation strategy when estimating optimal DTRs in sequential multiple assignment randomized trials. \cite{Dong_et_al_2020} applied the augmented inverse probability weighting method to Q-learning and a generalized version of outcome-weighted learning when dealing with missing data due to patients' dropout. In simpler settings of estimating single-stage optimal treatment regimes, 
\cite{Shen_Hubbard_Linn_2023} further discussed the multiple imputation method proposed by \cite{Shortreed_et_al_2014}  for estimating and evaluating optimal individualized treatment rules that were not directly observed in the design. Also for the single-stage setting with missing data,  \cite{Huang_Zhou_2020} investigated the performance of an augmented inverse probability weighted estimator in the direct-optimization framework.  However, these existing methods all focus on ignorable missingness and could potentially yield sub-optimal DTRs when missing data for covariates, treatments or outcomes are nonignorable  \citep{Little_Rubin_2014}.

In this research, we aim to address the issue of nonignorable missing covariates in the estimation of optimal DTRs, which is prominent in settings where patients are informatively monitored, e.g., in electronic medical records (EMR) databases or disease clinics. Our research was motivated by a recent investigation on optimal fluid strategy for sepsis patients in intensive care units (ICUs) \citep{Speth_et_al_2022}. 
Using EMR data from the Medical Information Mart for Intensive Care (MIMIC-III) \citep{Johnson_et_al_2016}, \cite{Speth_et_al_2022} estimated an optimal two-stage DTR for treating sepsis patients with fluid resuscitation in ICUs by incorporating baseline covariates such as demographics, time-varying covariates such as mechanical ventilation usage and vasopressor administration,  and the outcome measure of multi-organ failure at the first stage. However, according to the recommendations of the `Surviving Sepsis Campaign', hemodynamic variables should also be taken into account when reassessing the need for additional fluid administration following the initial resuscitation \citep{Rhodes_et_al_2017}. Unfortunately,  within the MIMIC-III database, certain hemodynamic variables, including blood pressure, respiratory rate, and body temperature, are subject to missingness at the conclusion of the initial resuscitation. Furthermore,  as highlighted by \cite{Awad_et_al_2017},  patients in the MIMIC-III database were informatively monitored since patients with more severe conditions, who typically presented with more abnormal physiological indicators than others, received more intensive monitoring.  Consequently, their measurements of hemodynamic variables were less likely to be missing. In this case, the missingness probabilities of the hemodynamic variables were directly related to their own values, indicating that the missing data mechanism was nonignorable missing. This characteristic could potentially lead to sub-optimal estimation of DTRs if existing methods are employed.

The identification and estimation of optimal DTRs become much more challenging when covariates are nonignorable missing. In the literature on handling  nonignorable missing covariates, it has been concluded that the parameters of an outcome regression model can be consistently estimated using complete case analysis as long as the missingness of the covariates is not directly related to the outcome \citep{Yang_Wang_Ding_2019}. Therefore, considering that Q-learning relies on consistently estimated outcome models and is both flexible and straightforward to implement, it offers a natural approach for accommodating nonignorable missing covariates when estimating optimal DTRs. 
However, when implementing the step-wise outcome regression algorithm in  Q-learning,  challenges arise not only from the missingness of the covariates but also from the missingness of the \emph{pseudo-outcomes}, even if  outcome variables are fully observed. This is because the Q-learning algorithm computes the pseudo-outcome in the preceding stage using covariates from the subsequent stage. As a result, if there are nonignorable missing covariates in the later stages, it can lead to nonignorable missing pseudo-outcomes in the earlier stages.  We provide a detailed illustration of this unique missing data problem with DTRs caused by backward-induction-induced nonignorable missing pseudo-outcomes in Section 2.3. 

It is well known that with nonignorable missing outcomes, outcome regression models are generally not identifiable; see examples provided by \cite{Miao_Ding_Geng_2016}. Inspired by the nonignorable missing outcome literature, we propose two weighted Q-learning approaches to tackling nonignorable missing  pseudo-outcomes in optimal DTR estimation. The first approach is to employ estimating equations to estimate a working model for missingness probability of the pseudo-outcome with valid nonresponse instrumental variables  \citep{Shao_Wang_2016}.  The pseudo-outcome estimates based on observed data are then weighted by the inverse of the estimated missingness probabilities in the Q-learning algorithm. In scenarios where nonresponse instrumental variables are not available, we specify a sensitivity parameter to quantify the residual association between the missing pseudo-outcome at stage $t$ and its conditional missingness probability given observed information up to stage $t$. Thus the missingness probabilities of the pseudo-outcomes are functions of the sensitivity parameter. We develop a practical approach to calibrating the sensitivity parameter  and then perform a sensitivity analysis to evaluate the impact of the pre-specified sensitivity parameter on the estimated optimal DTRs.

\section{Method}

\subsection{Setting and notation}

We consider an observational cohort with $n$ patients and finite $T$ treatment stages. Let subscripts $i=1,\ldots,n$, $t=1,\ldots,T$ denote patients and stages, respectively. Unless specified otherwise, we use capital letters to denote random variables and lowercase letters to indicate specific realizations of random variables.  Complete data for the patients are assumed independent and identically distributed and thus we suppress the patient-specific subscript $i$. Let $A_t\in \{-1, 1\}$ be the assigned treatment at stage $t$ for $t = 1, \ldots, T$.   $\bm{X}_t \in \mathbb{R}^{p_t}$ is a  $1 \times p_t$ vector of covariates measured before $A_t$, and $Y_t$ denotes the longitudinal 
outcome measured after $A_t$ at stage $t$. 

The final outcome of interest is a pre-specified function (e.g., sum or maximum) of the longitudinal outcome in $T$ stages, $Y = f(Y_1,\ldots, Y_T)$. Higher values of $Y$ are assumed to be better. We define $\bm{H}_1=\bm{X}_1$, and $\bm{H}_t=\left(\bm{H}_{t-1}, A_{t-1}, Y_{t-1}, \bm{X}_t\right)$ for $t=2, \ldots, T$. Thus, $\bm{H}_t$ represents the information available before making the treatment decision at stage $t$. Let $\mathcal{H}_t$ denote the support of $\bm{H}_t$. A DTR $\bm{d}$ consists of a set of decision rules $(d_1, \ldots, d_T)$, where $d_t: \mathcal{H}_t \to \{-1, 1\}$ is a function that takes the observed history $\bm{h}_t$ as input and outputs a treatment decision at stage $t$. An optimal DTR is the set of decision rules $\left(d^{\mathrm{opt}}_1, \ldots, d^{\mathrm{opt}}_T\right)$ that maximizes the expectation of the final outcome. We formalize this definition using the potential outcome framework.

For $t=1,\ldots,T$, let $Y_t^*(\overline{\bm{a}}_t)$  denote the potential longitudinal outcome at stage $t$ and $\bm{X}_{t+1}^*(\overline{\bm{a}}_t)$  denote the potential covariates at stage $t+1$  if a patient, possibly contrary to fact, had received the treatment sequence $\overline{\bm{a}}_t = (a_1,\ldots,a_t)$  by stage $t$. We then define the set of potential outcomes under $\overline{\bm{a}}_t$ as $\bm{O}_t^*(\overline{\bm{a}}_t) = \{Y_1^*(a_1), \bm{X}_2^*(a_1),\ldots,Y_{t}^*(\overline{\bm{a}}_{t}),\bm{X}_{t+1}^*(\overline{\bm{a}}_{t}) \}$  for $t=1,\ldots,T-1$. The potential final outcome under a regime $\bm{d}$ is
\begin{align*}
Y^*(\bm{d})=\sum_{\overline{\bm{a}}_T} Y^*\left(\overline{\bm{a}}_T\right) \prod_{t=1}^T \mathbb I (d_t\left[\{\bm{X}_1,a_1,Y_1^*(a_1),\ldots,a_{t-1},Y_{t-1}^*(\overline{\bm{a}}_{t-1}),\bm{X}_{t}^*(\overline{\bm{a}}_{t-1})\}\right]=a_t),
\end{align*}
where $Y^*\left(\overline{\bm{a}}_T\right)=f\left\{Y_1^*\left(a_1\right), Y_2^*\left(\overline{\bm{a}}_2\right), \ldots, Y_T^*\left(\overline{\bm{a}}_T\right)\right\}$. 
Define the value of a regime $\bm{d}$ to be $V(\bm{d}) = E\{Y^*(\bm{d})\}$. Let $\mathcal{D}$ denote all feasible regimes. An optimal DTR, $\bm{d}^{\mathrm{opt}}\in \mathcal{D}$, satisfies that $V(\bm{d}^{\mathrm{opt}}) \geq V(\bm{d})$ for all $\bm{d} \in \mathcal{D}$.

\subsection{Q-learning with complete data}

In this section, we briefly review the Q-learning algorithm for estimating optimal DTRs with complete observational data. To estimate the value of candidate regimes from observed data, it is necessary to express the value of regimes solely in terms of observables rather than potential outcomes. This becomes feasible under the following, now standard, causal inference assumptions:
\begin{description}
    \item {\em Assumption} 1 (Consistency). $Y_t^*(\overline{\bm{a}}_t) = Y_t$ for $t = 1,\ldots,T$ and $\bm{X}_{t+1}^*(\overline{\bm{a}}_t) = \bm{X}_{t+1}$ for $t = 1,\ldots,T-1$ when $\overline{\bm{a}}_t$ are actually received.
    \item {\em Assumption} 2 (Sequential ignorability). $\{\bm{O}_{T-1}^*(\overline{\bm{a}}_{T-1}), Y_T^*(\overline{\bm{a}}_T) : \overline{\bm{a}}_T \in \bigotimes_{t=1}^{T} \{-1,1\}\} \upmodels A_t \mid \bm{H}_t$ for $t=1,\ldots,T$, where $\bigotimes$ denotes the Cartesian product.
    \item {\em Assumption} 3 (Positivity). $P(A_t=a_t\mid \bm{h}_t) > \epsilon >0$ for $a_t \in \{-1,1\}$, $\bm{h}_t \in  \mathcal{H}_t$, and $t=1,\ldots,T$, where $\epsilon$ is a positive constant. 
\end{description}

Under Assumptions 1-3, the value of a DTR can be identified from the observed data. Specifically, in the 2-stage scenario we have $E\{Y^*(\bm{d})\} = E(E[\{Y\mid \bm{H}_2,A_2 = d_2(\bm{H}_2)\}\mid \bm{H}_1,A_1 = d_1(\bm{H}_1)] )$.
This repeated expectation form suggests that the optimal DTR can be computed by a backward-induction procedure, which is based on the following recursively defined Q-functions. 
At the final stage $T$, $Q_T\left(\bm{h}_T, a_T\right) = E\left(Y \mid \bm{H}_T=\bm{h}_T, A_T=a_T\right)$. For $ t=T-1,\ldots,1$, $Q_t\left(\bm{h}_t, a_t\right) = E\{\max_{a_{t+1} \in \{-1,1\}} Q_{t+1}(\bm{H}_{t+1}, a_{t+1}) \mid \bm{H}_t=\bm{h}_t, A_t=a_t\}.$

The true Q-functions are unknown and must be estimated from the data. Since Q-functions represent conditional expectations, it is natural to estimate them using regression models. Note that in these models, the response variable is $\max _{a_{t+1} \in \{-1,1\}} Q_{t+1}\left(\bm{H}_{t+1}, a_{t+1}\right)$ instead of $Y_t$ ($t = 1,\ldots,T-1$). These response variables are commonly referred to as pseudo-outcomes, and we define $Y_{pse,T} = Y$ and $Y_{pse,t} = \max _{a_{t+1} \in \{-1,1\}} Q_{t+1}\left(\bm{H}_{t+1}, a_{t+1}\right)$ for $t = 1,\ldots,T-1$. Subsequently, $Q_t(\bm{h}_t,a_t) = E\left( Y_{pse,t}\mid \bm{h}_t,a_t \right)$ for $t = 1,\ldots,T$. Since $a_t \in \{-1,1\}$, any specified model for $Q_t(\bm{h}_t,a_t)$ can be decomposed as $Q_t(\bm{h}_t,a_t;\bm{\theta}_t) = q_{t,0}(\bm{h}_t; \bm{\beta}_t) + a_t q_{t,1}(\bm{h}_t;\bm{\psi}_t)$, where $\bm{\theta}_t = (\bm{\beta}_t, \bm{\psi}_t)$, $q_{t,0}(\bm{h}_t) = \frac{Q_t(\bm{h}_t,1) + Q_t(\bm{h}_t,-1)}{2}$, and $q_{t,1}(\bm{h}_t)  = \frac{Q_t(\bm{h}_t,1) - Q_t(\bm{h}_t,-1)}{2}$ for $t=1\ldots,T$. As a result, the Q-function is divided into two components: (1) a stage $t$ treatment-free component $ q_{t,0}(\bm{h}_t; \bm{\beta}_t)$ which depends on (a subset of) the history $\bm{h}_t$  before stage $t$ but not on the stage $t$ treatment $a_t$, and (2) a stage $t$ treatment effect component $a_tq_{t,1}(\bm{h}_t;\bm{\psi}_t)$ which depends on (a potentially different subset of) $\bm{h}_t$ and specifically includes the main effect of treatment $a_t$ and its interactions with tailoring variables (i.e., effect modifiers). The blip function, as defined in \cite{Chakraborty_Moodie_2013}, equals to $a_t q_{t,1}(\bm{h}_t;\bm{\psi}_t) + q_{t,1}(\bm{h}_t;\bm{\psi}_t)$ under this setting. Therefore, we refer to $\bm{\psi}_t$ as the blip function parameters. Using this modeling approach, we have $d_t^{\mathrm{opt}}(\bm{h}_t) = 2\mathbb I \left\{q_{t,1}(\bm{h}_t;\bm{\psi}_t)>0\right\} -1$. 
Under Assumptions 1-3, the estimator $\widehat{\bm{\theta}}_t$ of $\bm{\theta}_t$ can be recursively constructed by regressing $Y_{pse,t}$ on $\bm{H}_t$ and $A_t$ . Then, the Q-learning estimator of $d^{\mathrm{opt}}_t$ can be obtained by $\widehat{d}_t^{\mathrm{opt}}\left(\bm{h}_t\right) = 2\mathbb{I}\left\{q_{t,1}(\bm{h}_t;\widehat{\bm{\psi}}_t)>0\right\} -1$ for $t=1,\ldots,T$.

\subsection{Q-learning for DTRs with  nonignorable missing covariates: why complete case analysis fails at earlier stages while working at the final stage?}

We focus on the scenario that the treatment and longitudinal outcome, $(A_t, Y_t)$, are fully observed while the covariates, $\bm{X}_t$, contain missing values. This scenario is frequently encountered in practice because treatments are typically well-documented and the longitudinal outcomes of interest are usually key clinical disease activity indices which are regularly measured. Moreover, the probability of a covariate being missing is likely to depend on the actual value of the covariate itself due to informative monitoring of patients. 
For instance, in the MIMIC-III data, patients who displayed symptoms of abnormal heart rates such as chest pain were more likely to have their heart rates recorded. Therefore, the absence of heart rate measurements could indicate nonignorable missingness, since patients with normal heart rates were less likely to be frequently monitored.

\begin{figure}[htbp]
    \centering
    \includegraphics[scale=0.45]{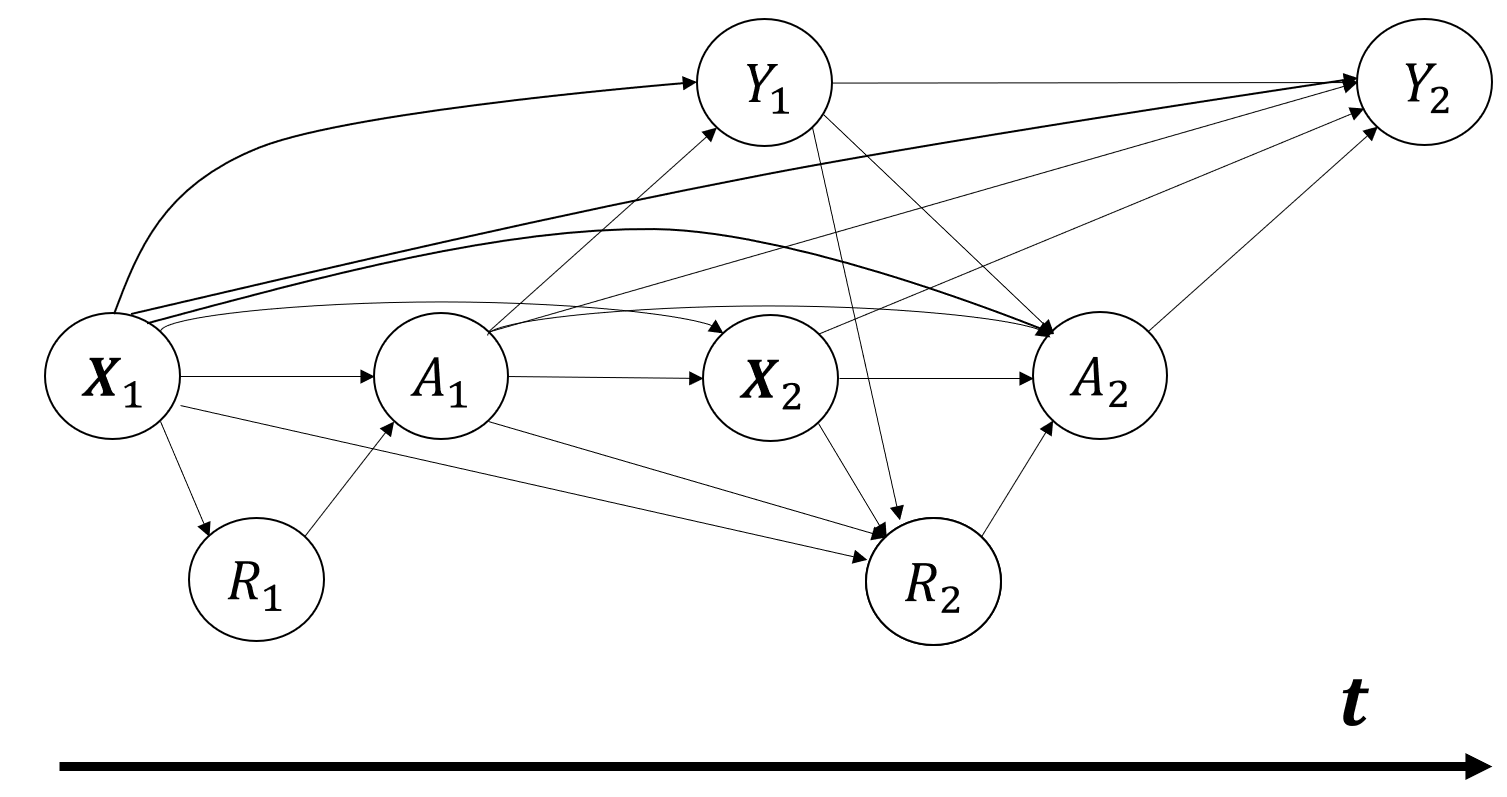}
    \caption{A direct acyclic graph illustrating Assumptions 2 and 4}
    \label{DAG}
\end{figure}

Let $R_t$ represent the missingness indicator for $\bm{X}_t$ such that $R_{t} = 1$ if all variables in $\bm{X}_t$ are fully observed and $R_{t} = 0$ if at least one component of $\bm{X}_t$ is missing. Define $\overline{\bm{R}}_t = (R_1,\ldots, R_t)$ as the history of missingness indicators up to stage $t$. We use $\bm{1}_{t}$ to denote a $1 \times t$ vector in which all elements are equal to $1$ for $t=1, \ldots, T$. In ICU settings, the effect of a treatment is commonly evaluated by the patient's longitudinal outcomes (e.g., physiological symptoms) observed several hours after treatment administration, while the treatment decision for a patient is based on the patient's medical history. Because of the temporal ordering of these covariates, it is plausible that the vector of missingness indicators $\overline{\bm{R}_t}$ does not directly depend on the patient's variables measured after receiving the current stage treatment $A_t$. Thus we make the following assumption:
\begin{description}
    \item {\em Assumption} 4 (Future-independent missingness). $\overline{\bm{R}_t} \upmodels (Y_t, \bm{X}_{t+1}, R_{t+1},  \ldots, \bm{X}_T,R_T,A_T,Y_T) \mid (\bm{H}_t, A_t)$  for $t=1,\ldots,T$.\label{A4}
\end{description}
Figure \ref{DAG} is a directed acyclic graph  illustrating Assumptions 2 and 4 in a 2-stage scenario. In this example, $\bm{H}_1 = \bm{X}_1$, $\bm{H}_2 = (\bm{X}_1,A_1,Y_1,\bm{X}_2)$. $R_1$ and $R_2$ are missing indicators for $\bm{X}_1$ and $\bm{X}_2$, respectively. Graphically, $A_1$ has no common parents with ($Y_1, \bm{X}_2, A_2, Y_2$) except for $\bm{H}_1$, and $A_2$ has no common parents with $Y_2$ except for $\bm{H}_2$, encoding Assumption 2. Besides, $R_1$ is d-separated from $(Y_1, \bm{X}_2,R_2, A_2, Y_2)$ by $(\bm{H}_1,A_1)$, and $(R_1, R_2)$ is d-separated from $Y_2$ by $(\bm{H_2},A_2)$ \citep{VanderWeele_Robins_2007}, encoding Assumption 4. 

In the single-stage scenario, Assumption 4 degenerates to the outcome-independent missingness assumption \citep{Yang_Wang_Ding_2019}, which posits that the missingness probability of covariates or confounders is conditionally independent of the outcome, given all variables except for the outcome, i.e., $R_1 \upmodels Y_1\mid (\bm{X}_1, A_1)$. Under this assumption, we have $P(Y_1\mid \bm{X}_1, A_1) = P(Y_1\mid \bm{X}_1, A_1, R_1=1)$, so that we can obtain unbiased parameter estimates of the outcome model with complete case analysis. Consequently, in the single-stage scenario, Q-learning based on complete case analysis yields consistently optimal individualized treatment rules. In this research, considering the plausibility of  conditional independence ensured by temporal ordering of the variables, we adapt the outcome-independent missingness assumption to longitudinal settings with multiple stages and make Assumption 4. 

Given $\bm{H}_t$ and $A_t$, the pseudo-outcome $Y_{pse,t}$ only depends on $Y_t$ and the variables measured after stage $t$. Therefore, under Assumption 4, $Y_{pse,t} \upmodels \overline{\bm{R}}_t \mid (\bm{H}_t, A_t$) and we have $P(Y_{pse,t} \mid \bm{H}_t, A_t) = P(Y_{pse,t} \mid \bm{H}_t, A_t, \overline{\bm{R}}_t = \bm{1}_t)$ for $t = 1,\ldots,T$.
Thus, we can obtain consistent estimates of $Q_t\left(\bm{h}_t, a_t\right)$ by utilizing data from patients with complete data up to stage $t$ if there are no missing values in  $Y_{pse,t}$. For example, at stage $T$, $Y_{pse,T}=Y$ is fully observed. Thus we can use complete case analysis to estimate $Q_T(\bm{h}_T, a_T)$.

\begin{figure}[htbp]
    \centering
    \includegraphics[scale=0.27]{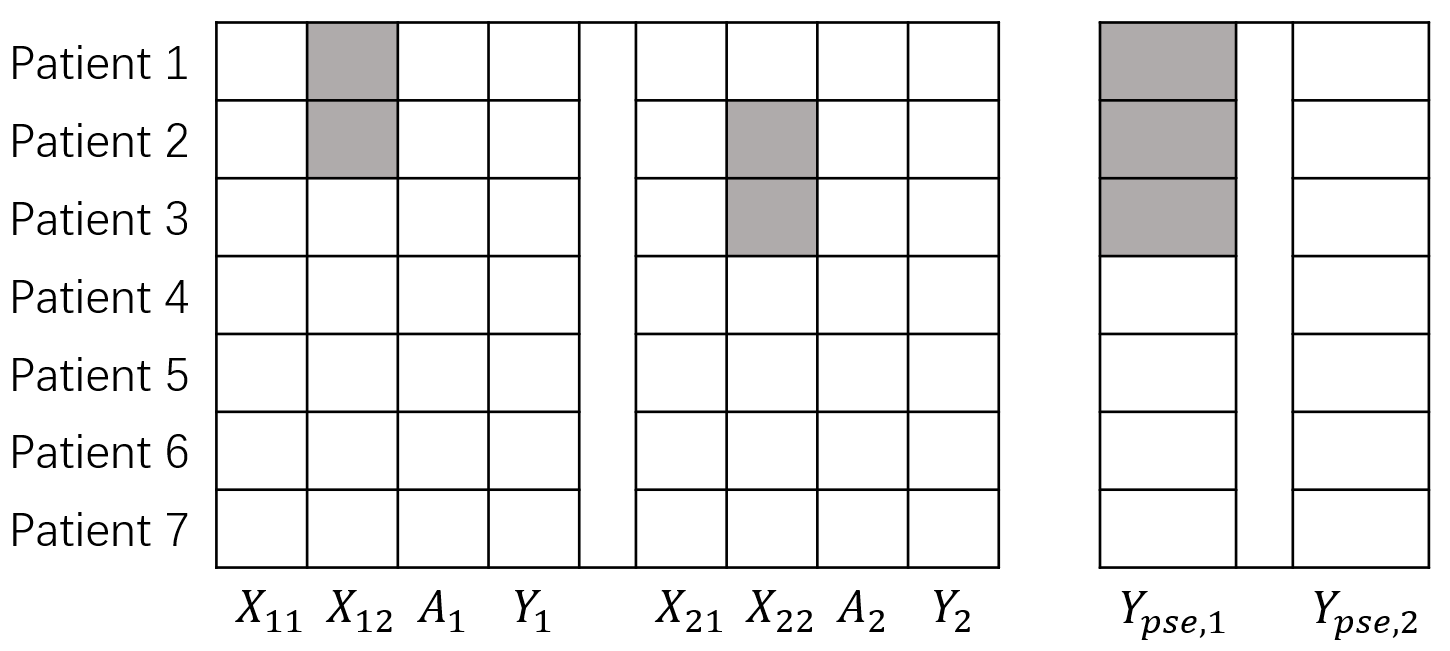}
    \caption{A hypothetical example illustrating the relationship between the missing data pattern of covariates and that of pseudo-outcomes. Gray areas represent missing information. The setting of the variables is the same as that in Simulation 1 in Web Appendix H.}
    \label{pseudo-outcome missing pattern}
\end{figure}

However, conditional on $\overline{\bm{R}}_t=\bm{1}_t$ (i.e., for complete cases up to stage $t$), although there is no missingness in $\bm{H}_t$, the pseudo-outcome $Y_{pse,t}$ may contain missing values because $Y_{pse,t}$  depends on $\bm{X}_{t+1}$, which can be missing  (i.e.,  $R_{t+1} =0$) when $\overline{\bm{R}}_t=\bm{1}_t$. We illustrate this backward-induction-induced missing pseudo-outcome problem in Figure \ref{pseudo-outcome missing pattern} with a hypothetical example coming from the same setting of our simulation study in Section 3.  In this example, all variables for Patient 3 at stage one are fully observed. Nevertheless, $Y_{pse,1}$ for Patient 3 is unavailable because $X_{22}$, a covariate required for computing $Y_{pse,1}$, is missing. 

Let $R_{pse,t}$ denote the missingness indicator of $Y_{pse,t}$ for $t=1,\ldots,T$. Because the missingness probabilities of $\bm{X}_{t+1}$ depend on their own values, and $Y_{pse,t}$ is computed based on $\bm{X}_{t+1}$, it is likely that $Y_{pse,t}$ is nonignorably missing. This backward-induction-induced nonignorable missing pseudo-outcome problem would make identification and consistent estimation of Q-functions challenging. A complete case analysis for Q-learning would likely result in suboptimal DTR. In Section 3, we will use simulations to demonstrate this phenomenon. 

\subsection{Weighted Q-learning for DTRs with nonignorable missing pseudo-outcomes: a nonresponse instrument variable approach}

We propose two approaches for inverse probability weighted Q-learning to tackle the nonignorable missing pseudo-outcome problem. The first involves utilizing valid nonresponse instrumental variables. Specifically, we consider a semiparametric model for $R_{pse,t}$:
\begin{description}
    \item {\em Assumption} 5 (semiparametric working model for missingness probability of the pseudo-outcomes). For $t = 1,\ldots,T-1$, 
\begin{align}\label{missingnessmodel}
    P(R_{pse,t}=1\mid \bm{h}_t,a_t,y_{pse,t},\overline{\bm{R}}_t=\bm{1}_t) = \pi(\bm{u}_t,y_{pse,t};\gamma_t)= \frac{1}{1+\exp\{s_t(\bm{u}_t)+\gamma_t y_{pse,t}\}},
\end{align}
\end{description}where $\bm{U}_t \subset (\bm{H}_t,A_t)$ is a proper subset of $(\bm{H}_t,A_t)$, $P(Y_{pse,t}\mid \bm{H}_t, A_t) \neq P(Y_{pse,t}\mid \bm{U}_t)$, and $s_t(\cdot)$ is an unknown and unspecified function of $\bm{u}_t$. 

This semiparametric model assumption does not impose any parametric constraint on the function of $\bm{u}_t$ in working model \eqref{missingnessmodel}. Instead, it restricts the parametric form of $y_{pse,t}$ and requires that there is no interaction between $\bm{u}_t$ and $y_{pse,t}$ in working model \eqref{missingnessmodel}. 
It is noteworthy that the variables observed after stage $t$ may exhibit an association with $R_{pse,t}$ even after conditioning on ($\bm{h}_t,a_t,y_{pse,t},\overline{\bm{R}}_t=\bm{1}_t$) and therefore model \eqref{missingnessmodel} may not be the true data-generating model for $R_{pse,t}$. However,  weighting by the inverse of  $P(R_{pse,t}=1\mid \bm{h}_t,a_t,y_{pse,t},\overline{\bm{R}}_t=\bm{1}_t)$ would suffice to balance the distribution of the variables between the patients with and without missing $y_{pse,t}$ for the purpose of estimating $Q_t(\bm{h}_t,a_t)$ and thus lead to consistent estimation of the Q-functions after weighting. 

Assumption 5 also implies that $(\bm{H}_t,A_t)$ can be decomposed as $(\bm{H}_t,A_t) = (\bm{U}_t,\bm{Z}_t)$ such that $\bm{Z}_t \nupmodels Y_{pse,t} \mid \bm{U}_t$ and $\bm{Z}_t \upmodels R_{pse,t} \mid  \bm{U}_t, Y_{pse,t}$ when $\overline{\bm{R}}_t=\bm{1}_t$. That is, conditional on $\bm{U}_t$, $\bm{Z}_t$ is a predictor of the pseudo-outcome $Y_{pse,t}$, but given $\bm{U}_t$ and $Y_{pse,t}$, $\bm{Z}_t$ is conditionally independent of the missingness indicator $R_{pse,t}$.
$\bm{Z}_t$ is referred to as the nonresponse instrumental variables at stage $t$. The nonresponse instrumental variables, sometimes known as the `shadow variables', are commonly employed in the literature for nonignorable missing outcomes.  It has been demonstrated in previous studies of nonignorable missing outcomes \citep{Shao_Wang_2016, Miao_Tchetgen_2016} that  nonresponse instrumental variables facilitate the identification of the working model \eqref{missingnessmodel} and the outcome model. In certain scenarios, it may not be difficult to find valid nonresponse instrumental variables. For example, gender and weight are associated with the severity of sepsis in ICU patients \citep{kalani2020sepsis, sakr2013influence}. It is plausible that they are independent of the missing data mechanism of hemodynamic variables, given other pre-treatment covariates of the patients  (e.g., comorbidities) and the severity of sepsis represented by $Y_{pse,t}$. These characteristics render gender and weight suitable candidates for nonresponse instrumental variables.

If Assumptions 5 holds, to estimate the working  model~\eqref{missingnessmodel}, we have to overcome the challenge that $s_t(\cdot)$ is nonparametric. Following \cite{Shao_Wang_2016}, we consider the idea of profiling and obtain the following kernel estimate of $s_t(\cdot)$ for a given $\gamma_t$.
\begin{align}
    \exp\{\widehat{s}_{\gamma_t,t}(\bm{u}_t) \} =  \frac{\sum_{i=1}^{n} \left(1-r_{pse,t,i}\right)  K_{c_t}\left(\bm{u}_t-\bm{u}_{t,i}\right) \mathbb{I}(\overline{\bm{R}}_{t,i}=\bm{1}_t) }{\sum_{i=1}^{n}  r_{pse,t,i}   \exp \left(\gamma_t y_{pse,t,i}\right) K_{c_t}\left(\bm{u}_t-\bm{u}_{t,i}\right) \mathbb{I} (\overline{\bm{R}}_{t,i}=\bm{1}_t)}, \label{2.1}
\end{align}
where $K_{c_t}(\cdot) = c_t^{-1}K(\cdot/c_t)$, with $K(\cdot)$ being a symmetric kernel function and $c_t$ a bandwidth. 
For a fixed $\gamma_t$, the working model \eqref{missingnessmodel} can be expressed as $\widehat{\pi}_t(\bm{u}_t,y_{pse,t};\gamma_t) = [1+\exp\{\widehat{s}_{\gamma_t,t}(\bm{u}_t) + \gamma_t y_{pse,t}\} ]^{-1}$.  Then, we can estimate $\gamma_t$ with the following estimating equation,
\begin{align}
    \widehat{E} \left[ \mathbb{I} (\overline{\bm{R}}_t=\bm{1}_t) \bm{l}_t(\bm{z}_t)\left\{\frac{r_{pse,t}}{\widehat{\pi}_t(\bm{u}_t,y_{pse,t}; \gamma_t)} -1 \right\} \right]  = \bm{0}, \label{2.2} 
\end{align}
where $\widehat{E}[\cdot]$ denotes the empirical mean, $\bm{l}_t(\bm{z}_t) = (l_{t,1}(\bm{z}_t), \ldots, l_{t, L}(\bm{z}_t))$ is a user-specified differentiable vector function of the nonresponse instrumental variables of dimension $L$, $L\geq2$. 

Note that in estimating equations \eqref{2.2}, if $\bm{z}_t$ are constants, then the above equations are not solvable because they are under-identified. Nevertheless, when $L > 2$ and the equations are over-identified, we can employ the two-step generalized method of moments (GMM) to estimate $\gamma_t$. Let $b_{t,j}(r_{pse,t}, \bm{u}_t, y_{pse,t};\gamma_t) = \mathbb{I} (\overline{\bm{R}}_t=\bm{1}_t) l_{t,j}(\bm{z}_t)\left\{\frac{r_{pse,t}}{\widehat{\pi}_t(\bm{u}_t,y_{pse,t}; \gamma_t)} -1 \right\}$ for $j=1,\ldots,L$ and $\bm{B}_t(\gamma_t) = (\widehat{E}[b_{t,1}],\widehat{E}[b_{t,2}],\ldots,\widehat{E}[b_{t,L}] )$. Then, the first-step GMM estimator of $\gamma_t$ is $\widehat{\gamma}^{(1)}_t=\underset{\gamma_t}{\arg\min } \; \bm{B}_t(\gamma_t) \bm{B}_t(\gamma_t)^{\mathrm{T}}$.

Based on the first-step GMM estimator, we can build a more efficient estimator of $\gamma_t$ with a weight matrix $\bm{W}_t$. Let $\bm{W}_{t}$ be the inverse matrix of the $L \times L$ matrix with the $\left(j, j^{\prime}\right)$ element as $\frac{1}{n} \sum_{i=1}^{n} b_{t,j}(r_{pse,t}, \bm{u}_t, y_{pse,t};\widehat{\gamma}^{(1)}_t) b_{t,j'}(r_{pse,t}, \bm{u}_t, y_{pse,t};\widehat{\gamma}^{(1)}_t)$.
Then the second-step GMM estimator of $\gamma_t$ is $\widehat{\gamma}_t = \underset{\gamma_t}{\arg\min }  \; \bm{B}_t(\gamma_t) \bm{W}_{t} \bm{B}_t(\gamma_t)^{\mathrm{T}}$.

With the estimated working model for the missingness probability, we can employ the weighted least square method to obtain consistent estimates of the Q-function parameters. Subsequently, the optimal treatment at stage $t$ can be determined based on the estimated parameters. We henceforth refer to this approach as the estimating equation (EE) method.

\subsection{Weighted Q-learning for DTRs with nonignorable missing pseudo-outcomes: a sensitivity analysis approach}

In real data applications, it might be challenging to find valid nonresponse instrument variables for every stage based on domain knowledge. 
In such scenarios, we develop a sensitivity analysis approach 
for the conditional missingness probability of $Y_{pse,t}$. 
Specifically, we slightly modify Assumption 5 and introduce the following assumption.
\begin{description}
    \item {\em Assumption} 6 (Semiparametric working model for missingness probability  without nonresponse instrumental variables).
    \begin{align} \label{missingmodel2}
        P(R_{pse,t}=1\mid \bm{h}_t, a_t, y_{pse,t},\overline{\bm{R}}_t = \bm{1}_t) =\pi(\bm{h}_t,a_t,y_{pse,t};\gamma_t')= \frac{1}{1+\exp\{s'_t(\bm{h}_t,a_t)+\gamma'_t y_{pse,t}\}},
    \end{align}
    where $s'_t(\cdot)$ is an unknown and unspecified function of $(\bm{h}_t,a_t)$ and $\gamma'_t$ is  the pre-specified sensitivity parameter. Note that here we include all historical variables in~\eqref{missingmodel2}. 
\end{description}



The sensitivity parameter $\gamma'_t$ quantifies the residual effect of $Y_{pse,t}$ on $R_{pse,t}$ after adjusting for $\bm{h}_t$ and $a_t$ for patients with complete covariates up to stage $t$. \cite{Kim_Yu_2011} proposed a similar semiparametric exponential tilting model for nonignorable missing outcomes in cross-sectional settings, where external data were used to estimate their sensitivity parameter. In more general scenarios where external data are unavailable, the specification of sensitivity parameters in  nonignorable missingness models (and  in related problems of sensitivity analysis for unmeasured confounding) remains an active  area of research. In Web Appendix A, we first discuss how to interpret the sign of $\gamma'_t$ using domain knowledge, and then develop a simulation-based approach to calibrating the magnitude of  $\gamma'_t$, building upon the methods by \cite{yin2019simulation}.

After $\gamma_t'$ is specified, we can obtain the kernel estimate of $\exp\{s'_{\gamma'_t,t}(\bm{h}_t,a_t)\}$ in a similar manner as in equation \eqref{2.1}:
\begin{align}
    \exp\{\widehat{s}'_{\gamma'_t,t}(\bm{h}_t,a_t) \} =  \frac{\sum_{i=1}^{n} \left(1-r_{pse,t,i}\right)  K'_{c'_t}\left(\bm{g}_t-\bm{g}_{t,i}\right) \mathbb{I}(\overline{\bm{R}}_{t,i}=\bm{1}_t) }{\sum_{i=1}^{n}  r_{pse,t,i}   \exp \left(\gamma'_t y_{pse,t,i}\right) K'_{c'_t}\left(\bm{g}_t-\bm{g}_{t,i}\right) \mathbb{I} (\overline{\bm{R}}_{t,i}=\bm{1}_t)}, \label{2.3} 
\end{align}
where $\bm{g}_t = (\bm{h}_t,a_t)$ and $K'_{c'_t}(\cdot) = {c'_t}^{-1}K'(\cdot/c'_t)$, with $K'(\cdot)$ being a symmetric kernel function and $c'_t$ a bandwidth.

Subsequently, we can employ inverse probability weighting  to obtain consistent estimates of the Q-functions and the optimal DTRs. If the working model~\eqref{missingmodel2} is correctly specified (including the sensitivity parameter $\gamma'_t$)  and certain regularity conditions for $\exp\{s'_{\gamma'_t,t}(\bm{h}_t)\}$ are satisfied, along with correctly specified Q-functions, the parameter estimators in the Q-functions are consistent. As the true value of  $\gamma'_t$ is unknown, we conduct a sensitivity analysis by varying the value of  $\gamma'_t$ in the determined range to assess its  impact  on the Q-learning results. 
 We refer to this approach as the SA method.

\subsection{Summary of the estimation steps}\label{steps}
We summarize the steps of the proposed weighted Q-learning approach as follows:

\begin{enumerate}
    \item  Compute $y_{pse,T} = f(y_{1},y_{2}, \ldots, y_{T})$. Let $\bm{M}_T(\bm{\theta}_T) =\widehat{E}\left[ \mathbb{I}(\overline{\bm{R}}_{T}=\bm{1}_T) \frac{\partial \left[ \left\{  
  Q_T(\bm{h}_T,a_T;\bm{\theta}_T) - y_{pse,T}\right\} ^2\right]}{\partial \bm{\theta}_T} \right]$. Then obtain $\widehat{\bm{\theta}}_T$ by solving the estimating equation $\bm{M}_T(\bm{\theta}_T) = \bm{0}$
    \item  For $t = T-1,\ldots, 1$, repeat the following steps recursively:
    \begin{enumerate}
        \item Obtain stage $t$ pseudo-outcome estimate as $\widehat{y}_{pse,t} = \max\limits_{a_{t+1}} \; Q_{t+1}(\bm{h}_{t+1},a_{t+1};\widehat{\bm{\theta}}_{t+1})  $, denote the
        missingness indicator of $\widehat{y}_{pse,t}$ as $\widehat{R}_{pse,t}$.
        \item Estimate $P(\widehat{R}_{pse,t}=1\mid \bm{h}_t,a_t, \widehat{y}_{pse,t}, \overline{\bm{R}}_t=\bm{1}_t)$  using the EE or the SA method and  denote the estimated probability as $\widehat{\pi}_t$.
        \item Let $\bm{M}_t(\bm{\theta}_t) = \widehat{E} \left(\frac{\widehat{r}_{pse,t}}{\widehat{\pi}_{t}} \mathbb{I}(\overline{\bm{R}}_{t}=\bm{1}_t) \frac{\partial \left[ \left\{ Q_t(\bm{h}_t,a_t;\bm{\theta}_t) - \widehat{y}_{pse,t}  \right\}^2\right]}{\partial \bm{\theta}_t} \right)$. Solve the estimating equation \\$\bm{M}_t(\bm{\theta}_t) = \bm{0}$ to obtain $\widehat{\bm{\theta}}_t$.
    \end{enumerate}
    \item  Obtain the optimal  regime at stage $t$  by $\widehat{d}_t^{\mathrm{opt}}\left(\bm{h}_t\right) =  \argmax\limits_{a_t}Q_t(\bm{h}_t,a_t;\widehat{\bm{\theta}}_t)$ for $t=1,\ldots,T$. 
\end{enumerate}

{\em Remark 1.} Recall that, since $a_t\in\{-1,1\}$,  $Q_t(\bm{h}_t,a_t;\widehat{\bm{\theta}}_t)$ can be written as $ q_{t,0}(\bm{h}_t; \widehat{\bm{\beta}}_t) + a_t q_{t,1}(\bm{h}_t;\widehat{\bm{\psi}}_t)$. Subsequently,  $\widehat{y}_{pse,t}$ can be calculated by $q_{t+1,0}(\bm{h}_{t+1}; \widehat{\bm{\beta}}_{t+1}) + |q_{t+1,1}(\bm{h}_{t+1};\widehat{\bm{\psi}}_{t+1})|$ and $\widehat{d}_t^{\mathrm{opt}}\left(\bm{h}_t\right)$ is determined by the sign of $q_{t,1}(\bm{h}_t;\widehat{\bm{\psi}}_t)$.

{\em Remark 2.} The pseudo-outcome $Y_{pse,t}$ is determined by the specified parametric form of $Q_{t+1}(\cdot)$ and the true values of $\bm{\theta}_{t+1}$ for $t=1,\ldots,T-1$. Since the  true value of $\bm{\theta}_{t+1}$ is unknown in practice, we use $\widehat{Y}_{pse,t}$ calculated by the parameter estimates $\widehat{\bm{\theta}}_{t+1}$ in  step (2). In this scenario,  $\widehat{Y}_{pse,t}$ differs slightly from $Y_{pse,t}$, which leads $\widehat{\pi}_t$ to differ slightly from $\pi(\bm{u}_t,y_{pse,t})$ and $\pi(\bm{h}_t,a_t,y_{pse,t})$, when the EE and SA methods are used respectively. Nevertheless, if all the models are correctly specified, $\widehat{\bm{\theta}}_{t+1}$ serves as a consistent estimator for $\bm{\theta}_{t+1}$. Subsequently, $\widehat{Y}_{pse,t}$  converges to $Y_{pse,t}$. It follows that the estimated optimal DTR $\widehat{\bm{d}}$ will also converge to the true optimal DTR $\bm{d}$. We formalize these asymptotic properties in the following section. 

Remark 3.
$Y_{pse,t}=\max_{a_{t+1}\in \{-1,1\}} Q_{t+1}(\bm{H}_{t+1},a_{t+1})$ implies that 
the missingness of $Y_{pse,t}$ is determined by the model for $Q_{t+1}(\cdot)$ and the missingness of $\bm{H}_{t+1}$.
If all members of $\bm{H}_{t+1}$ are employed in the calculation of $Y_{pse,t}$, then $R_{pse,t} = \mathbb{I}(\overline{\bm{R}}_{t+1} = \bm{1}_{t+1} )$. Therefore, the missingness of covariates at stage $t+2, \ldots, T$ does not directly affect $R_{pse,t}$. Nevertheless, if nonignorable missing covariates in the later stages are not handled properly, it not only directly impacts the estimation at those stages but also has an influence on earlier stages. This is because bias would propagate and accumulate during the backward-induction procedure through the pseudo-outcomes calculated with biased Q-functions. 
Therefore, in this scenario increasing the total number of stages could lead to worse performance of the estimated optimal DTRs. 
We investigate this issue further in Simulation 4  in Web Appendix H.

We provide  R Markdown tutorials for implementing the proposed weighted Q-learning approaches  using simulated data in the Supplementary Materials. In Web Appendices B-E, we establish the consistency and asymptotic normality of the estimators in Section~\ref{steps} in the 2-stage scenario.
The form of the asymptotic covariance matrix of the estimated parameters at stage 1 is complicated. 
Therefore, following \cite{Shao_Wang_2016}, we suggest using the bootstrap method to estimate variances and construct confidence intervals of the parameters in practice. Additionally, when 
non-regularity of stage 1 parameter estimators is of concern, the $m$-out-of-$n$ bootstrap should be used for the inference (see details in Web Appendix G).

\section{Simulation}

We conducted four simulation studies to evaluate the finite-sample performance of the proposed weighted Q-learning method with conditional missingness probabilities of pseudo-outcomes estimated by  the EE or the SA method, which we denote as the WQ-EE method or the WQ-SA method, respectively. 
In Simulation 1, we focused on the scenario where Assumptions 1-5 were satisfied. In Simulation 2, we evaluated the robustness of the WQ-EE method against the violations of Assumption 5 when (1) the chosen instrumental variable was weakly associated with the missingness indicator and (2)  there were interactions between $\bm{u}_t$  and $y_{pse,t}$ in the true conditional missingness probability model. 
In Simulation 3, we demonstrated our simulation-based approach to calibrating the  sensitivity parameter and the performance of the WQ-SA method based on the specified sensitivity parameter values. In Simulation 4, we considered a 3-stage setting where the data generation mechanism  and  the true Q-functions in the first two stages were the same as those in Simulation 1. By comparing the results in Simulations 1 and 4, we investigated the impact of an increased number of total stages on the performance of the proposed methods.
For comparison, we also assessed three other methods for handling partially missing covariate data in Q-learning: a naive method which ignored partially observed covariates (`naive'), complete case analysis (`CC'), and multiple imputation (`MI') \citep{Little_Rubin_2014}. Q-learning when all the covariates are fully observed (`All') was used as the benchmark. More details 
can be found in Web Appendix H.

All methods were evaluated in two aspects: Firstly, we compared their performances for estimating the blip function parameters $\bm{\psi}_t$. Secondly, we assessed their abilities to identify the true optimal DTR and compared their values of the estimated optimal DTRs. In Simulation 1, the WQ-EE method and CC analysis provided consistent estimates of blip function parameters at stage 2 (i.e., the final stage) due to the future-independent missingness assumption being satisfied and fully observed pseudo-outcomes. In contrast, naive and MI methods resulted in significant biases at this stage due to the misspecified Q-function models that excluded partially missing covariates in the naive method and the violation of ignorable missingness assumption required by the MI method, respectively.
At stage 1, WQ-EE estimators remained consistent, while the CC method showed substantial biases that did not decrease with increasing sample size. This was because, at stage 1, not only were there nonignorable missing covariates, but the pseudo-outcomes were also nonignorably missing.
The WQ-EE method demonstrated the highest proportion of correctly identifying the optimal treatments in both stages. In contrast, the DTRs obtained from CC and MI methods had much lower correct classification rates, and the mean final outcomes under the estimated DTRs were lower than that from the WQ-EE method. In Simulation 2, the WQ-EE method demonstrated robustness to mild model misspecification and consistently outperformed the other three methods, yielding higher mean final outcomes and correct classification rates when Assumption 5 was violated. 
In Simulation 3, the proposed simulation-based approach determined plausible ranges for the sensitivity parameter $\gamma_1'$ that covered the true values of $\gamma_1'$ across simulations, and the WQ-SA method with varied values of $\gamma_1'$ was robust and outperformed the naive, CC, and MI methods. In Simulation 4, 
 we compared the values and  overall correct classification rates of the estimated DTRs from Simulations 1 and 4. These performance metrics decreased with the increased number of  total stages,  confirming that the impact of nonignorable missing covariates  propagated over time.

\section{Application to the MIMIC-III data}

We applied the proposed methods to the MIMIC-III data \citep{Johnson_et_al_2016} for investigating the optimal 2-stage fluid strategy to treat sepsis patients in ICUs. 

Sepsis is a life-threatening syndrome caused by the body’s response to infection, resulting in damage to its own tissues and organs. Timely and effective fluid resuscitation is crucial for stabilizing tissue hypoperfusion or septic shock in sepsis patients. The `Surviving Sepsis Campaign' guideline strongly recommends administrating at least 30 mL/kg of intravenous (IV) fluid within the first 3 hours of ICU admission \citep{Rhodes_et_al_2017}.  To address the lack of robust evidence for the optimal fluid resuscitation strategy in the early hours of treatment, \cite{Speth_et_al_2022} utilized MIMIC-III data to assess the optimal DTR during the 0-3 hours and 3-24 hours following admission to the medical ICU. Building upon \cite{Speth_et_al_2022} and in line with the recommendation of the `Surviving Sepsis Campaign' that additional fluid administration after initial resuscitation should be guided by reassessing the patient's hemodynamic status, we incorporated hemodynamic variables in determining the optimal fluid resuscitation strategy during the 3-24 hours period after ICU admission (stage 2). 

We followed similar patient selection criteria as outlined by \cite{Speth_et_al_2022}, focusing on adult septic patients admitted to the medical ICU after initially presenting to the emergency department. The detailed cohort eligibility criteria can be found in Web Appendix I. The treatments in both stages, the baseline covariates, and the final outcome were defined according to \cite{Speth_et_al_2022}. Specifically, the stage 1 and stage 2 treatments were categorized as either fluid restrictive ($<$30 ml/kg) or fluid liberal ($\geq$ 30 ml/kg) strategies in 0-3 hours and 3-24 hours post-ICU admission, respectively. The baseline covariates included gender, age, weight, racial groups, and Elixhauser comorbidity score. The outcome of interest was the Sequential Organ Failure Assessment (SOFA) score evaluated at 24 hours post-admission. The SOFA score is a clinical tool used to assess the severity of organ dysfunction by assigning scores to various organ systems. A higher SOFA score indicates a more severe impairment. In our analysis, we used the negative value of the SOFA score at 24 hours post-admission as the final outcome to be maximized by the optimal fluid resuscitation strategy.

For the intermediate variables considered before stage 2 treatment, we included mechanical ventilation and vasopressors within the first 3-hour period, as well as the patients' SOFA score evaluated at 3 hours post-admission. Additionally, we also considered heart rate, blood pressure, Saturation of Peripheral Oxygen (SpO2), respiratory rate, temperature at 3 hours post-admission and urine output within 0-3 hours after admission. These hemodynamic variables contained missing values, which were likely to be nonignorable due to the informative monitoring of the patients in the MIMIC-III database, as discussed previously.
The missing proportions of the hemodynamic variables are provided in Web Table 10.

There were 973 patients in the selected cohort, among them $53.2\%$ were male and $78.7\%$ were Caucasian. 45.5\% of the patients received the fluid liberal strategy in 0-3 hours post-admission while $53.5\%$ received the fluid liberal strategy in 3-24 hours post-admission. Overall, 67.1\% of patients had fully observed covariates at stage 2, while the rest of the patients had at least one covariate with missing values.

As discussed in Section 2.3, it was plausible that Assumption 4 holds, that is, the missing data mechanism of the hemodynamic variables at 3 hours post-admission was unrelated to the patient’s SOFA score at 24 hours post-admission, conditioning on the patient's medical history and treatments up to stage 2. Denote the negative of SOFA score at 24 hours post-admission under the optimal stage 2 treatment as $Y_{pse,1}$. It was plausible that weight and gender were independent of the missingness of hemodynamic variables given baseline covariates and $Y_{pse,1}$. Since weight and gender were associated with $Y_{pse,1}$,  they were suitable candidates for nonresponse instrumental variables. Additionally, the Elixhauser comorbidity score can be regarded as a mismeasured proxy of $Y_{pse,1}$, which may also serve as a candidate for the nonresponse instrumental variable,  according to \cite{Miao_Tchetgen_2018}. We investigated the impact of using different sets of candidates for nonresponse instrumental variables in Table~\ref{RDAtable1}. Besides, we applied CC, MI, and the naive method that ignored hemodynamic variables for comparison. The improvement of SOFA at 24 hours was then calculated as the difference between the means of the observed  SOFA scores at 24 hours and the expected  SOFA scores at 24 hours under the estimated optimal DTR. 
We performed cross-validation and $m$-out-of-$n$ bootstrap to assess the performance of the estimated optimal DTRs. More details of the estimation procedure are given in Web Appendix I.

\begin{table}[htbp]
\caption{Improvement of SOFA at 24 hours post-admission for sepsis patients in the selected cohort of  the MIMIC-III database under the estimated optimal fluid strategies with different approaches to handling nonignorable missing covariates.}\label{RDAtable1}
  \centering
  \begin{threeparttable}
    \begin{tabular}{llccc}
    \hline
     & Mean & $m$ & Bootstrap std & Bootstrap 95\% CI \\
    \hline 
    WQ-EE (ECS)    &  1.297  & 823 & 0.186 & (0.929, 1.667) \\
    WQ-EE (weight)   & 1.432  & 823 & 0.179 &  (1.083, 1.783) \\
    WQ-EE (ECS, weight)   & 1.338 & 823 & 0.181 & (0.976, 1.681) \\
    WQ-EE (gender)   & 1.430 & 823 &  0.210 & (1.012, 1.841) \\
    WQ-EE (gender, ECS)   & 1.338  & 823 & 0.194 &  (0.930, 1.696)\\
    WQ-EE (gender, weight)    & 1.459 & 823 & 0.193 & (1.083, 1.833) \\
    WQ-EE (gender, weight, ECS)    & 1.345 & 823 & 0.197 & (0.949, 1.730)  \\
    CC &  1.190  & 823 &  0.168  &  (0.861, 1.523)     \\
    MI &  1.092  & 827 &  0.123  &  (0.853, 1.340)     \\
    naive &  0.909 & 831 & 0.116   & (0.685, 1.135)  \\
    \hline
    \end{tabular}%
\begin{tablenotes}
    \small
    \item Note: 
    Mean, the average of estimated improvement  of SOFA based on cross-validation using 1000 random splits;
    $m$, the  resample size in the $m$-out-of-$n$ bootstrap selected using the double bootstrap method described in Web Appendix G;
    Bootstrap Std, the standard deviation of the $m$-out-of-$n$ bootstrap estimates;
    Bootstrap 95\% CI, 95\% confidence interval constructed by $m$-out-of-$n$ bootstrap percentiles;
    ECS, Elixhauser comorbidity score; WQ-EE ($\cdot$), the WQ-EE method based on the corresponding nonresponse instrumental variables listed in parenthesis.
\end{tablenotes}
    \end{threeparttable}
\end{table}%

Table \ref{RDAtable1} presents the estimated mean, standard error, and 95\% confidence interval of the improvement of SOFA score at 24 hours post-admission. Notably, all three methods that considered hemodynamic variables outperformed the naive method, which aligns with the recommendation from the Surviving Sepsis Campaign.
Furthermore, the estimated improvement of SOFA at 24 hours post-admission based on the WQ-EE method were notably larger than those based on the CC and MI methods. Importantly, while the WQ-EE method exhibited robustness to the choice of the three candidate nonresponse instrumental variables, the estimated improvement of SOFA at 24 hours post-admission would be lower if the Elixhauser comorbidity score was included in the set of  nonresponse instrumental variables. This could be attributed to the association between the Elixhauser comorbidity score and the missingness of hemodynamic variables. In Simulation 2, we have demonstrated that this association can affect the performance of the WQ-EE method. Therefore, we recommend excluding the Elixhauser comorbidity score from the set of nonresponse instrumental variables. The remaining three combinations of nonresponse instrumental variables showed similar means and standard errors for the estimated improvement of the final outcome. 

However, we cannot rule out that gender and weight could be weakly associated with the missingness of hemodynamic variables, even after conditioning on other covariates and the pseudo-outcome. Therefore, we also applied the proposed  WQ-SA method  in case none of the aforementioned variables were valid nonresponse instrument variables. We assumed $\gamma'_1 \geq 0$ because patients with higher SOFA scores (i.e., lower final outcomes) were more severely ill, and they may undergo more intensive monitoring of their hemodynamic status. 
To calibrate the magnitude of $\gamma'_1$, we applied our simulation-based approach described in Web Appendix A.     Table \ref{RDA_sen_spec} presented the medians of 1000   Wilcoxon rank-sum test p-values for checking the similarity between the `observed' pseudo-outcomes estimates and the \textit{replications} of `observed' pseudo-outcomes simulated under the working missingness model ~\eqref{missingmodel2} with a specified value of $\gamma'_1$. A large median of the p-values indicated that 
the replications of  observed pseudo-outcomes were similar to the observed  pseudo-outcome estimates in the MIMIC-III data, thus 
the corresponding value of $\gamma'_1$ was considered to be plausible. From Table \ref{RDA_sen_spec}, we can  see that the medians of   the p-values were larger than 0.05 when $\gamma_t'$ lied in the range of $[0,1]$ and it was smaller than 0.05 when $\gamma_t' \geq 1.25$. For the sake of prudence, we therefore considered a range of values for $\gamma'_1$ within the interval $[0, 1.25]$.
\begin{table}[htbp]
    \centering
    \caption{Results for the simulation-based approach to determining  a plausible range of the sensitivity parameter for the WQ-SA method  in the  MIMIC-III data application. }
    \begin{tabular}{lccccccccc}
    \hline
   $\gamma_1'$  & 0 & 0.25 & 0.5 & 0.75 & 1 & 1.25 & 1.5  & 1.75  \\
    \hline
     Median of  p-values & 0.207 & 0.230 & 0.190 & 0.129 & 0.063 & 0.039 & 0.021 & 0.016  \\
    \hline
    \end{tabular}
    \label{RDA_sen_spec}
\end{table}

Table \ref{RDA_sen} shows that when $\gamma'_1=0$, the estimated improvement in SOFA at 24 hours post-admission was nearly identical to that of the CC method, which also assumed the future-independent missingness assumption and ignored the relationship between $Y_{pse,1}$ and $R_{pse,1}$. The estimated improvement in SOFA at 24 hours post-admission increased rapidly as $\gamma'_1$ changed from $0$ to $0.75$. When $\gamma'_1$ changes from $0.75$ to $1.25$, the estimated improvement in SOFA at 24 hours post-admission stabilized at around 1.47.   
\begin{table}[htbp]
\centering
  \caption{The improvement of SOFA at 24 hours post-admission for sepsis patients in the selected cohort of  the MIMIC-III database under the estimated optimal fluid strategies  with different values of the  sensitivity parameter  $\gamma_1'$ in the WQ-SA method}
  \centering
  \begin{threeparttable}
  \label{RDA_sen}
    \centering
    \begin{tabular}{rcccc}
    \hline
    $\gamma'_1$ & Mean & $m$ &  Bootstrap std & Bootstrap 95\% CI \\
    \hline
    0 & 1.184  & 823 & 0.178 & (0.837, 1.534) \\
    0.25 & 1.288  & 823 &  0.181 & (0.930, 1.642) \\
    0.5 & 1.369  & 823 & 0.185 & (1.003, 1.726) \\
    0.75 & 1.436  & 823 & 0.198 & (1.044, 1.821) \\
    1 & 1.475  & 823 & 0.201 & (1.082, 1.873) \\
    1.25 & 1.486  & 823 & 0.204 & (1.085, 1.887) \\
    \hline 
    \end{tabular} 
    \begin{tablenotes}
    \small
    \item Note: 
    Mean, the average of estimated improvement  of SOFA based on cross-validation using 1000 random splits;
    $m$, the  resample size in the $m$-out-of-$n$ bootstrap selected with the double bootstrap method described in Web Appendix G;
    Bootstrap Std, the  standard deviation of the $m$-out-of-$n$ bootstrap estimates;
    Bootstrap 95\% CI, 95\% confidence interval constructed by $m$-out-of-$n$ bootstrap percentiles.
    \end{tablenotes}
\end{threeparttable}   
\end{table}%


While the challenge of optimizing the management of septic patients is intricate and multifaceted, we employed the proposed methods to investigate whether incorporating partially missing hemodynamic variables could influence the fluid administration after the initial resuscitation to enhance the overall outcome of the septic patients. In line with the best practice recommendations of the Surviving Sepsis Campaign, our analysis results indicated that certain hemodynamic variables should be taken into account when formulating the fluid resuscitation strategy during the 3-24 hours following admission to the medical ICU because accounting for these nonignorable missing covariates with the proposed methods increased the estimated improvement of SOFA at 24 hours post-admission.

\section{Discussion}

In this article, we proposed two weighted Q-learning approaches to estimating optimal DTRs with nonignorable missing covariates and consequent nonignorable missing pseudo-outcomes. To the best of our knowledge, this is the first attempt to estimate optimal DTRs in these challenging nonignorable missingness scenarios. Specifically, we utilized the future-independent missingness assumption and developed a semi-parametric estimating equation-based approach to estimating the conditional missingness probability of the pseudo-outcomes with valid nonresponse instruments. After obtaining the estimated conditional  missingness probabilities, we applied inverse probability weighting to estimate the Q-function parameters. When nonresponse instrumental variables were not available, we developed a practical  approach to calibrating the sensitivity parameter in the WQ-SA method and conducted a sensitivity analysis to assess the impact of varied sensitivity parameter values on the estimated optimal DTRs.  
Our simulation results showed that the WQ-EE and WQ-SA estimators were consistent when their corresponding assumptions hold. They were also robust and outperformed the competing methods under mild model misspecifications. 
In the application to the MIMIC-III data,  we showed that incorporating hemodynamic variables into optimal fluid strategy estimation and handling nonignorable missing covariates with the proposed methods can improve the SOFA score for sepsis patients in ICUs.

In practice, the ignorable missingness assumption is unverifiable from observed data. Therefore, when researchers are not certain about missingness assumptions based on substantive knowledge,  we suggest using the proposed WQ-EE or WQ-SA method to accommodate possible departures from ignorable missingness and  the uncertainty brought by non-ignorable missingness when estimating optimal DTRs.  Notably, the proposed methods are still valid under ignorable missingness (i.e., $\gamma_t=0$ in Assumption 5, $\gamma'_t=0$ in Assumption 6).

There are several directions for future work. First, this article exclusively addresses the issue of missing pseudo-outcomes due to MNAR covariates in optimal DTR estimation with Q-learning. However, in observational studies or EMR databases, patients may not progress through all stages of clinical intervention due to experiencing a terminal event or being censored before the end of the study. In such cases, it is desirable to integrate the terminal event and longitudinal outcome measurements for estimating optimal DTRs while accounting for the censoring and MNAR covariates. Second, the dimension of the covariates is often high in practice, especially in multiple-stage settings. In such scenarios, the kernel regression method is susceptible to the `curse of dimensionality' problem. To address this challenge, one may consider dimension reduction techniques proposed in \cite{Tang_Zhao_Zhu_2014}. 
Third, our methods utilize inverse probability weighting on the  sample with complete covariates at different stages, whose size can be small  when the majority of patients have at least one covariate with missing values. 
In this case, we could explore the multiple imputation method under nonignorable missingness \citep{tompsett2018use} as a more practical alternative.  
Last, the proposed approach can be extended to other  DTRs methods, such as A-learning.

\vspace{-8pt}
\bibliographystyle{biom} 
\bibliography{DTR_MNAR_bib}

\end{document}


\maketitle

\section{Web Appendix A: Calibrating the sensitivity parameter in the WQ-SA method}

Although the importance of sensitivity analysis with respect to missing data has been recognized in research and  regulatory agencies' guidelines for analysis and reporting of clinical trials \citep{little2012prevention,european2010guideline,european2020CHMP},
calibrating sensitivity parameters in  nonignorable missingness models (and  in related problems of sensitivity analysis for unmeasured confounding) is still a challenging task and an active  area of research \citep{yin2019simulation, franks2019flexible, Sjolander2022}. We first discuss how to determine the sign of the sensitivity parameter using domain knowledge. Then we describe how to calibrate the magnitude of the sensitivity parameter in the WQ-SA methods. 

\subsection{Determining the sign of the sensitivity parameter}

 In practice, prior knowledge and domain expertise often can  be used to reason about some  characteristics of the missing data mechanism and determine the sign of   sensitivity parameters. In the working model (4) of the main text for the WQ-SA method, the sign of the sensitivity parameter $\gamma_t'$ has a relatively straightforward interpretation: for fixed $\bm{h}_t$ and $a_t$, if $\gamma_t' >0$, then a smaller value of $y_{pse,t}$ leads to a smaller missingness probability, and vice versa. For example, in the MIMIC-III data, given the covariates and treatments up to stage $t$ (i.e., $\bm{h}_t$ and $a_t$), patients with worse disease conditions (i.e., smaller values of $y_{pse,t}$) were more likely to be monitored frequently and  less likely to have missing values of $y_{pse,t}$. Therefore, it is reasonable to assume that the sensitivity parameter $\gamma_t'$ is positive in this scenario.

\subsection{The simulation-based sensitivity analysis approach by Yin and Shi (2019)}
We now turn to the more challenging task of specifying the plausible magnitude of the sensitivity parameters. The research on specifying the sensitivity parameters in nonignorable missingness models   is relatively sparse in the literature. \cite{yin2019simulation} developed a simulation-based  approach to searching for plausible values of the sensitivity parameters in  logistic models for nonignorable missingness probabilities. This simulation-based method was initially introduced by \cite{gelman1996posterior} for assessing the goodness-of-fit with a Bayesian posterior predictive model. It was further employed for model checking with missing and latent data \citep{gelman2005multiple,copas2000meta,copas2005local,daniels2012bayesian}. 
Using K-nearest-neighbor (KNN) distance metrics, \cite{yin2019simulation} compared the similarity between the \textit{observed} outcome  data with \textit{{replications}} of \textit{{observed outcomes}} simulated from the estimated full data distribution and the estimated  missingness probabilities with fixed values of their sensitivity parameters. If the KNN distance metrics had small values, it indicated that the estimated full data distribution and the missingness probability model (including the specified values of the sensitivity parameters) were \textit{compatible} with the observed outcome data, thus the specified sensitivity parameter values were considered to be plausible.  
Inspired by their methods, we propose a simulation-based approach to determining the plausible range of our sensitivity parameters in the WQ-SA method.


\subsection{Distinctions between the nonignorable missing outcome settings in Yin and Shi (2019)  and the setting with nonignorable missing pseudo-outcomes}
Before we proceed to describe the proposed approach, we emphasize some distinctions between the nonignorable missing outcome settings examined by \cite{yin2019simulation} and our DTR setting with nonignorable missing pseudo-outcomes.  In typical nonignorable missing outcome settings in \cite{yin2019simulation}, the observed outcomes are directly available from the raw data. However, due to the potential outcome framework used in the DTR setting, the  pseudo-outcomes before the final stage are conditional expectations of  potential outcomes, thus not directly available from the raw data. Rather, we use estimated Q-functions to obtain estimates of the pseudo-outcomes in the backward-induction procedure. However, due to the nonignorable missing covariates in future stages, the estimates of pseudo-outcomes at the current  stage are partially missing as well and we need to apply missing data methods to address the selection bias from this missingness.  In analogy to the typical non-ignorable outcome settings, we can treat the `observed' pseudo-outcome \textit{estimates} as the \textit{`raw observed outcomes'}  in the DTR setting and  compare their distribution with the distribution of the \textit{replications} of  `\textit{observed}' pseudo-outcomes  in our simulated-based approach. This is reasonable because,  as discussed in Section 2.6 of the main text, the pseudo-outcome estimates converge to the pseudo-outcomes when all the models for the Q functions and the missingness probabilities are correctly specified. When describing  the implementation steps in the next section, we still employ the pseudo-outcome notation and  their models as described in the main text.

\subsection{A simulation-based approach to calibrating the magnitude of the sensitivity parameter}
Let $Y_{pse,t}^{obs}$ and $Y_{pse,t}^{mis}$ denote the observed and unobserved values of $Y_{pse,t}$, respectively, in the sample with complete covariates until stage $t$ (i.e., the patients with $\overline{\bm{R}}_t = \bm{1}_t$). Similarly, let $\widehat{Y}_{pse,t}^{obs}$ denote the observed values of $\widehat{Y}_{pse,t}$ in the sample with $\overline{\bm{R}}_t = \bm{1}_t$.  We propose to  evaluate the plausibility of  each pre-specified value of the sensitivity parameter $\gamma_t'$ by comparing the distribution of the \textit{observed} pseudo-outcome estimates $\widehat{Y}_{pse,t}^{obs}$ with that of $Y_{pse,t}^{obs,\gamma_t'}$, which are the \textit{replications}  of \textit{observed} pseudo-outcomes simulated based on the estimated Q-function at stage $t$ and the estimated working missingness model (4) corresponding to   a specified value of $\gamma_t'$.

Let $\pr(\cdot)$ and $\pr( \cdot\mid\cdot )$ denote the probability density function and the conditional probability density function, respectively. To simulate data with a specified value of   $\gamma_t'$, we first evaluate the density $\pr(y_{pse,t}\mid \bm{h}_t,a_t,\overline{\bm{R}}_t = \bm{1}_t,R_{pse,t}=0)$, which can be based on  the following Lemma. 
\begin{lemma}
    In the sample with complete covariates until stage $t$ (i.e., $\overline{\bm{R}}_t = \bm{1}_t$), the conditional density of the missing part of $Y_{pse,t}$ (i.e., with $R_{pse,t}=0$) can be expressed by
    \begin{align}
    \frac{\pr(y_{pse,t}\mid \bm{h}_t,a_t,\overline{\bm{R}}_t = \bm{1}_t,R_{pse,t}=0)}{\pr(y_{pse,t}\mid \bm{h}_t,a_t,\overline{\bm{R}}_t = \bm{1}_t,R_{pse,t}=1)} = \frac{\exp(\gamma'_t y_{pse,t})}{E\{\exp(\gamma'_t y_{pse,t}) \mid \bm{h}_t,a_t,\overline{\bm{R}}_t = \bm{1}_t, R_{pse,t}=1\}}. \label{pseudo-obs-mis}
\end{align}
\end{lemma}

In equation~\eqref{pseudo-obs-mis}, $\pr(y_{pse,t}\mid \bm{h}_t,a_t,\overline{\bm{R}}_t = \bm{1}_t,R_{pse,t}=1)$ and $E\{\exp(\gamma'_t y_{pse,t}) \mid \bm{h}_t,a_t,\overline{\bm{R}}_t = \bm{1}_t, R_{pse,t}=1\}$ can be estimated by some nonparametric methods such as kernel regressions based on $\widehat{Y}_{pse,t}^{obs}$. 
Then for any given $\gamma_t'$, we can generate samples from  the estimated $\pr(y_{pse,t}\mid \bm{h}_t,a_t,\overline{\bm{R}}_t = \bm{1}_t,R_{pse,t}=0)$   and impute the missing values in $Y_{pse,t}$ for the patients with complete covariates until stage $t$ (i.e., $\overline{\bm{R}}_t=\bm{1}_t$). We denote these imputed  pseudo-outcomes as $Y_{pse,t}^{\gamma_t',imp}$.
Thus we have complete data ${Y}_{pse,t}^{\gamma_t',*} = (\widehat{Y}_{pse,t}^{obs}, Y_{pse,t}^{\gamma_t',imp})$ for the patients with complete covariates until stage $t$.
Under the future-independence missingness assumption (Assumption 4), we have $E(Y_{pse,t}\mid \bm{h}_t,a_t) = E(Y_{pse,t}\mid \bm{h}_t,a_t,\overline{\bm{R}}_t=\bm{1}_t)$, that is, the Q function model applies to the patients with $\overline{\bm{R}}_t=\bm{1}_t$.  Therefore, we  fit   the assumed Q-function model to ${Y}_{pse,t}^{\gamma_t',*}$ and   estimate $E(Y_{pse,t}\mid \bm{h}_t,a_t,\overline{\bm{R}}_t=\bm{1}_t)$.
The density of the residuals of  this conditional mean  model can also be estimated by nonparametric methods. Combining the estimated conditional mean model and the estimated density of the residuals,  we can obtain the conditional distribution $\pr(y_{pse,t}^{\gamma_t',*}\mid \bm{h}_t,a_t,\overline{\bm{R}}_t=\bm{1}_t)$.  

Next, we take a sample  $Y_{pse,t}^{\gamma_t',**}$ from the estimated $\pr(y_{pse,t}^{\gamma_t',*}\mid \bm{h}_t,a_t, \overline{\bm{R}}_t=\bm{1}_t)$  for the patients with complete covariates until stage $t$ (i.e., $\overline{\bm{R}}_t=\bm{1}_t$).
The missingness indicator $R_{pse,t}^{\gamma_t',*}$ given $Y_{pse,t}^{\gamma_t',**}$ and $\gamma_t'$ would be generated  with the probability $\widehat{\pi}(\bm{h}_t,a_t,y_{pse,t}^{\gamma_t',**};\gamma_t') = \left[1+\exp\left\{\widehat{s}'_{\gamma'_t,t}(\bm{h}_t,a_t) +\gamma'_t y_{pse,t}^{\gamma_t',**}\right\}\right]^{-1}$, where $\widehat{s}'_{\gamma'_t,t}(\bm{h}_t,a_t)$ is estimated by equation (5) in the main text. Drop the values of $Y_{pse,t}^{\gamma_t',**}$ for those with $R_{pse,t}^{\gamma_t',*}=0$ and denote the remaining part of $Y_{pse,t}^{\gamma_t',**}$ as $Y_{pse,t}^{\gamma_t',obs}$, which we refer to as the replications of  observed pseudo-outcomes.
Finally, we compare the distribution of $Y_{pse,t}^{\gamma_t',obs}$ with that of $\widehat{Y}_{pse,t}^{obs}$ using the Wilcoxon rank-sum test.  A large  p-value of the Wilcoxon rank-sum test indicates that the replications of  observed pseudo-outcomes are similar to the observed pseudo-outcome estimates $\widehat{Y}_{pse,t}^{obs}$. Therefore, the specified value of $\gamma_t'$ is considered to be plausible.

For each candidate value of $\gamma_t'$, we formalize the proposed simulation-based approach in the following steps:
\begin{enumerate}
    \item Obtain $\exp\{\widehat{s}'_{\gamma'_t,t}(\bm{h}_t,a_t) \}$ using equation (5) in the main text. 
    \item Based on the observed pseudo-outcome estimates $\widehat{Y}_{pse,t}^{obs}$, estimate $\pr(y_{pse,t}\mid \bm{h}_t,a_t,\overline{\bm{R}}_t = \bm{1}_t, R_{pse,t}=1)$ and $E\{\exp(\gamma'_t y_{pse,t}) \mid \bm{h}_t,a_t,\overline{\bm{R}}_t = \bm{1}_t, R_{pse,t}=1\}$ nonparametrically. Afterwards, we can estimate $\pr(y_{pse,t}\mid \bm{h}_t,a_t,\overline{\bm{R}}_t = \bm{1}_t,R_{pse,t}=0)$ using equation~\eqref{pseudo-obs-mis}.
    \item Impute $Y_{pse,t}^{\gamma_t',imp}$ for the patients with $(\overline{\bm{R}}_t = \bm{1}_t,R_{pse,t}=0)$  by simulating from the estimated $\pr(y_{pse,t}\mid \bm{h}_t,a_t,\overline{\bm{R}}_t = \bm{1}_t,R_{pse,t}=0)$. Denote $Y_{pse,t}^{\gamma_t',*} = (\widehat{Y}_{pse,t}^{obs}, Y_{pse,t}^{\gamma_t',imp})$.
    \item Based on $Y_{pse,t}^{\gamma_t',*}$, estimate $\pr(y_{pse,t}^{\gamma_t',*}\mid \bm{h}_t,a_t,\overline{\bm{R}}_t=\bm{1}_t)$ with the assumed Q-function model and the nonparametric residual density estimator.
    \item Generate $Y_{pse,t}^{\gamma_t',**}$ from the estimated  $\pr(y_{pse,t}^{\gamma_t',*}\mid \bm{h}_t,a_t, \overline{\bm{R}}_t=\bm{1}_t)$ for the patients with $\overline{\bm{R}}_t=\bm{1}_t$. Generate $R_{pse,t}^{\gamma_t',*}$ from $\widehat{\pi}(\bm{h}_t,a_t,y_{pse,t}^{\gamma_t',**};\gamma_t')$. Let $Y_{pse,t}^{\gamma_t',obs}$ denote  $Y_{pse,t}^{\gamma_t',**}$ with $R_{pse,t}^{\gamma_t',*}=1$.
    \item Compare the distribution of $Y_{pse,t}^{\gamma_t',obs}$ and $\widehat{Y}_{pse,t}^{obs}$ using the Wilcoxon rank-sum test.
    \item Repeat Steps (3)-(6) for many Monte Carlo replicates, report the median of the p-values. Let MCR denote the number of Monte Carlo replicates.
\end{enumerate}

  {\em Remark S1.} In the above procedure, Steps (1)-(5) corresponds to Step (ii) in Section 2.2 of \cite{yin2019simulation} but they have been tailored to our proposed WQ-SA method. It involves the specifications for the sensitivity parameter values, the parametric model for the Q-functions, and the exponential tilting form of the working missingness model. Hence, the similarity between $Y_{pse,t}^{\gamma_t',obs}$ and $\widehat{Y}_{pse,t}^{obs}$ can be used to assess both the plausibility of $\gamma_t'$ and the assumed models in the WQ-SA method.     

 {\em Remark S2.} In Step (6) of the above procedure, we evaluated the similarity between $Y_{pse,t}^{\gamma_t',obs}$ and $\widehat{Y}_{pse,t}^{obs}$ using the Wilcoxon rank-sum test. The Wilcoxon rank-sum test, which is also known as the Mann–Whitney $U$ test, is widely used to determine if 
 a two group sample follows the same distribution. It is a nonparametric test that does not depend upon any particular distributional form (or parameters). As suggested by  Remark 3 in \cite{yin2019simulation}, various methods can be employed to measure the similarity between $Y_{pse,t}^{\gamma_t',obs}$ and $\widehat{Y}_{pse,t}^{obs}$. The Wilcoxon rank-sum test performed well in Simulation 3 in Web Appendix G and in  the MIMIC-III data example,  and it can offer an interpretable measure of the similarity between $Y_{pse,t}^{\gamma_t',obs}$ and $\widehat{Y}_{pse,t}^{obs}$. In contrast,  the KNN distance was computationally expensive, and results relied on tuning parameters in our numerical studies. Therefore, we chose  the Wilcoxon rank-sum test as the similarity measure. 
 
 {\em Remark S3.} To reduce the impact of random errors on the p-values caused by the generation of $Y_{pse,t}^{\gamma_t',imp}$ and $Y_{pse,t}^{\gamma_t',**}$, we repeated Steps (3)-(6) for MCR times. We reported the median of the resulting p-values since they had a skewed distribution.   


 {\em Remark S4.} In analogy to the conventional significance level $0.05$ for hypothesis testing based on p-values, we considered a threshold of the median p-value  0.05 to determine a plausible range in Simulation 3 and MIMIC-III data analysis. Specifically, a value of $\gamma_t'$  would be included in the plausible range if its corresponding  median p-value was greater than 0.05.

 {\em Remark S5.} The simulation-based approach is for determining a plausible range of the sensitivity parameter, rather than serving as a method for obtaining its point estimate. Following the discussion in \cite{yin2019simulation}, we recommend using a set of sensitivity parameters within a range determined by a tolerance level of p-value medians,  instead of a single value of sensitivity parameter,  in the WQ-SA method.

We evaluated the above approach in Simulation 3 in Web Appendix G and applied it to the MIMIC-III data analysis.

\section{Web Appendix B: Asymptotic properties}

The consistency and asymptotic normality of the estimators in Section 2.6 of the main text  can be established under suitable regularity conditions for estimating equations and an additional assumption that the optimal treatment is unique for all subjects at all stages. Note that the parameters are estimated separately using estimating equations recursively at each stage. Therefore, it is natural to establish the asymptotic properties of the stage-specific parameters recursively as well. For simplicity, we will focus on the two-stage setting, but extensions to the general $T$-stage setting follow directly. 

\begin{theorem}
In the 2-stage scenario, if Assumptions 1-4 hold, the Q-functions at stage 2 are Lipschitz continuous and correctly specified, and $\bm{M}_2(\bm{\theta}_2) = \bm{0}$ has a unique solution, 
\begin{align*}
    \widehat{\bm{\theta}}_2 - \bm{\theta}_{2}^* &\stackrel{d}{\longrightarrow} N(\bm{0}, \Sigma_{\bm{\theta}_2}) , \\
    \widehat{d}_2^{\mathrm{opt}}(\bm{h}_2) & \stackrel{p}{\longrightarrow} d_2^{\mathrm{opt}}(\bm{h}_2) \text{ for all } \bm{h}_2 \in \mathcal{H}_2,\\
    \widehat{Y}_{pse,1}  &\stackrel{p}{\longrightarrow} Y_{pse,1},
\end{align*}
where $\bm{\theta}_{2}^*$ stands or the true values of $\bm{\theta}_{2}$. $\widehat{\bm{\theta}}_{2}$, $\widehat{d}_{2}^{\mathrm{opt}}(\cdot)$, and $\widehat{Y}_{pse,1}$ denote the estimators of $\bm{\theta}_{2}$, $d_2^{\mathrm{opt}}(\cdot)$, and $Y_{pse,1}$, respectively. Notations $\stackrel{d}{\longrightarrow}$ and  $\stackrel{p}{\longrightarrow}$ stand for converge in distribution and converge in probability, respectively. The detailed form of $\Sigma_{\bm{\theta}_2}$ is given in Web Appendix  C.
\end{theorem}

\begin{theorem}
In the two-stage scenario, if Assumptions 1-5 hold, all models are Lipschitz continuous and correctly specified, and the estimating equations for the Q-functions and the estimating equation (3)  in the main text each have a unique solution, then under certain additional regularity conditions, 
\begin{align*}
    \widehat{\bm{\theta}}_{1,EE} - \bm{\theta}_{1}^* &\stackrel{d}{\longrightarrow} N(\bm{0}, \Sigma_{\bm{\theta}_1}) , \\
    \widehat{d}_{1,EE}^{\mathrm{opt}}(\bm{h}_1) & \stackrel{p}{\longrightarrow} d_1^{\mathrm{opt}}(\bm{h}_1) \text{ for all } \bm{h}_1 \in \mathcal{H}_1,
\end{align*}
where $\bm{\theta}_{1}^*$ stands or the true values of $\bm{\theta}_{1}$. $\widehat{\bm{\theta}}_{1,EE}$ and $\widehat{d}_{1,EE}^{\mathrm{opt}}(\cdot)$ denote the estimators of $\bm{\theta}_{1}$ and $d_1^{\mathrm{opt}}(\cdot)$ with the working model (1) in the main text estimated by the EE method, respectively. The regularity conditions and the detailed form of $\Sigma_{\bm{\theta}_1}$ are given in Web Appendix  D.
\end{theorem}

\begin{theorem}
In the two-stage scenario, if Assumptions 1-4 and 6 hold, all models are Lipschitz continuous and correctly specified, the sensitivity parameter $\gamma'$ is correctly specified, and the estimating equations for the Q-functions each have a unique solution, then under certain additional regularity conditions, 
\begin{align*}
    \widehat{\bm{\theta}}_{1,SA} - \bm{\theta}_{1}^* &\stackrel{d}{\longrightarrow} N(\bm{0}, \Sigma'_{\bm{\theta}_1}) , \\
    \widehat{d}_{1,SA}^{\mathrm{opt}}(\bm{h}_1) & \stackrel{p}{\longrightarrow} d_1^{\mathrm{opt}}(\bm{h}_1) \text{ for all } \bm{h}_1 \in \mathcal{H}_1,
\end{align*}
where $\bm{\theta}_{1}^*$ stands or the true values of $\bm{\theta}_{1}$. 
$\widehat{\bm{\theta}}_{1,SA}$ and $\widehat{d}_{1,SA}^{\mathrm{opt}}(\cdot)$ denote the estimators of $\bm{\theta}_{1}$ and $d_1^{\mathrm{opt}}(\cdot)$ with the working model (4) in the main text estimated by the SA method, respectively. The regularity conditions and the detailed form of $\Sigma'_{\bm{\theta}_1}$ are given in  Web Appendix  E.
\end{theorem}

\section{Web Appendix C: Proof of Theorem 1}

At stage 2, the parameters $\bm{\theta}_2$ are estimated by solving $\bm{M}_2(\bm{\theta}_2) = \bm{0}$, where $\bm{M}_2(\bm{\theta}_2) =\widehat{E}\left\{\mathbb{I} (\overline{\bm{R}}_{2}=\bm{1}_2) \frac{\partial \left[ \left\{  
  Q_2(\bm{h}_2,a_2;\bm{\theta}_2) - y_{pse,2}\right\} ^2\right]}{\partial \bm{\theta}_2} \right\}$. Under the conditions specified in Theorem 1, $\bm{M}_2(\bm{\theta}_2)$ is an unbiased estimating equation for $\bm{\theta}_2$. With the unbiasedness of these estimating equations, we can get the asymptotic property following \citet[Thm. 6.1]{Newey_McFadden_1994}. Using a first order Taylor expansion centered around $\bm{\theta}_2^*$, we find
\begin{align*}
    \bm{M}_2(\widehat{\bm{\theta}}_2) &= M(\bm{\theta}_2^*) + E\left\{\frac{\partial \bm{M}_2(\bm{\theta}_2^*)}{\partial \bm{\theta}_2}\right\}(\widehat{\bm{\theta}}_2 - \bm{\theta}_2^*) + o(1),\\
    \widehat{\bm{\theta}}_2 - \bm{\theta}_2^* &= -E\left\{\frac{\partial \bm{M}_2(\bm{\theta}_2^*)}{\partial \bm{\theta}_2}\right\}^{-1}\bm{M}_2(\bm{\theta}_2^*)+ o(1).
\end{align*}
For simplicity, we use $\frac{\partial f(\alpha^*, \beta^*)}{\partial \alpha}$ to denote the value of $\frac{\partial f(\alpha,\beta)}{\partial \alpha}$ when $\alpha = \alpha^*, \beta = \beta^*$ for any function $f(\cdot)$ and parameter values $(\alpha^*,\beta^*)$ in the Web Appendix.
\begin{align*}
    E\left[E\left\{\frac{\partial \bm{M}_2(\bm{\theta}_2^*)}{\partial \bm{\theta}_2}\right\}^{-1}\bm{M}_2(\bm{\theta}_2^*)\right] &= \bm{0},\\
    Var\left[E\left\{\frac{\partial \bm{M}_2(\bm{\theta}_2^*)}{\partial \bm{\theta}_2}\right\}^{-1}\bm{M}_2(\bm{\theta}_2^*)\right] &=E\left(\left[E\left\{\frac{\partial \bm{M}_2(\bm{\theta}_2^*)}{\partial \bm{\theta}_2}\right\}^{-1}\bm{M}_2(\bm{\theta}_2^*) \right]^{\otimes 2}\right).
\end{align*}
Besides, $\bm{M}_2(\bm{\theta}_2)$ is defined as an empirical mean of a function of i.i.d. variable. According to the Central Limit Theorem, $(\widehat{\bm{\theta}}_2 - \bm{\theta}_2^*) \stackrel{d}{\longrightarrow} N\left(\bm{0}, E\left\{\left[E\left\{\frac{\partial \bm{M}_2(\bm{\theta}_2^*)}{\partial \bm{\theta}_2}\right\}^{-1}\bm{M}_2(\bm{\theta}_2^*) \right]^{\otimes 2}\right\} \right)$.

The detailed form of $\frac{\partial \bm{M}_2(\bm{\theta}_2^*)}{\partial \bm{\theta}_2}$ can be derived by
\begin{align}
    &\frac{\partial \bm{M}_2(\bm{\theta}_2^*)}{\partial \bm{\theta}_2} = \widehat{E}\left[ \mathbb{I}(\overline{\bm{R}}_{2}=\bm{1}_2) \frac{\partial^2 \left\{Q_2(\bm{h}_2,a_2;\bm{\theta}_2^*) - y_{pse,2}\right\} ^2}{\partial \bm{\theta}_2^2} \right] \notag \\
  &= \widehat{E}\left( 2 \mathbb{I} (\overline{\bm{R}}_{2}=\bm{1}_2)  \frac{\partial \left[ \left\{  
  Q_2(\bm{h}_2,a_2;\bm{\theta}_2^*) - y_{pse,2}\right\} \frac{\partial Q_2(\bm{h}_2,a_2;\bm{\theta}_2^*)}{\partial \bm{\theta}_2} \right]}{\partial \bm{\theta}_2} \right) \notag \\
  &= \widehat{E}\left( 2 \mathbb{I} (\overline{\bm{R}}_{2}=\bm{1}_2) \left[ \left\{ \frac{\partial Q_2(\bm{h}_2,a_2;\bm{\theta}_2^*)}{\partial \bm{\theta}_2} \right\}^2  +  \left\{  
  Q_2(\bm{h}_2,a_2;\bm{\theta}_2^*) - y_{pse,2}\right\} \frac{\partial^2 Q_2(\bm{h}_2,a_2;\bm{\theta}_2^*)}{\partial \bm{\theta}_2^2}   \right] \right), \label{eq1}
\end{align}
where $\frac{\partial Q_2(\bm{h}_2,a_2;\bm{\theta}_2^*)}{\partial \bm{\theta}_2}$ and $ \frac{\partial^2 Q_2(\bm{h}_2,a_2;\bm{\theta}_2^*)}{\partial \bm{\theta}_2^2}$ depend on the specific form of $Q_2(\bm{h}_2,a_2;\bm{\theta}_2)$.

Given that $E(\widehat{\bm{\theta}}_2 - \bm{\theta}_2^*) = 0$, according to the Law of Large Numbers, $\widehat{\bm{\theta}}_2$ converges to $\bm{\theta}_2^*$ in probability. That is, for any $\epsilon > 0$, $P(|\widehat{\bm{\theta}}_2 - \bm{\theta}_2^*| > \epsilon) = 0$ as $n \to \infty$. Furthermore, the Lipschitz continuity of $Q_2(\cdot)$ implies the existence of a Lipschitz constant $L_p > 0$ such that for all ($\bm{\theta}'_2$, $\bm{\theta}_2$), $|Q_2(\bm{h}_2, a_2; \bm{\theta}'_2) - Q_2(\bm{h}_2, a_2; \bm{\theta}_2)| \leq L_p |\bm{\theta}'_2 - \bm{\theta}_2 |$. Therefore, for all $\bm{h}_2 \in \mathcal{H}_2$ and $a_2 \in \{-1, 1\}$, we have
\begin{align*}
    \lim\limits_{n\to\infty}P\{|Q_2(\bm{h}_2,a_2;\widehat{\bm{\theta}}_2) - Q_2(\bm{h}_2,a_2;\bm{\theta}_2^*)|\geq \epsilon\} &\leq  \lim\limits_{n\to\infty} P( L_p |\widehat{\bm{\theta}}_2-\bm{\theta}_2^* | \geq \epsilon)\\
    & =0
\end{align*}
That is, $Q_2(\bm{h}_2,a_2;\widehat{\bm{\theta}}_2)$ converges in probability to $Q_2(\bm{h}_2,a_2;\bm{\theta}_2^*)$ for all fixed $(\bm{h}_2,a_2)$. Since $Q_2(\bm{h}_2,a_2;\bm{\theta}_2)$ can be decomposed as $q_{2,0}(\bm{h}_2; \bm{\beta}_2) + a_2 q_{2,1}(\bm{h}_2;\bm{\psi}_2)$, $q_{2,0}(\bm{h}_2; \widehat{\bm{\beta}}_2)$ and $q_{2,1}(\bm{h}_2;\widehat{\bm{\psi}}_2)$ converge in probability to $q_{2,0}(\bm{h}_2; \bm{\beta}_2^*)$ and $q_{2,1}(\bm{h}_2;\bm{\psi}_2^*)$ for all fixed $\bm{h}_2$, respectively.

Besides, $\widehat{d}_2^{\mathrm{opt}}\left(\bm{h}_2\right) = 2\mathbb{I}\left\{q_{2,1}(\bm{h}_2;\widehat{\bm{\psi}}_2)>0\right\} -1$. Under the standard regularity assumption that $P\left\{q_{2,1}(\bm{h}_2;\widehat{\bm{\psi}}_2)=0\right\} = 0$, we have
\begin{align*}
    P&\left\{\widehat{d}_2^{\mathrm{opt}}(\bm{h}_2) \neq d_2^{\mathrm{opt}}(\bm{h}_2)\right\} = P\left[\mathbb{I}\{q_{2,1}(\bm{h}_2;\widehat{\bm{\psi}}_2)>0\} \neq \mathbb{I}\{q_{2,1}(\bm{h}_2;\bm{\psi}^*_2)>0\}\right]\\
    &= P\{q_{2,1}(\bm{h}_2;\widehat{\bm{\psi}}_2)>0, q_{2,1}(\bm{h}_2;\bm{\psi}_2^*)<0\} +  P\{q_{2,1}(\bm{h}_2;\widehat{\bm{\psi}}_2)<0, q_{2,1}(\bm{h}_2;\bm{\psi}_2^*)>0\}
\end{align*}
Because $\lim\limits_{n\to\infty}q_{2,1}(\bm{h}_t;\widehat{\bm{\psi}}_t) = q_{2,1}(\bm{h}_t;\bm{\psi}_t^*)$, we have $\lim\limits_{n\to\infty}P\left\{q_{2,1}(\bm{h}_2;\widehat{\bm{\psi}}_2)>0, q_{2,1}(\bm{h}_2;\bm{\psi}_2^*)<0\right\} = 0$ and $\lim\limits_{n\to\infty}P\left\{q_{2,1}(\bm{h}_2;\widehat{\bm{\psi}}_2)<0, q_{2,1}(\bm{h}_2;\bm{\psi}_2^*)>0\right\} = 0$. Therefore, for all $\bm{h}_2 \in \mathcal{H}_2$,
$$
\lim\limits_{n\to\infty}P\left\{\widehat{d}_2^{\mathrm{opt}}(\bm{h}_2) \neq d_2^{\mathrm{opt}}(\bm{h}_2)\right\} = 0, \quad 
$$
which means that $\widehat{d}_2^{\mathrm{opt}}(\bm{h}_2) \stackrel{p}{\longrightarrow} d_2^{\mathrm{opt}}(\bm{h}_2)$ for any fixed $\bm{h}_2$. 

Since $Y_{pse,1} = \max _{a_{2} \in \{-1,1\}} Q_{2}\left(\bm{H}_{2}, a_{2};\bm{\theta}_2^*\right)$, we have $y_{pse,1} =  Q_{2}\left\{\bm{h}_{2}, d_2^{\mathrm{opt}}(\bm{h}_2);\bm{\theta}_2^*\right\}$ and $\widehat{y}_{pse,1} =  Q_{2}\left\{\bm{h}_{2}, \widehat{d}_2^{\mathrm{opt}}(\bm{h}_2);\widehat{\bm{\theta}}_2\right\}$ for a given $\bm{h}_2$. Based on the convergence of $\widehat{d}_2^{\mathrm{opt}}(\bm{h}_2)$ and $Q_2(\bm{h}_2,a_2;\widehat{\bm{\theta}}_2)$, we have
\begin{align*}
    \lim\limits_{n\to \infty}& P(|\widehat{y}_{pse,1}-y_{pse,1}|\geq \epsilon) = \lim\limits_{n\to \infty} P\left[\left|Q_2\left\{\bm{h}_2,\widehat{d}_2^{\mathrm{opt}}(\bm{h}_2);\widehat{\bm{\theta}}_2\right\}-Q_2\left\{\bm{h}_2,d_2^{\mathrm{opt}}(\bm{h}_2);\bm{\theta}_2^*\right\}\right|\geq \epsilon \right]\\
    &= \lim\limits_{n\to \infty} P\left[\left|Q_2\left\{\bm{h}_2,\widehat{d}_2^{\mathrm{opt}}(\bm{h}_2);\widehat{\bm{\theta}}_2\right\}-Q_2\left\{\bm{h}_2,d_2^{\mathrm{opt}}(\bm{h}_2);\bm{\theta}_2^*\right\}\right|\geq \epsilon, \widehat{d}_2^{\mathrm{opt}}(\bm{h}_2) \neq d_2^{\mathrm{opt}}(\bm{h}_2) \right] \\
    &\qquad + \lim\limits_{n\to \infty} P\left[ \left|Q_2\left\{\bm{h}_2,\widehat{d}_2^{\mathrm{opt}}(\bm{h}_2);\widehat{\bm{\theta}}_2\right\}-Q_2\left\{\bm{h}_2,d_2^{\mathrm{opt}}(\bm{h}_2);\bm{\theta}_2^*\right\}\right|\geq \epsilon, \widehat{d}_2^{\mathrm{opt}}(\bm{h}_2) = d_2^{\mathrm{opt}}(\bm{h}_2) \right]\\
    &= \lim\limits_{n\to \infty} P\left[ \left|Q_2\left\{\bm{h}_2,d_2^{\mathrm{opt}}(\bm{h}_2);\widehat{\bm{\theta}}_2\right\}-Q_2\left\{\bm{h}_2,d_2^{\mathrm{opt}}(\bm{h}_2);\bm{\theta}_2^*\right\}\right|\geq \epsilon \right]\\
    &=0.
\end{align*}
Therefore, $\widehat{y}_{pse,1} \stackrel{p}{\longrightarrow} y_{pse,1}$ for any fixed $\bm{h}_2$. Subsequently, $\widehat{Y}_{pse,1} \stackrel{p}{\longrightarrow} Y_{pse,1}$.

\section{Web Appendix D: Proof of Theorem 2}

At stage 1, the parameters $\bm{\theta}_1$ are estimated by solving the estimating equation $\bm{M}_1(\bm{\theta}_1) = \bm{0}$, where $\bm{M}_1(\bm{\theta}_1) = \widehat{E} \left(\frac{\widehat{r}_{pse,1}}{\widehat{\pi}_{1}} \mathbb{I}(R_{1}=1) \frac{\partial \left[ \left\{ Q_1(\bm{h}_1,a_1;\bm{\theta}_1) - \widehat{y}_{pse,1}  \right\}^2\right]}{\partial \bm{\theta}_1} \right)$. Following \cite{wallace2015doubly}  and \cite{Simoneau_et_al_2020}, we establish the asymptotic normality of $\widehat{\bm{\theta}}_1$ under the regularity condition that $P(\{q_{2,1}(\bm{H}_2; \widehat{\bm{\psi}}_2) = 0\})=0$. 
Since $(\widehat{Y}_{pse,t},\widehat{R}_{pse,t},\widehat{\pi}_t)$ are calculated with $(\widehat{\bm{\theta}}_2,\widehat{\gamma}_1)$, $\bm{M}_1$ depends $(\widehat{\bm{\theta}}_2,\widehat{\gamma}_1)$ and thus can also be denoted as $\bm{M}_1(\bm{\theta}_1, \widehat{\bm{\theta}}_2,\widehat{\gamma}_1)$. The variance of $\bm{\theta}_1$ must adjust for the plug-in estimates of nuisance parameters $(\bm{\theta}_2,\gamma_1)$. By performing a first-order Taylor expression of $\bm{M}_1(\bm{\theta}_1, \widehat{\bm{\theta}}_2,\widehat{\gamma}_1)$ around $(\bm{\theta}_2^*,\gamma_1^*)$, we have
\begin{align*}
    M_{\mathrm{adj}, 1}(\bm{\theta}_1) = \bm{M}_1(\bm{\theta}_1, \widehat{\bm{\theta}}_2,\widehat{\gamma}_1) + E\left\{ \frac{\partial \bm{M}_1(\bm{\theta}_1, \bm{\theta}_2^*,\gamma_1^*)}{\partial (\bm{\theta}_2,\gamma_1)} \right\}
    \left\{\left(\begin{array}{l}
\widehat{\bm{\theta}}_2 \\
\widehat{\gamma}_1
\end{array}\right)-\left(\begin{array}{l}
\bm{\theta}_2^* \\
\gamma_1^*
\end{array}\right)\right\}+ o(1),
\end{align*}
where $
    E\left\{ \frac{\partial \bm{M}_1(\bm{\theta}_1, \bm{\theta}_2^*,\gamma_1^*)}{\partial (\bm{\theta}_2,\gamma_1)} \right\} = E\left\{ \left( \frac{\partial \bm{M}_1(\bm{\theta}_1, \bm{\theta}_2^*,\gamma_1^*)}{\partial \bm{\theta}_2} , \frac{\partial \bm{M}_1(\bm{\theta}_1, \bm{\theta}_2^*,\gamma_1^*)}{\partial \gamma_1}  \right) \right\}
$, and we assume that $\bm{M}_1$ is infinitely differentiable for $(\bm{\theta}_1,\bm{\theta}_2,\gamma_1)$. Therefore, 
\begin{align*}
    \bm{M}_{adj,1}(\widehat{\bm{\theta}}_1) &= \bm{M}_{adj,1}(\bm{\theta}_1^*) + E\left\{\frac{\partial \bm{M}_{adj,1}(\bm{\theta}_1^*)}{\partial \bm{\theta}_1}\right\}(\widehat{\bm{\theta}}_1 - \bm{\theta}_1^*) + o(1),\\
    \widehat{\bm{\theta}}_1 - \bm{\theta}_1^* &= -E\left\{\frac{\partial \bm{M}_{adj,1}(\bm{\theta}_1^*)}{\partial \bm{\theta}_1}\right\}^{-1}\bm{M}_{adj,1}(\bm{\theta}_1^*)+ o(1).
\end{align*}

We need the following conditions to make sure that $\lim\limits_{n\to \infty}\widehat{\gamma}_1 - \gamma_1^* = 0$ and subsequently $\lim\limits_{n\to \infty}\bm{M}_{adj,1}(\bm{\theta}_1^*) = \bm{0}$.

Condition 1. The kernel $K(\bm{u}_1)$ has bounded derivatives of order $\delta$, satisfies $\int K(\bm{u}_1) \mathrm{d} \bm{u}_1=1$, and has zero moments of order up to $m-1$ and nonzero $m$ th-order moment.

Condition 2. The true function of $s_1(\bm{u}_1)$ is continuously differentiable and bounded on an open set containing the support of $\bm{u}_1$.

Condition 3. The moment $E\{\exp (4 \gamma_1 \widehat{y}_{pse,1})\}$ is finite and the function $E\{\exp (4 \gamma_1 \widehat{y}_{pse,1}) \mid \bm{u}_1\} \pr(\bm{u}_1)$ is bounded, where $\pr(\bm{u}_1)$ is the marginal density of $\bm{u}_1$.

Condition 4. The bandwidth $c_1=c_{n_1}$ is such that $c_{n_1} \rightarrow 0, n_1 c_{n_1}^{p_1} \rightarrow \infty, n_1^{1 / 2} c_{n_1}^{p_1+2 \delta} / \log n_1 \rightarrow \infty$, and $n_1 c_{n_1}^{2 m} \rightarrow 0$ as $n_1 \rightarrow \infty$, where $p_1$ is the dimension of $\bm{u}_1$, $n_1$ is the number of units with fully-observed covariates at stage 1.

Let $\bm{B}_1^{\prime}(\bm{\theta}_2,\gamma_1) = \frac{\partial \bm{B}_1(\bm{\theta}_2,\gamma_1)^{\mathrm{T}}W_1\bm{B}_1(\bm{\theta}_2,\gamma_1)}{\partial \gamma_1}$, with conditions 1-4 and the convergence of $\bm{\theta}_2$, we have $\lim\limits_{n\to \infty}E\{\bm{B}_1^{\prime}(\widehat{\bm{\theta}}_2,\gamma_1^*)\}\stackrel{p}{\longrightarrow}0$ and $\lim\limits_{n\to \infty} \bm{W}_1 \stackrel{p}{\longrightarrow} \bm{W}_1^*$. 

By Taylor expansion, we have:
\begin{align*}
    &\left(\begin{array}{l}\widehat{\bm{\theta}}_2\\\widehat{\gamma}_1\end{array}\right) - \left(\begin{array}{l}
    \bm{\theta}_2^* \\ \gamma_1^* \end{array}\right) = - E\left[ \frac{\partial\{\bm{M}_2(\bm{\theta}_2^*),\bm{B}_1^{\prime}(\bm{\theta}_2^*,\gamma_1^*)\}}{\partial (\bm{\theta}_2, \gamma_1)}  \right]^{-1} \left(\begin{array}{l}
    \bm{M}_2(\bm{\theta}_2^*) \\ \bm{B}_1^{\prime}(\bm{\theta}_2^*,\gamma_1^*) \end{array}\right) + o(1)\\
    &= -\left( E\left[ \begin{array}{ll}
    \frac{\partial \bm{M}_2(\bm{\theta}_2^*)}{\partial \bm{\theta}_2} & \frac{\partial \bm{M}_2(\bm{\theta}_2^*)}{\partial \gamma_1}\\ \frac{\partial \bm{B}_1^{\prime}(\bm{\theta}_2^*,\gamma_1^*)}{\partial \bm{\theta}_2} & \frac{\partial \bm{B}_1^{\prime}(\bm{\theta}_2^*,\gamma_1^*)}{\partial \gamma_1}\end{array} \right]\right)^{-1}\left(\begin{array}{l}
    \bm{M}_2(\bm{\theta}_2^*) \\ \bm{B}_1^{\prime}(\bm{\theta}_2^*,\gamma_1^*) \end{array}\right) + o(1)\\
    & = -\left(\left[ \begin{array}{ll}
    E\left\{\frac{\partial \bm{M}_2(\bm{\theta}_2^*)}{\partial \bm{\theta}_2}\right\} & \bm{0}\\ E\left\{\frac{\partial \bm{B}_1^{\prime}(\bm{\theta}_2^*,\gamma_1^*)}{\partial \bm{\theta}_2}\right\} & E\left\{\frac{\partial \bm{B}_1^{\prime}(\bm{\theta}_2^*,\gamma_1^*)}{\partial \gamma_1}\right\} \end{array} \right]\right)^{-1} \left(\begin{array}{l}
    \bm{M}_2(\bm{\theta}_2^*) \\ \bm{B}_1^{\prime}(\bm{\theta}_2^*,\gamma_1^*) \end{array}\right) + o(1)\\
    &= \left[ \begin{array}{ll}
    -\left[E\left\{\frac{\partial \bm{M}_2(\bm{\theta}_2^*)}{\partial \bm{\theta}_2}\right\}\right]^{-1} & \bm{0}\\ \left[E\left\{ \frac{\partial \bm{B}_1^{\prime}(\bm{\theta}_2^*,\gamma_1^*)}{\partial \gamma_1}\right\}\right]^{-1} E\left\{\frac{\partial \bm{B}_1^{\prime}(\bm{\theta}_2^*,\gamma_1^*)}{\partial \bm{\theta}_2}\right\} \left[E\left\{\frac{\partial \bm{M}_2(\bm{\theta}_2^*)}{\partial \bm{\theta}_2}\right\}\right]^{-1} & -\left[E\left\{\frac{\partial \bm{B}_1^{\prime}(\bm{\theta}_2^*,\gamma_1^*)}{\partial \gamma_1}\right\}\right]^{-1} \end{array} \right] \left(\begin{array}{l} \bm{M}_2(\bm{\theta}_2^*) \\ \bm{B}_1^{\prime}(\bm{\theta}_2^*,\gamma_1^*) \end{array}\right) + o(1).
\end{align*}
Therefore, $\widehat{\gamma}_1 - \gamma_1^* = E\left\{\frac{\partial \bm{B}_1^{\prime}(\bm{\theta}_2^*,\gamma_1^*)}{\partial \bm{\theta}_2}\right\} \left[E\left\{\frac{\partial \bm{M}_2(\bm{\theta}_2^*)}{\partial \bm{\theta}_2}\right\}\right]^{-1}\bm{M}_2(\bm{\theta}_2^*) - \left[E\left\{\frac{\partial \bm{B}_1^{\prime}(\bm{\theta}_2^*,\gamma_1^*)}{\partial \gamma_1}\right\}\right]^{-1} \bm{B}_1^{\prime}(\bm{\theta}_2^*,\gamma_1^*)$,
\begin{align*}
    &M_{\mathrm{adj}, 1}(\bm{\theta}_1) = \bm{M}_1(\bm{\theta}_1, \bm{\theta}_2^*,\gamma_1^*) - E\left\{\frac{\partial \bm{M}_1(\bm{\theta}_1, \bm{\theta}_2^*,\gamma_1^*)}{\partial \bm{\theta}_2}\right\} \left[E\left\{\frac{\partial \bm{M}_2(\bm{\theta}_2^*)}{\partial \bm{\theta}_2}\right\}\right]^{-1} \bm{M}_2(\bm{\theta}_2^*)\\
    &+E\left\{\frac{\partial \bm{M}_1(\bm{\theta}_1, \bm{\theta}_2^*,\gamma_1^*)}{\partial \gamma_1}\right\}\left[E\left\{\frac{\partial \bm{B}_1^{\prime}(\bm{\theta}_2^*,\gamma_1^*)}{\partial \gamma_1}\right\}\right]^{-1}E\left\{\frac{\partial \bm{B}_1^{\prime}(\bm{\theta}_2^*,\gamma_1^*)}{\partial \bm{\theta}_2}\right\} \left[E\left\{\frac{\partial \bm{M}_2(\bm{\theta}_2^*)}{\partial \bm{\theta}_2}\right\}\right]^{-1}\bm{M}_2(\bm{\theta}_2^*)\\
    &-E\left\{\frac{\partial \bm{M}_1(\bm{\theta}_1, \bm{\theta}_2^*,\gamma_1^*)}{\partial \gamma_1}\right\}\left[E\left\{\frac{\partial \bm{B}_1^{\prime}(\bm{\theta}_2^*,\gamma_1^*)}{\partial \gamma_1} \right\} \right]^{-1} \bm{B}_1^{\prime}(\bm{\theta}_2^*,\gamma_1^*)
     + o(1).
\end{align*}

And we have $\widehat{\bm{\theta}}_{1,EE} - \bm{\theta}_1^* \rightarrow N(\bm{0},\Sigma_{\bm{\theta}_1} )$, where
$\Sigma_{\bm{\theta}_1} = E\left\{\left(\left[E\left\{\frac{\partial M_{adj,1}(\bm{\theta}_1^*)}{\partial \bm{\theta}_1} \right\}\right]^{-1} M_{adj,1}(\bm{\theta}_1^*) \right)^{\otimes 2}\right\} $.


The detailed form of $\frac{\partial \bm{M}_2(\bm{\theta}_2^*)}{\partial \bm{\theta}_2}$ is provided in equation \eqref{eq1} in Web Appendix A. Assume that  $\frac{\partial \bm{B}_1^{\prime}(\bm{\theta}_2^*,\gamma_1^*)}{\partial \bm{\theta}_2}$ and $\frac{\partial \bm{B}_1^{\prime}(\bm{\theta}_2^*,\gamma_1^*)}{\partial \gamma_1}$ exist. To obtain $M_{\mathrm{adj}, 1}(\bm{\theta}_1)$ from observed data, we give the detailed form of $E\left\{\frac{\partial \bm{M}_1(\bm{\theta}_1, \bm{\theta}_2^*,\gamma_1^*)}{\partial \bm{\theta}_2}\right\}$, $E\left\{\frac{\partial \bm{M}_1(\bm{\theta}_1, \bm{\theta}_2^*,\gamma_1^*)}{\partial \gamma_1}\right\}$, $E\left\{\frac{\partial \bm{B}_1^{\prime}(\bm{\theta}_2^*,\gamma_1^*)}{\partial \bm{\theta}_2}\right\}$, and $E\left\{\frac{\partial \bm{B}_1^{\prime}(\bm{\theta}_2^*,\gamma_1^*)}{\partial \gamma_1}\right\}$ in the following text. Since the missing indicator of $Y_{pse,1}$ is mainly determined by the parametric form $Q_{2}$, under the assumption that the variables employed in the computation of $Y_{pse,1}$ remain unchanged, we have $\widehat{r}_{pse,1} = r_{pse,1}$. 

Since $\widehat{y}_{pse,1}$ is calculated with $\bm{\theta}_2$, we denote it as $\widehat{y}_{pse,1}(\widehat{\bm{\theta}}_2)$. Note that $\widehat{y}_{pse,1}(\bm{\theta}_2^*) = y_{pse,1}$. Consequently,
\begin{align*}
    &\frac{\partial \bm{M}_1(\bm{\theta}_1, \bm{\theta}_2^*,\gamma_1^*)}{\partial \bm{\theta}_2}  = \frac{\partial \bm{M}_1(\bm{\theta}_1, \widehat{y}_{pse,1}(\bm{\theta}_2^*),\gamma_1^*)}{\partial \left\{\widehat{y}_{pse,1}(\bm{\theta}_2^*)\right\}} \frac{\partial \left\{\widehat{y}_{pse,1}(\bm{\theta}_2^*)\right\}}{\partial \bm{\theta}_2}\\
    &= \frac{\partial \bm{M}_1(\bm{\theta}_1, y_{pse,1},\gamma_1^*)}{\partial y_{pse,1}} \frac{\partial y_{pse,1}}{\partial \bm{\theta}_2}\\
    &= \frac{\partial \widehat{E} \left(\frac{r_{pse,1}}{\widehat{\pi}_{1}}r_{1} \frac{\partial \left[ \left\{ Q_1(\bm{h}_1,a_1;\bm{\theta}_1) - y_{pse,1}  \right\}^2\right]}{\partial \bm{\theta}_1} \right)}{\partial y_{pse,1}} \frac{\partial y_{pse,1}}{\partial \bm{\theta}_2}\\
    &= \widehat{E} \left( 2 r_{pse,1} r_1 \frac{\partial Q_1(\bm{h}_1,a_1;\bm{\theta}_1)}{\partial \bm{\theta}_1} \frac{\partial [ \widehat{\pi}_1\{ Q_1(\bm{h}_1,a_1;\bm{\theta}_1) - y_{pse,1}  \}] }{\partial y_{pse,1}} \right) \frac{\partial y_{pse,1}}{\partial \bm{\theta}_2}\\
    &= \widehat{E} \left( 2 r_{pse,1} r_1 \frac{\partial Q_1(\bm{h}_1,a_1;\bm{\theta}_1)}{\partial \bm{\theta}_1} \frac{\partial [ 1/\widehat{\pi}_1\{ Q_1(\bm{h}_1,a_1;\bm{\theta}_1) - y_{pse,1}  \}] }{\partial y_{pse,1}} \right) \frac{\partial y_{pse,1}}{\partial \bm{\theta}_2}\\  
    &= \widehat{E} \left( 2 r_{pse,1} r_1 \frac{\partial Q_1(\bm{h}_1,a_1;\bm{\theta}_1)}{\partial \bm{\theta}_1} \left[ - \frac{1}{\widehat{\pi}_1} + \{ Q_1(\bm{h}_1,a_1;\bm{\theta}_1) - y_{pse,1}\} \frac{\partial ( 1/\widehat{\pi}_1)}{\partial y_{pse,1}} \right]\right) \frac{\partial y_{pse,1}}{\partial \bm{\theta}_2}\\
    &= \widehat{E} \left( 2 r_{pse,1} r_1 \frac{\partial Q_1(\bm{h}_1,a_1;\bm{\theta}_1)}{\partial \bm{\theta}_1} \left[ \{ \gamma_1^* Q_1(\bm{h}_1,a_1;\bm{\theta}_1) - \gamma_1^* y_{pse,1}-1\} \exp\{\widehat{s}_{\gamma_1^*}(\bm{u}_1) + \gamma_1^*y_{pse,1}\} -1 \right]\right) \frac{\partial y_{pse,1}}{\partial \bm{\theta}_2},
\end{align*}
where $\frac{\partial Q_1(\bm{h}_1,a_1;\bm{\theta}_1)}{\partial \bm{\theta}_1}$ and $\frac{\partial y_{pse,1}}{\partial \bm{\theta}_2}$ depends on the specified form of the Q-functions.

Similarly, we obtain $\frac{\partial \bm{M}_1(\bm{\theta}_1, \bm{\theta}_2^*,\gamma_1^*)}{\partial \gamma_1}$ using the following equations
\begin{align}
    &\frac{\partial \bm{M}_1(\bm{\theta}_1, \bm{\theta}_2^*,\gamma_1^*)}{\partial \gamma_1}  = \frac{\partial \bm{M}_1\{\bm{\theta}_1, \widehat{y}_{pse,1}(\bm{\theta}_2^*),\gamma_1^*\}}{\partial \left[1/\widehat{\pi}_1\{\bm{u}_1,\widehat{y}_{pse,1}(\bm{\theta}_2^*),\gamma_1^*\}\right]} \frac{\partial \left[1/\widehat{\pi}_1\{\bm{u}_1,\widehat{y}_{pse,1}(\bm{\theta}_2^*),\gamma_1^*\}\right]}{\partial \gamma_1} \notag \\
    &=  \frac{\partial \bm{M}_1(\bm{\theta}_1, y_{pse,1},\gamma_1^*)}{\partial \left\{1/\widehat{\pi}_1(\bm{u}_1,y_{pse,1},\gamma_1^*)\right\}} \frac{\partial \left\{1/\widehat{\pi}_1(\bm{u}_1,y_{pse,1},\gamma_1^*)\right\}}{\partial \gamma_1} \notag \\
    &= \frac{\partial \widehat{E} \left(\frac{r_{pse,1}}{\widehat{\pi}_{1}}r_{1} \frac{\partial \left[ \left\{ Q_1(\bm{h}_1,a_1;\bm{\theta}_1) - y_{pse,1}  \right\}^2\right]}{\partial \bm{\theta}_1} \right)}{\partial (1/\widehat{\pi}_1)} \frac{\partial \left\{1/\widehat{\pi}_1(\bm{u}_1,y_{pse,1},\gamma_1^*)\right\}}{\partial \gamma_1} \notag \\
    &= \widehat{E} \bigg[2r_{pse,1}r_{1}\left\{ Q_1(\bm{h}_1,a_1;\bm{\theta}_1) - y_{pse,1}\right\} \frac{\partial  Q_1(\bm{h}_1,a_1;\bm{\theta}_1) }{\partial \bm{\theta}_1} \bigg] \bigg[ \frac{\partial \left\{1/\widehat{\pi}_1(\bm{u}_1,y_{pse,1},\gamma_1^*)\right\}}{\partial \gamma_1}\bigg]. \label{eq2}
\end{align}

The explicit expression for $\frac{\partial ( 1/\widehat{\pi}_1)}{\partial \gamma_{1}}$ can be derived with the following equations
\begin{align}
    &\frac{\partial \{ 1/\widehat{\pi}_1(\bm{u}_1,y_{pse,1},\gamma_1^*)\}}{\partial \gamma_{1}} = \frac{\partial \left[1+\exp\{\widehat{s}_{\gamma_1^*}(\bm{u}_1)\}\exp(\gamma_1^*y_{pse,1})\right]}{\partial \gamma_1} \notag \\
    & = \gamma_1^*\exp\{\widehat{s}_{\gamma_1^*}(\bm{u}_1)\}\exp(\gamma_1^*y_{pse,1}) + \frac{\partial \exp\{\widehat{s}_{\gamma_1^*}(\bm{u}_1)\}}{\partial \gamma_1} \exp(\gamma_1^*y_{pse,1}) \notag \\
    & = \gamma_1^*\exp\{\widehat{s}_{\gamma_1^*}(\bm{u}_1)\}\exp(\gamma_1^*y_{pse,1}) + \frac{\partial \left\{\frac{\sum_{i=1}^{n} \left(1-r_{pse,1,i}\right)  K_{c_1}\left(\bm{u}_1-\bm{u}_{1,i}\right) r_{1,i} }{\sum_{i=1}^{n}  r_{pse,1,i}   \exp \left(\gamma_1^* y_{pse,1,i}\right) K_{c_1}\left(\bm{u}_1-\bm{u}_{1,i}\right) r_{1,i}} \right\}}{\partial \gamma_1} \exp(\gamma_1^*y_{pse,1}) \notag \\
    & =\exp\{\widehat{s}_{\gamma_1^*}(\bm{u}_1)\}\exp(\gamma_1^*y_{pse,1}) \left\{\gamma_1^*+ \frac{\sum_{i=1}^{n} r_{pse,1,i} y_{pse,1,i}  \exp \left(\gamma_1^* y_{pse,1,i}\right) K_{c_1}\left(\bm{u}_1-\bm{u}_{1,i}\right) r_{1,i} }{\sum_{i=1}^{n}  r_{pse,1,i}   \exp \left(\gamma_1^* y_{pse,1,i}\right) K_{c_1}\left(\bm{u}_1-\bm{u}_{1,i}\right) r_{1,i}}\right\}. \label{eq3}
\end{align}

Combining equations \eqref{eq2} and \eqref{eq3}, we can obtain the detailed form of $\frac{\partial \bm{M}_1(\bm{\theta}_1, \bm{\theta}_2^*,\gamma_1^*)}{\partial \gamma_1}$.


The detailed form of $E\{\frac{\partial \bm{B}_1^{\prime}(\bm{\theta}_2^*,\gamma_1^*)}{\partial \gamma_1}\}$ and $E\{\frac{\partial \bm{B}_1^{\prime}(\bm{\theta}_2^*,\gamma_1^*)}{\partial \bm{\theta}_2}\}$ can be derived following \cite{Shao_Wang_2016} when the following regularity conditions hold.

Condition 5. There is a vector of the functional $G(r_{pse,1}, \bm{u}_1, \bm{\omega})$ which is linear in $\bm{\omega}=\left(\omega_{1}, \omega_{2}\right)^{\mathrm{T}}$ and such that:

(i) for small enough $\left\|\bm{\omega}-\bm{\omega}^*\right\|$, $\| \widetilde{m}\left(r_{pse,1}, \bm{u}_1, \widehat{y}_{pse,1}, \bm{\omega}, \gamma_1\right)-\widetilde{m}\left(r_{pse,1}, \bm{u}_1, \widehat{y}_{pse,1}, \bm{\omega}^*, \gamma_1\right)-G(r_{pse,1},\\ \bm{u}_1, \bm{\omega}-\bm{\omega}^*) \| \leqslant c(r_{pse,1}, \bm{u}_1)\left(\left\|\bm{\omega}-\bm{\omega}^*\right\|\right)^2$, where $\widetilde{m}(r_{pse,1}, \bm{u}_1, \widehat{y}_{pse,1}, \bm{\omega}, \gamma_1)$ is the $L$-dimensional vector. And its $l$-th component $\widetilde{m}_l(r_{pse,1}, \bm{u}_1, \widehat{y}_{pse,1}, \bm{\omega}, \gamma_1)$, is defined as $\mathbb{I}(\bm{z}_1=l)[r_{pse,1}\{1+\exp (\gamma_1 \widehat{y}_{pse,1}) \omega_1(\bm{u}_1) / \omega_2(\bm{u}_1)\}-1]$, where $\bm{\omega}^* = [E(1-r_{pse,1} \mid \bm{u}_1), E(r_{pse,1} \exp (\gamma_1^* y_{pse,1}) \mid \bm{u}_1)]^{\mathrm{T}}$, and $E\{c(r_{pse,1}, \bm{u}_1)\}<\infty$;

(ii) $\|G(r_{pse,1}, \bm{u}_1, \bm{\omega})\| \leqslant e(r_{pse,1}, \bm{u}_1)\| \bm{\omega}\|$ and $E\{e(r_{pse,1}, \bm{u}_1)^2\}<\infty$;

(iii) there exists an almost everywhere continuous function $v_1(\bm{u}_1)$ with $\int\|v_1(\bm{u}_1)\| \mathrm{d} \bm{u}_1<\infty$, $E\{G(r_{pse,1}, \bm{u}_1, \bm{\omega})\}=\int v_1(\bm{u}_1) \bm{\omega}(\bm{u}_1) \mathrm{d} \bm{u}_1$ for all $\|\bm{\omega}\| \leqslant \infty$, and $E\{\sup _{\|\zeta\| \leqslant \epsilon}\|v_1(\bm{u}_1+\zeta)\|^4\}<\infty$ for some $\epsilon>0$.

Condition 6. For small enough $\left\|\bm{\omega}-\bm{\omega}_0\right\|, \widetilde{m}(r_{pse,1}, \bm{u}_1, \widehat{y}_{pse,1}, \bm{\omega}, \gamma_1)$ is continuously differentiable in $\gamma_1$ in a neighbourhood of $(y_{pse,1},\gamma_{1}^*)$. And there is a function $k(r_{pse,1}, \bm{u}_1)$ satisfying $\left\| \frac{\partial \widetilde{m}(r_{pse,1}, \bm{u}_1, \widehat{y}_{pse,1}, \bm{\omega}, \gamma_1)}{\partial \gamma_1}- \frac{\partial \widetilde{m}\left(r_{pse,1}, \bm{u}_1, y_{pse,1}, \bm{\omega}^*, \gamma_{1}^*\right)}{\partial \gamma_1}\right\| \leq k(r_{pse,1}, \bm{u}_1)(\left\|\gamma_{1}^*-\gamma_1\right\|^\epsilon+\left\|\bm{\omega}-\bm{\omega}^*\right\|^\epsilon+\left\|\widehat{y}_{pse,1}-y_{pse,1}\right\|^\epsilon)$ for an $\epsilon>0$, $E\{k(r_{pse,1}, \bm{u}_1)\}<\infty$. Besides, $  \bm{\Gamma}_{\gamma_1} = E\left\{\frac{\partial \widetilde{m}\left(r_{pse,1}, \bm{u}_1, y_{pse,1}, \bm{\omega}^*, \gamma_{1}^*\right)}{\partial \gamma_1}\right\}$ and $\bm{\Gamma}_{y_{pse,1}} = E\left\{\frac{\partial \widetilde{m}\left(r_{pse,1}, \bm{u}_1, y_{pse,1}, \bm{\omega}^*, \gamma_{1}^*\right)}{\partial \widehat{y}_{pse,1}}\right\}$ exist and are of full rank.

Assume that Conditions $1-6$ hold. Then, as $n \rightarrow \infty$, $E\left\{\frac{\partial \bm{B}_1'(\bm{\theta}_2^*,\gamma_1^*)}{\partial \gamma_1}\right\} \to 2\bm{\Gamma}_{\gamma_1}^{\mathrm{T}}\bm{W}_1^*\bm{\Gamma}_{\gamma_1}$, $E\left\{\frac{\partial \bm{B}_1'(\bm{\theta}_2^*,\gamma_1^*)}{\partial \bm{\theta}_2}\right\} \to 2\bm{\Gamma}_{y_{pse,1}}^{\mathrm{T}}\bm{W}_1^*\bm{\Gamma}_{y_{pse,1}}\frac{\partial y_{pse,1}}{\partial \bm{\theta}_2}$. 

As mentioned by \cite{Shao_Wang_2016}, the form of the asymptotic variance of this semi-parameteric estimator is complicated. Consequently, we suggest using the bootstrap method to estimate it.

Besides, according to the Law of Large Numbers, $\widehat{\bm{\theta}}_{1,EE}$ converges to $\bm{\theta}_1$ in probability. Therefore, following the proof of Theorem 1, we can similarly prove the convergence in probability of $\widehat{d}^{\mathrm{opt}}_1$ by proving that $Q_1(\bm{h}_1,a_1;\widehat{\bm{\theta}}_{1,EE}) \stackrel{p}{\longrightarrow} Q_1(\bm{h}_1,a_1;\bm{\theta}_{1})$ for any fixed $(\bm{h}_1,a_1)$.

\section{Web Appendix E: Proof of Theorem 3}

We need the following conditions to guarantee that $\lim\limits_{n\to \infty}\exp\{\widehat{s}'_{\gamma'_1,1}(\bm{g}_1) \} = \exp\{s'_{\gamma'_1,1}(\bm{g}_1) \}$:

Condition 7. The kernel $K'(\bm{g}_1)$ has bounded derivatives of order $\delta'$, satisfies \\ $\int K'(\bm{g}_1) \mathrm{d} \bm{g}_1=1$, and has zero moments of order up to $m-1$ and nonzero $m$ th-order moment.

Condition 8. The true function of $s_1'(\bm{g}_1)$ is continuously differentiable and bounded on an open set containing the support of $\bm{g}_1$.

Condition 9. The moment $E[\exp (4 \gamma_1 \widehat{y}_{pse,1})]$ is finite and the function $E[\exp (4 \gamma_1 \widehat{y}_{pse,1}) \mid \bm{g}_1] \pr(\bm{g}_1)$ is bounded, where $\pr(\bm{g}_1)$ is the marginal density of $\bm{g}_1$.

Condition 10. The bandwidth $c'_1=c'_{n_1}$ is such that $c'_{n_1} \rightarrow 0, n_1 {c'_{n_1}}^{p_1'} \rightarrow \infty, n_1^{1/2} {c'_{n_1}}^{p_1'+2 \delta'} / \log n_1 \rightarrow \infty$, and $n_1 {c'_{n_1}}^{2 m} \rightarrow 0$ as $n_1 \rightarrow \infty$, where $p_1'$ is the dimension of $\bm{g}_1$, $n_1$ is the number of units with fully-observed covariates at stage 1.

Under these conditions, $\exp\{\widehat{s}'_{\gamma'_1,1}(\bm{h}_1)\}$ is a consistent estimator of $\exp\{s'_{\gamma'_1,1}(\bm{h}_1)\}$. Therefore, the estimated missingness probability model is consistent. And $\bm{M}_1(\bm{\theta}_1)$ is an unbiased estimation equation for $\bm{\theta}_1$ \citep{Newey_McFadden_1994}. Similar to Web Appendix D, where we provided the asymptotic properties of $\widehat{\bm{\theta}}_1$ for the  WQ-EE method, we establish the asymptotic normality of $\widehat{\bm{\theta}}_1$ for the WQ-SA method under the regularity condition that $P(\{q_{2,1}(\bm{H}_2; \widehat{\bm{\psi}}_2) = 0\})=0$. 

Let  $
M_{\mathrm{adj}, 1}'(\bm{\theta}_1) = \bm{M}_1(\bm{\theta}_1, \bm{\theta}_2^*) + E\left\{ \frac{\partial \bm{M}_1(\bm{\theta}_1, \bm{\theta}_2^*)}{\partial \bm{\theta}_2} \right\}
(\widehat{\bm{\theta}}_2 -\bm{\theta}_2^*) + o(1)
$. Then we have
\begin{align*}
    M_{\mathrm{adj}, 1}'(\bm{\theta}_1) = \bm{M}_1(\bm{\theta}_1, \bm{\theta}_2^*) - E\left\{ \frac{\partial \bm{M}_1(\bm{\theta}_1, \bm{\theta}_2^*)}{\partial \bm{\theta}_2} \right\}\left\{ \frac{\partial \bm{M}_2(\bm{\theta}_2^*)}{\partial \bm{\theta}_2} \right\}^{-1}\bm{M}_2(\bm{\theta}_2^*).
\end{align*}

And $\widehat{\bm{\theta}}_{1,SA} - \bm{\theta}_1 \rightarrow N(\bm{0},\Sigma'_{\bm{\theta}_1} )$, where
$\Sigma'_{\bm{\theta}_1} = E\left\{\left(\left[E\left\{\frac{\partial M'_{adj,1}(\bm{\theta}_1)}{\partial \bm{\theta}_1} \right\}\right]^{-1} M'_{adj,1}(\bm{\theta}_1) \right)^{\otimes 2}\right\} $. The explicit mathematical expression for $\frac{\partial \bm{M}_2(\bm{\theta}_2^*)}{\partial \bm{\theta}_2}$ can be found in equation \eqref{eq1} in Web Appendix C. Employing a similar approach to that outlined in Web Appendix D, we obtain $\frac{\partial M'_{adj,1}(\bm{\theta}_1)}{\partial \bm{\theta}_1}  =\\ \widehat{E} \left( 2 r_{pse,1} r_1 \frac{\partial Q_1(\bm{h}_1,a_1;\bm{\theta}_1)}{\partial \bm{\theta}_1} \left[ \{ \gamma_1 Q_1(\bm{h}_1,a_1;\bm{\theta}_1) - \gamma_1 y_{pse,1}-1\} \exp\{\widehat{s}_{\gamma_1^*}(\bm{h}_1) + \gamma_1 y_{pse,1}\} -1 \right]\right) \frac{\partial y_{pse,1}}{\partial \bm{\theta}_2}$, where $\frac{\partial Q_1(\bm{h}_1,a_1;\bm{\theta}_1)}{\partial \bm{\theta}_1}$ and $\frac{\partial y_{pse,1}}{\partial \bm{\theta}_2}$ depend on the specified form of the Q-functions. Because of the complexity of the asymptotic variance, we recommend utilizing the bootstrap method for its estimation. Following the derivation of Theorem 1, we can prove the convergence in probability of $\widehat{d}^{\mathrm{opt}}_1$ by proving that $Q_1(\bm{h}_1,a_1;\widehat{\bm{\theta}}_{1, SA}) \stackrel{p}{\longrightarrow} Q_1(\bm{h}_1,a_1;\bm{\theta}_{1})$ for any fixed $(\bm{h}_1,a_1)$.

\section{Web Appendix F: Proof of Lemma 1}

Using  Bayes' Theorem, we have
\begin{align*}
    \pr(y_{pse,t}\mid \bm{h}_t,a_t,\overline{\bm{R}}_t = \bm{1}_t,R_{pse,t}=0) &= \frac{\pr(y_{pse,t},\bm{h}_t,a_t,\overline{\bm{R}}_t = \bm{1}_t,R_{pse,t}=0)}{\pr (\bm{h}_t,a_t,\overline{\bm{R}}_t = \bm{1}_t,R_{pse,t}=0)},\\
    \pr(y_{pse,t}\mid \bm{h}_t,a_t,\overline{\bm{R}}_t = \bm{1}_t,R_{pse,t}=1) &= \frac{\pr(y_{pse,t},\bm{h}_t,a_t,\overline{\bm{R}}_t = \bm{1}_t,R_{pse,t}=1)}{\pr (\bm{h}_t,a_t,\overline{\bm{R}}_t = \bm{1}_t,R_{pse,t}=1)} .
\end{align*}

This leads to the following equations:
\begin{align*}
    &\frac{\pr(y_{pse,t}\mid \bm{h}_t,a_t,\overline{\bm{R}}_t = \bm{1}_t,R_{pse,t}=0)}{\pr(y_{pse,t}\mid \bm{h}_t,a_t,\overline{\bm{R}}_t = \bm{1}_t,R_{pse,t}=1)} \\
    &\quad = \frac{\pr(y_{pse,t},\bm{h}_t,a_t,\overline{\bm{R}}_t = \bm{1}_t,R_{pse,t}=0)}{\pr (\bm{h}_t,a_t,\overline{\bm{R}}_t = \bm{1}_t,R_{pse,t}=0)}\frac{\pr (\bm{h}_t,a_t,\overline{\bm{R}}_t = \bm{1}_t,R_{pse,t}=1)}{\pr(y_{pse,t},\bm{h}_t,a_t,\overline{\bm{R}}_t = \bm{1}_t,R_{pse,t}=1)}\\
    &\quad = \frac{P(R_{pse,t}=0\mid y_{pse,t},\bm{h}_t,a_t,\overline{\bm{R}}_t = \bm{1}_t)}{P(R_{pse,t}=1\mid y_{pse,t},\bm{h}_t,a_t,\overline{\bm{R}}_t = \bm{1}_t)}\frac{P(R_{pse,t}=1\mid \bm{h}_t,a_t,\overline{\bm{R}}_t = \bm{1}_t)}{P(R_{pse,t}=0 \mid \bm{h}_t,a_t,\overline{\bm{R}}_t = \bm{1}_t)}.
\end{align*}

Under Assumption 6, we have $\frac{P(R_{pse,t}=0\mid y_{pse,t},\bm{h}_t,a_t,\overline{\bm{R}}_t = \bm{1}_t)}{P(R_{pse,t}=1\mid y_{pse,t},\bm{h}_t,a_t,\overline{\bm{R}}_t = \bm{1}_t)} = \exp\{s'_t(\bm{h}_t,a_t)+\gamma'_t y_{pse,t}\}$. Besides, 
\begin{align*}
    &\frac{P(R_{pse,t}=0 \mid \bm{h}_t,a_t,\overline{\bm{R}}_t = \bm{1}_t)}{P(R_{pse,t}=1\mid \bm{h}_t,a_t,\overline{\bm{R}}_t = \bm{1}_t)} \\
    & \qquad = \int \pr(y_{pse,t} \mid \bm{h}_t,a_t,\overline{\bm{R}}_t = \bm{1}_t,R_{pse,t}=0) \frac{P(R_{pse,t}=0 \mid \bm{h}_t,a_t,\overline{\bm{R}}_t = \bm{1}_t)}{P(R_{pse,t}=1\mid \bm{h}_t,a_t,\overline{\bm{R}}_t = \bm{1}_t)} \mathrm{d} y_{pse,t}\\
    & \qquad =\int \frac{\pr\left(y_{pse,t} \mid \bm{h}_t,a_t,\overline{\bm{R}}_t = \bm{1}_t\right)}{P\left(R_{pse,t}=1 \mid \bm{h}_t,a_t,\overline{\bm{R}}_t = \bm{1}_t\right)} P\left(R_{pse,t}=0 \mid \bm{h}_t,a_t,\overline{\bm{R}}_t = \bm{1}_t, y_{pse,t}\right) \mathrm{d} y_{pse,t}\\
    & \qquad =\int \frac{\pr\left(y_{pse,t} \mid \bm{h}_t,a_t,\overline{\bm{R}}_t = \bm{1}_t, R_{pse,t}=1\right)}{P\left(R_{pse,t}=1 \mid \bm{h}_t,a_t,\overline{\bm{R}}_t = \bm{1}_t, y_{pse,t}\right)} P\left(R_{pse,t}=0 \mid \bm{h}_t,a_t,\overline{\bm{R}}_t = \bm{1}_t, y_{pse,t}\right) \mathrm{d} y_{pse,t}\\
    & \qquad =\int \frac{P\left(R_{pse,t}=0 \mid \bm{h}_t,a_t,\overline{\bm{R}}_t = \bm{1}_t, y_{pse,t}\right)}{P\left(R_{pse,t}=1 \mid \bm{h}_t,a_t,\overline{\bm{R}}_t = \bm{1}_t, y_{pse,t}\right)} \pr\left(y_{pse,t} \mid \bm{h}_t,a_t,\overline{\bm{R}}_t = \bm{1}_t, R_{pse,t}=1\right) \mathrm{d} y_{pse,t} \\
    & \qquad = \int \exp\{s'_t(\bm{h}_t,a_t)+\gamma'_t y_{pse,t}\} \pr\left(y_{pse,t} \mid \bm{h}_t,a_t,\overline{\bm{R}}_t = \bm{1}_t, R_{pse,t}=1\right) \mathrm{d} y_{pse,t}\\
    & \qquad = E[\exp\{s'_t(\bm{h}_t,a_t)+\gamma'_t y_{pse,t}\} \mid \bm{h}_t,a_t,\overline{\bm{R}}_t = \bm{1}_t, R_{pse,t}=1].
\end{align*}

Therefore, 
\begin{align*}
    \frac{\pr(y_{pse,t}\mid \bm{h}_t,a_t,\overline{\bm{R}}_t = \bm{1}_t,R_{pse,t}=0)}{\pr(y_{pse,t}\mid \bm{h}_t,a_t,\overline{\bm{R}}_t = \bm{1}_t,R_{pse,t}=1)}&= \frac{\exp\{s'_t(\bm{h}_t,a_t)+\gamma'_t y_{pse,t}\}}{E[\exp\{s'_t(\bm{h}_t,a_t)+\gamma'_t y_{pse,t}\} \mid \bm{h}_t,a_t,\overline{\bm{R}}_t = \bm{1}_t, R_{pse,t}=1]}\\
    &= \frac{\exp(\gamma'_t y_{pse,t})}{E\{\exp(\gamma'_t y_{pse,t}) \mid \bm{h}_t,a_t,\overline{\bm{R}}_t = \bm{1}_t, R_{pse,t}=1\}}.
\end{align*}
This proves the equation \eqref{pseudo-obs-mis} in Lemma 1.

\section{Web Appendix G: The $m$-out-of-$n$ bootstrap method for parameter inference in non-regularity scenarios }
The asymptotic properties for stage-1 parameters in Theorem 2 and Theorem 3 in Web Appendix B are based on regularity conditions (i.e., $p_{n-regu} = P\{q_{2,1}(\bm{h}_2;\bm{\psi}_2)=0\} = 0$). However, the inference for the parameters in Q-learning for optimal DTRs may face the challenge of non-regularity \citep{chakraborty2010inference},
which can undermine the validity of the asymptotic distribution provided in Web Appendix B and the ordinary $n$-out-of-$n$ bootstrap method \citep{andrews2000inconsistency}. 
Specifically,  in the two-stage scenario, the stage-1 pseudo-outcome and estimators are  non-smooth functions of $\widehat{\bm{\psi}}_2$ when $P\{q_{2,1}(\bm{h}_2;\widehat{\bm{\psi}}_2)=0\} > 0$. As a result, the asymptotic distribution of $(\widehat{\bm{\theta}}_1 - \bm{\theta}_1)$ would be non-normal. Small perturbations in the marginal distribution of $\bm{H}_2$, the parameter $\bm{\psi}_2$, or both, can lead to abrupt differences in the limiting distribution of $\widehat{\bm{\theta}}_1$, and thus make them non-regular. 
In this case, we  followed \cite{chakraborty2013inference} and employed the $m$-out-of-$n$ bootstrap method, which is consistent in non-smooth scenarios,  to construct confidence intervals.

The $m$-out-of-$n$ bootstrap uses a resample size $m$ smaller than the original sample size $n$ to asymptotically avoid the non-negligible probability concentrated at the discontinuous points of $\widehat{Y}_{pse,1}$ and $\widehat{\bm{\theta}}_1$ as functions of $\widehat{\bm{\psi}}_2$ \citep{xu2015regularized}. Specifically, we would draw $B$  bootstrap samples, each with a sample size $m$, and obtain $B$ parameter estimates. The 95\% confidence interval is then constructed using the 2.5th and 97.5th percentiles of the $B$ parameter estimates. 


For the choice of resample size $m$, \cite{chakraborty2013inference} proposed a data adaptive method that is directly connected with a measure of non-regularity. 
Specifically, they estimated $p_{n-{regu}}$ by $\widehat{p}_{n-regu} = \widehat{E}(\mathbb{I}[n \{q_{2,1}(\bm{h}_2;\widehat{\bm{\psi}}_2)\}^2 \leq \bm{h}_2\widehat{\Sigma}_{21}\bm{h}_2^{\mathrm{T}} \chi^2_{1,1-\nu}])$, where $\widehat{\Sigma}_{21}$ was the plug-in estimator of $n\text{Cov}(\widehat{\bm{\psi}}_2,\widehat{\bm{\psi}}_2)$, $\chi^2_{1,1-\nu}$ was the $(1-\nu)\times 100$ percentile of the $\chi^2$ distribution with 1 degree of freedom, and $\nu$ was set as $0.001$ in their simulation. We used the same value of $\nu$ in our simulations and MIMIC-III data analysis. The resample size $m$ was set as  $m = \left\lceil n^{\frac{1+\alpha(1-\widehat{p}_{n-regu})}{1+\alpha}} \right\rceil$, where $\lceil x \rceil$ denotes the smallest integer $\geq x$ and $\alpha$ was selected from a grid of candidate values from 0 to 1 using a double bootstrap method.   
Following \cite{chakraborty2013inference}, we applied the following double bootstrap method for choosing $\alpha$ from a grid of candidate values in our simulation and the MIMIC-III data example.

\begin{enumerate}
    \item Denote the estimate of the parameter of interest from original data as $\bm{c}^{\mathrm{T}} \widehat{\bm{\psi}}_1$, where $\bm{c}$ is an arbitrary vector of the same dimension as $\bm{\psi}_1$.
    \item Draw $B_1$ $n$-out-of-$n$ first-level bootstrap samples from original  data and calculate the $\widehat{p}_{n-regu}^{(b_1)}$ with $\nu=0.001$ using the sample where $\overline{\bm{R}}_2^{(b_1)} = \bm{1}_2^{(b_1)}$ (i.e., the sample with complete covariates in the $b_1$th first-level bootstrap resampling)  
    for $b_1 = 1,\ldots,B_1$. Fix $\alpha$ at the smallest value in the grid.
    \item Compute the corresponding values of $\widehat{m}^{(b_1)}$ by $$
    \widehat{m}^{(b_1)} = \left\lceil n^{\frac{1+\alpha\left(1-\widehat{p}_{n-regu}^{(b_1)}\right)}{1+\alpha}} \right\rceil,
    $$
     $b_1 = 1,\ldots,B_1$. 
    \item Conditional on each first-level bootstrap sample, draw $B_2$ $m^{(b_1)}$-out-of-$n$ second-level (nested) bootstrap samples and estimate the parameter of interest $\bm{c}^{\mathrm{T}} \widehat{\bm{\psi}}^{(b_1b_2)}_{1,\widehat{m}^{(b_1)}}$, $b_1 = 1,\ldots,B_1$, $b_2 = 1,\ldots,B_2$.
    \item For $b_1=1,...,B_1$, compute the 2.5th and 97.5th percentiles of $\bm{c}^{\mathrm{T}} \widehat{\bm{\psi}}^{(b_1b_2)}_{1,\widehat{m}^{(b_1)}}$; say $\widehat{L}_{db}^{(b_1)}$ and $\widehat{U}_{db}^{(b_1)}$, respectively. 
    \item Estimate the coverage rate of the bootstrap confidence interval from all the first-level bootstrap datasets as
    $$
    \frac{1}{B_1} \sum_{b_1=1}^{B_1} \mathbb{I} \left\{ 
    \widehat{L}_{db}^{(b_1)} \leq \bm{c}^{\mathrm{T}} \widehat{\bm{\psi}}_1 \leq \widehat{U}_{db}^{(b_1)}
    \right\}.
    $$
    \item  If the above coverage rate is at or above the nominal
    level, then pick the current value of $\alpha$ as the final value. Otherwise, increment $\alpha$ to the next highest value in the grid.
    \item Repeat steps (3)–(7), until the coverage rate of the double bootstrap confidence interval attains the nominal coverage rate or the grid is exhausted.
\end{enumerate}

As discussed in \cite{chakraborty2013inference}, the number of bootstrap replications in the above double bootstrap procedure to select $\alpha$ (i.e., $B_1$ and $B_2$) does not  need to be equal to the number of  bootstrap replications used for inference (i.e., $B$) after $\alpha$ is selected. Since the kernel regression for estimating the missingness probability models  in the WQ-EE and WQ-SA methods is computationally expensive, we set $B_1=B_2=200$ in our simulation and the MIMIC-III data example.  

\section{Web Appendix H: Simulation}

We conducted four simulation studies to evaluate the finite-sample performance of the WQ-EE and the WQ-SA  methods. In Simulation 1, we focused on the scenario where Assumptions 1-5 were satisfied and demonstrated the performance of the WQ-EE method. In Simulation 2, we evaluated the robustness of the WQ-EE method against the violations of Assumption 5 when (1) the chosen instrumental variable was weakly associated with $R_{pse,t}$ and (2)  there were interactions between $\bm{u}_t$  and $y_{pse,t}$ in the true missingness probability model.  In Simulation 3, we evaluated the proposed simulation-based approach to specifying a plausible range of the sensitivity parameter and the performance of the WQ-SA method based on the specified
range. In Simulation 4, we investigated the influence of an increased number of total stages when covariates are nonignorably missing.  

For comparison, we also assessed three other methods for handling partially missing covariate data in Q-learning: a naive method which ignored partially observed covariates (`naive'), complete case analysis (`CC'), and multiple imputation (`MI') \citep{Little_Rubin_2014}. Q-learning when all the covariates were fully observed (`All') was used as the benchmark. In the naive method, we omitted all terms containing the partially observed covariates in the specified Q-functions. Consequently, the Q-functions were misspecified in this scenario as Assumption 2 was violated. In the CC method, we excluded all patients whose covariates were subject to missingness in estimation. In the MI method, we first imputed missing covariates using predictive mean matching multiple times, which was implemented through the R package \texttt{MICE}. Subsequently, we applied the standard Q-learning algorithm to the imputed data sets. Finally, we calculated the average of the estimated parameters from the imputed datasets to obtain the final estimates of the Q-function parameters. 

All methods were evaluated in two aspects. Firstly, we compared the performances of the methods for estimating the blip function parameters $\bm{\psi}_t$. Secondly, we compared their abilities to identify the true optimal dynamic treatment regime (DTR) and the values of the estimated optimal DTRs. Besides, we also examined the coverage rates of the bootstrap confidence intervals for blip function parameters when the models in the WQ-EE and the WQ-SA methods were correctly specified in Simulation 1 and Simulation 3, respectively.

\subsection{Simulation 1: comparative performance of the WQ-EE method} when Assumptions 1-5 were satisfied
We simulated data from an observational study with two stages of intervention. There were two covariates $X_{t,1}$ and $X_{t,2}$ at stage $t=1,2$. Among these covariates, only $X_{1,2}$ and $X_{2,2}$ were partially observed, and their missingness indicators were $R_1$ and $R_2$, respectively.  Denote $\text{expit}(b) = \exp(b)/\{1+\exp(b)\}$ for $b$ in $\mathcal{R}$. The variables were generated as follows:
\begin{align*}
    & \left( \begin{matrix}X_{1,1}\\X_{2,1}\end{matrix}\right)  \sim N\left(\left( \begin{matrix}0\\0\end{matrix}\right), \left(  \begin{matrix}1 & 0.5\\0.5 & 1 \end{matrix}\right) \right) ;\\
    & X_{1,2} \sim \text{Uniform}(0,2); R_1  \sim \text{Bernoulli} (\{1+\exp(-3+X_{1,2})\}^{-1});\\
    & A_1  \sim 2\times \text{Bernoulli}(\text{expit}(-1+X_{1,1}+X_{1,2}-R_1)) - 1;\\
    & Y_1  =  A_1(\psi_{10}+2+\psi_{11}X_{1,2})+\beta_{10}+\beta_{11} X_{1,1} + \beta_{12} X_{1,2} + \epsilon_1 ; \epsilon_1 \sim N(0,3);\\
    & X_{2,2} \sim \text{Uniform}(0,2);  R_2 \sim  \text{Bernoulli} (\{1+\exp(-1+0.5X_{1,2}-Y_{1}-X_{2,1}-0.5X_{2,2})\}^{-1}) ; \\
    & A_2\sim 2\times \text{Bernoulli} (\text{expit}(-1-X_{1,1}-X_{1,2}+Y_1+X_{2,1}-R_2) ) - 1;\\
    & Y_2 = A_2(\psi_{20}+\psi_{2A}A_1+\psi_{22}X_{2,2}) + \beta_{2A}A_1 + \beta_{22}X_{2,1}+ \beta_{23}X_{2,2} + \epsilon_2; \epsilon_2 \sim N(0,1). 
\end{align*}
The true values of the parameters $\bm{\theta}_1 = (\psi_{10},\psi_{11},\beta_{10},\beta_{11},\beta_{12})$ and $\bm{\theta}_2=(\psi_{20},\psi_{2A},\psi_{22},\beta_{2A},\beta_{22},\beta_{23})$ were  $(1,-1,1.5,0.5,-0.5)$ and $(1,-1,1,-1,1,-0.5)$, respectively. We generated $1000$  data sets with sample sizes of $500$ and $2000$.

We considered $Y = Y_1 + Y_2$ as the final outcome. Under this setting, the true optimal treatment rule at stage 2 was $ 2\mathbb{I}(\psi_{20}+\psi_{2A}a_1+\psi_{22}x_{2,2}\geq 0) - 1$. Because $\bm{\theta}_2 = (1,-1,1,-1,1,-0.5)$ and $1-a_1+x_{2,2}\geq 0$, the optimal choice of $A_2$ was always 1. Then, the pseudo-outcome at stage $1$, $Y_{pse,1}$, was equal to $ 1-2A_1+Y_1+X_{2,1}+0.5X_{2,2} $. Therefore, $Q_1(x_{1,1},x_{1,2},a_1) = 1.5 + \beta_{10} + a_1(\psi_{10}+\psi_{11}x_{1,2}) + (0.5+\beta_{11})x_{1,1} - \beta_{12}x_{1,2}$. And the optimal treatment rule at stage 1 was $2\mathbb{I}(\psi_{10}+\psi_{11}X_{1,2}\geq 0) - 1$. Note that the missingness indicators of $X_{1,2}$ and $X_{2,2}$ were directly associated with their own values, thus they were nonignorable missing. $R_1$ only depended on $X_{1,2}$, and $R_2$ only depended on $(X_{1,2},Y_1,X_{2,1},X_{2,2})$,
so the future-independent missingness assumption held. The missingness indicator for $Y_{pse,1}$ was the same as that for $X_{2,2}$, since $Y_{pse,1}=1-2A_1+Y_1+X_{2,1}+0.5X_{2,2}$ and $(A_1, Y_1, X_{2,1})$ were fully observed. Besides, since $P(R_{2}=1  \mid \bm{H}_2)=\{1+\exp(-1+0.5X_{1,2}-Y_{1}-X_{2,1}-0.5X_{2,2})\}^{-1}$, it follows that  $P(R_{pse,1}=1\mid \bm{H}_1,A_1,Y_{pse,1})=\left\{1+\exp(0.5X_{1,2}-2A_1-Y_{pse,1})\right\}^{-1}$, which indicates that $Y_{pse,1}$ was nonignorable missing and $X_{1,1}$ was conditional independent of $R_{pse,1}$ given $(X_{1,2},A_1,Y_{pse,1})$. Since $ X_{1,1}$ was also associated with $Y_{pse,1}$ after conditioning on $(X_{1,2}, A_1)$, we employed $X_{1,1}$ as a nonresponse instrumental variable.

\begin{figure}[htbp]
    \centering
    \includegraphics[scale = 0.5]{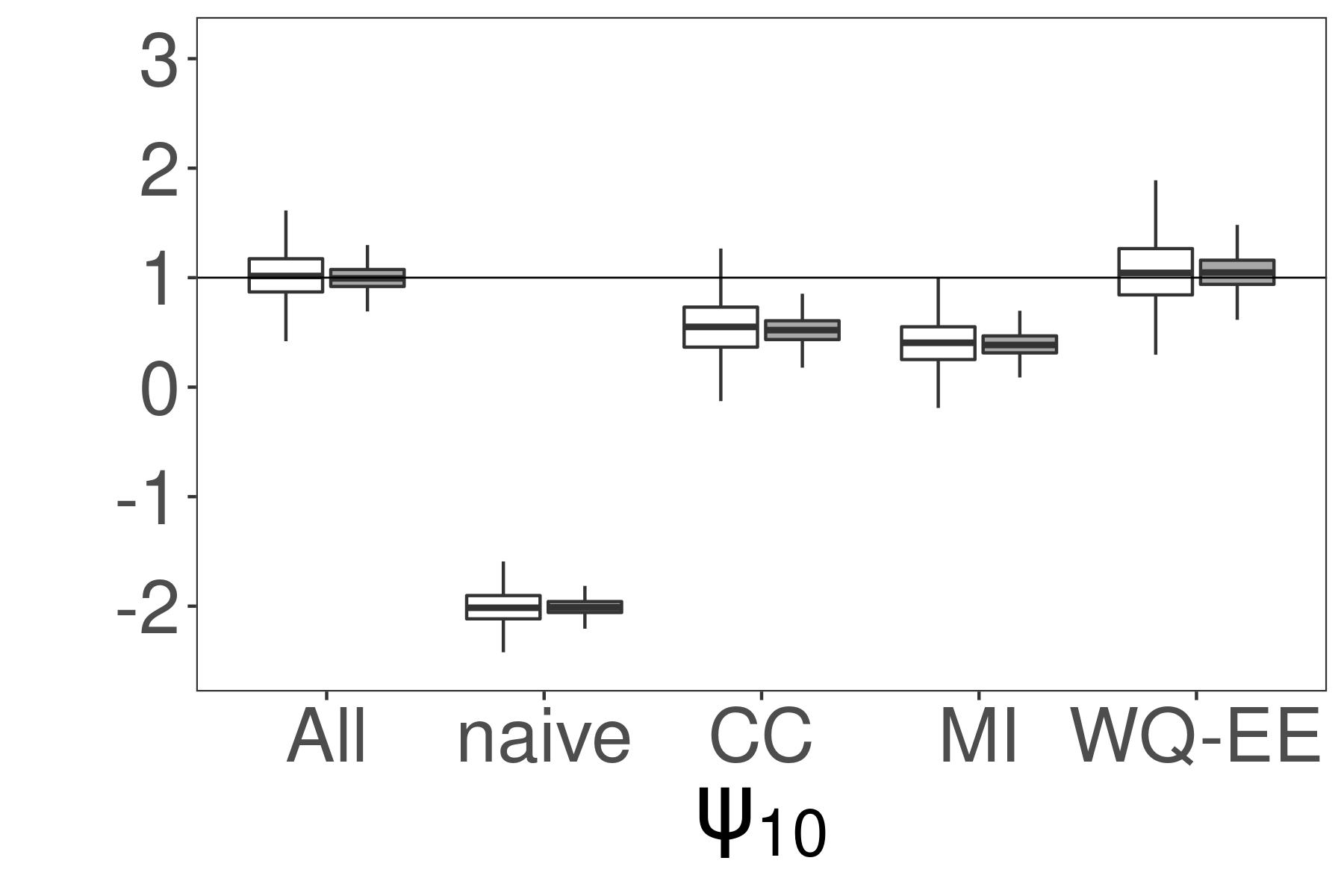}
    \includegraphics[scale = 0.5]{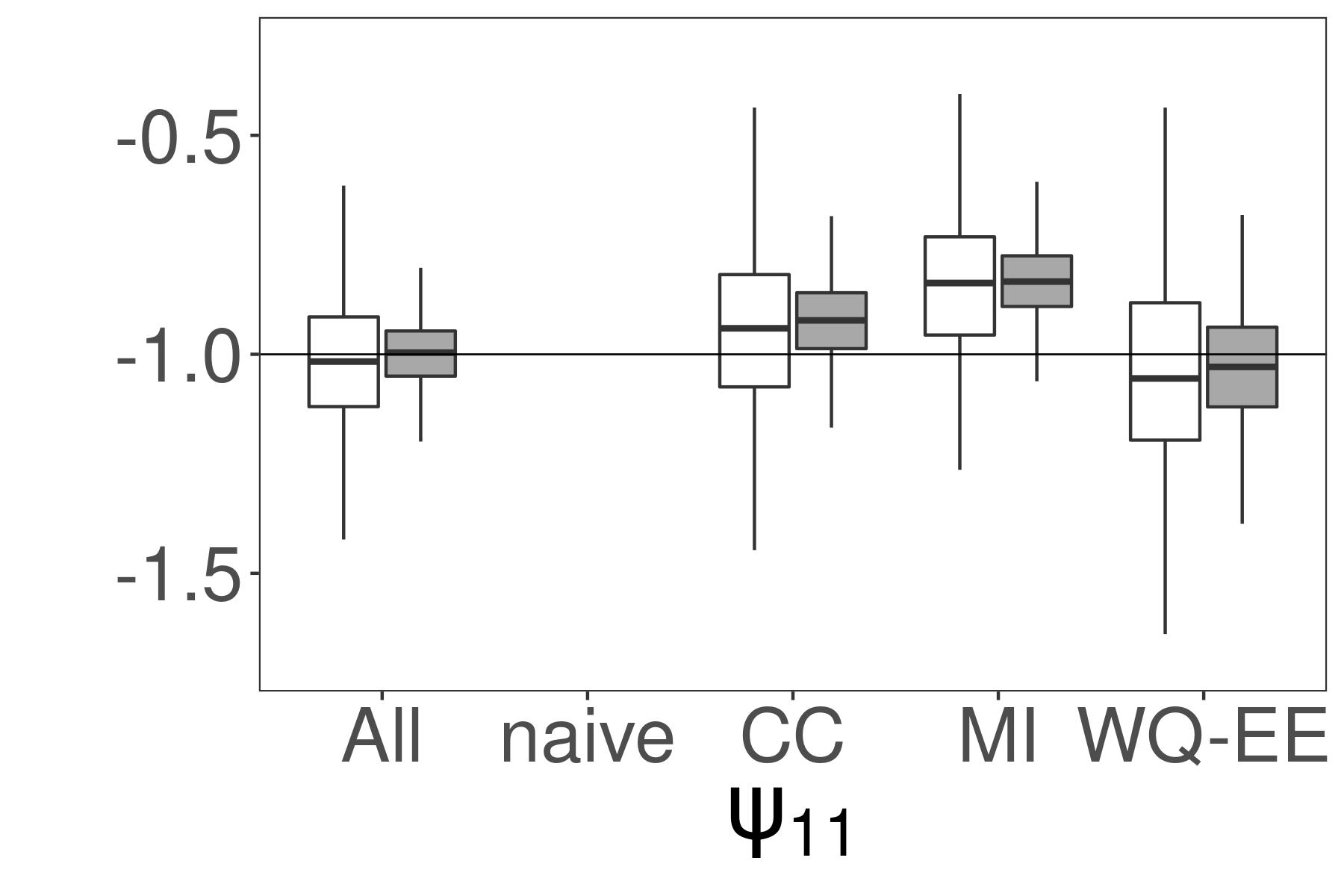}\\
    \includegraphics[scale = 0.5]{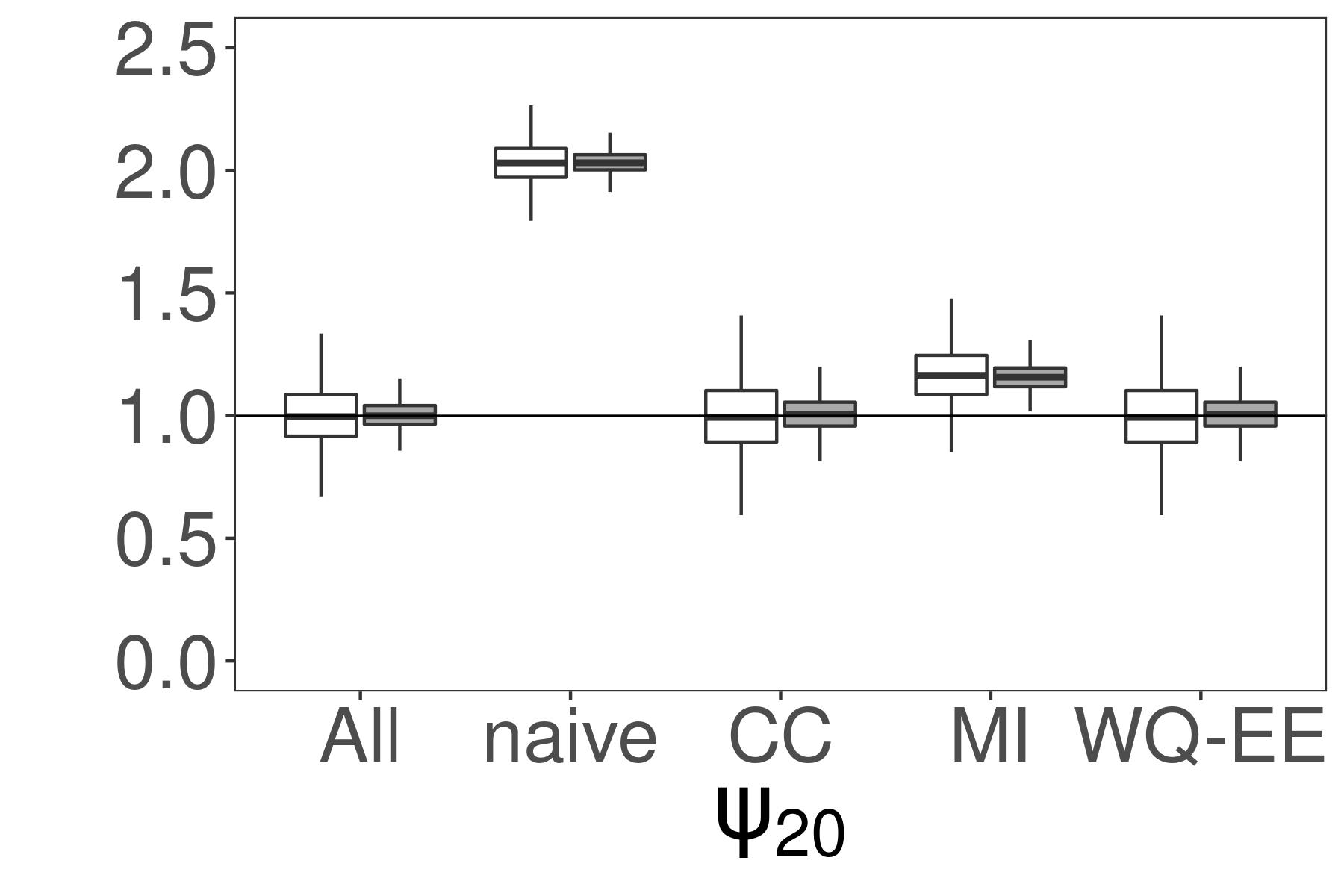}\includegraphics[scale = 0.5]{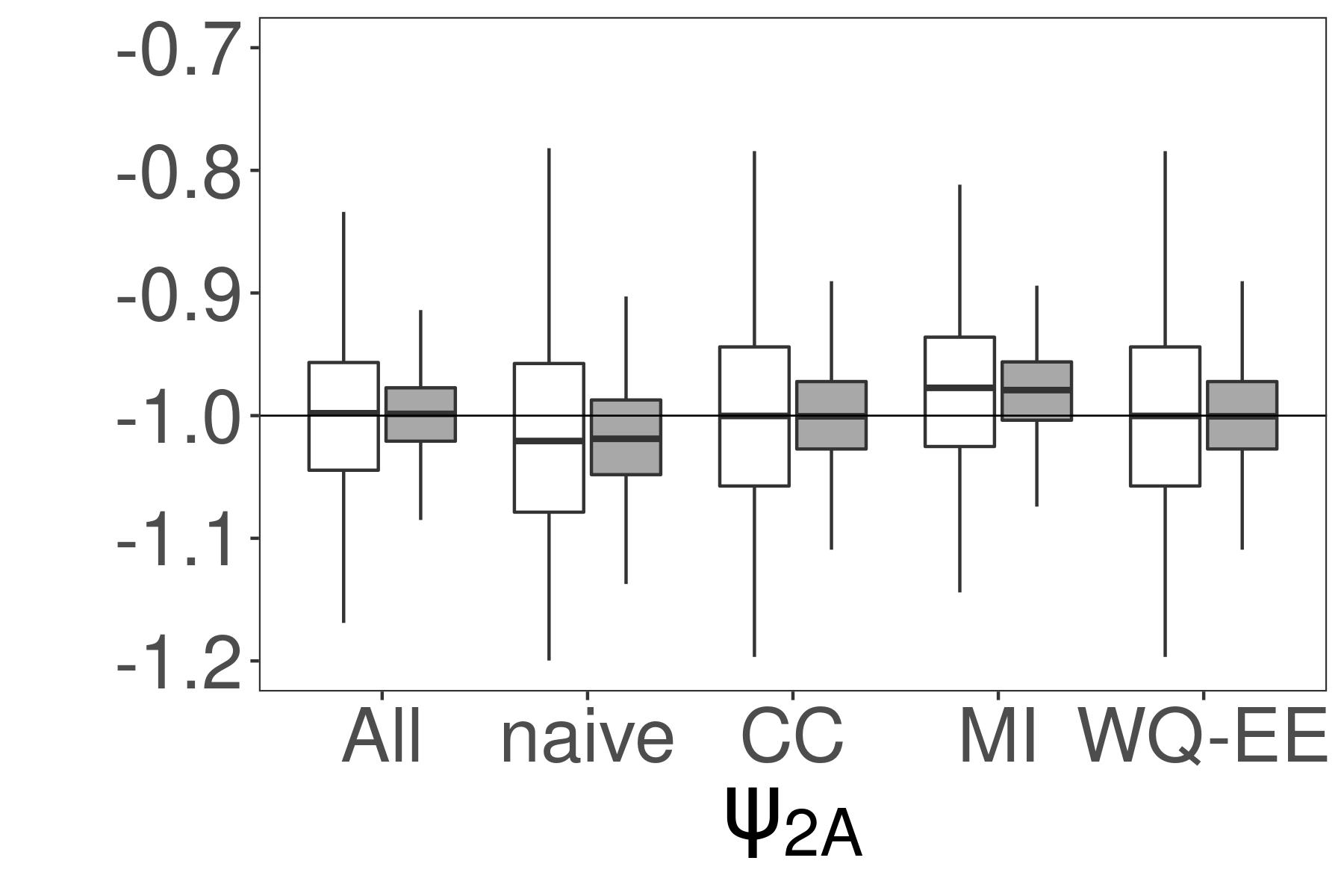}\\
    \includegraphics[scale = 0.5]{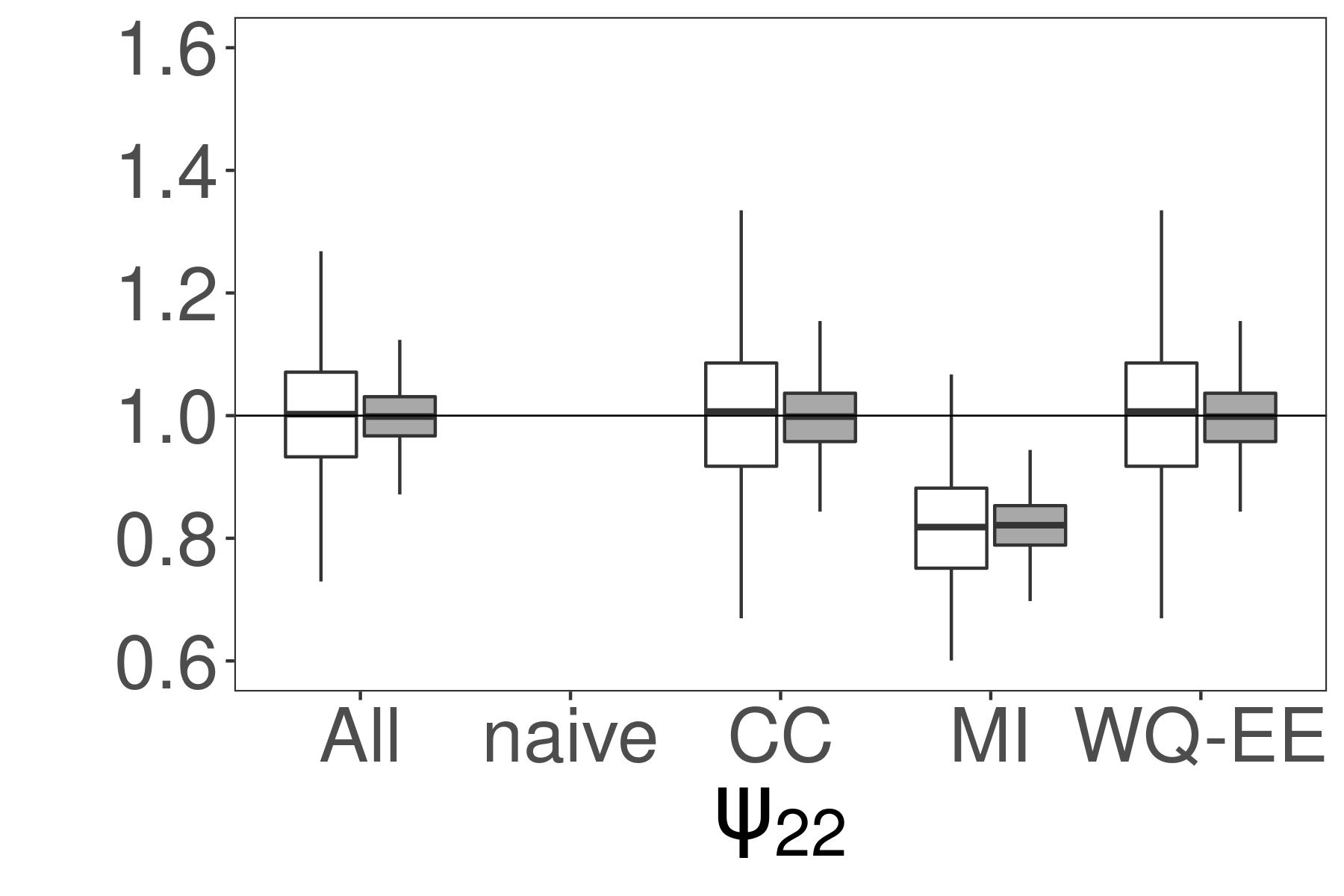}
    \caption{Boxplots of blip function parameter estimates from the compared methods in Simulation 1. In each boxplot, white and gray boxes are the results for the sample sizes 500 and 2,000, respectively. The horizontal line marks the true values of the blip function parameter estimates. Note that the naive method did not generate estimates of $\psi_{11}$ and $\psi_{22}$ as $\psi_{11}$ and $\psi_{22}$  are the coefficients of  partially missing $X_{1,2}$  and $X_{2,2}$ in the true treatment rules.  }
    \label{simufig1}
\end{figure}

Web Figure \ref{simufig1} shows the empirical distributions of the estimates of $\bm{\psi}_t= (\psi_{10},\psi_{11}, \psi_{20},\psi_{2A},\psi_{22})$, which directly influence the estimated optimal DTR. At stage 2, the proposed WQ-EE method and the CC method yielded unbiased estimates of the blip function parameters since the future-independent missingness assumption held and the outcome at stage 2 was fully observed. In contrast, the naive and MI methods led to non-negligible biases of the blip function parameters. At stage 1, the estimators in the WQ-EE method remained unbiased while the estimators in the naive, CC, and MI methods displayed relatively large biases, which were not mitigated as the sample size increased. Note that the naive estimates of $\psi_{11}$ and $\psi_{22}$ were unavailable, as we excluded the terms $A_1X_{1,2}$ and $A_2X_{2,2}$ from the specified Q-functions when implementing this method. And the misspecified Q-functions led to the relatively poor performance of the naive estimates. As for the bias of the CC estimates, 
it can be attributed to the violation of the ignorable missingness assumption of the pseudo-outcomes, which is necessary for the equation $E(Y_{pse,1}\mid \bm{h}_1,a_1) = E(Y_{pse,1}\mid \bm{h}_1,a_1,\overline{\bm{R}}_2 = \bm{1}_2)$ to hold. 
The bias of the MI estimates stemmed from the nonignorable missingness. It has been discussed that the standard MI method would be biased when data are nonignorably missing \citep{schafer2002missing}.  

The proportion of patients whose optimal treatments in both stages were correctly identified was high when employing the WQ-EE method, as demonstrated in Web Table \ref{simutable1}. With the smaller sample size ($n=500$), the WQ-EE method achieved an average correct classification rate of $89.9\%$ across the 1000 simulated datasets, with a mean final outcome of $2.98$. In general, larger sample sizes led to higher proportions of patients with their true optimal DTR correctly identified. In contrast, the DTRs obtained from the CC and MI methods had much lower correct classification rates and the mean final outcomes under the estimated DTRs were lower than that from the WQ-EE method.

\begin{table}[htbp]
  \centering
  \caption{Performance of estimated DTRs in Simulation 1}
  \begin{threeparttable}
    \begin{tabular}{lcccc}
    \hline
          & \multicolumn{2}{c}{n=500}     & \multicolumn{2}{c}{n=2000} \\
          \cmidrule(lr){2-3}\cmidrule(lr){4-5}
          & Value & Opt\% & Value & Opt\% \\
          \hline
    All  & 2.988 (0.057) & 0.928 (0.052) & 2.997 (0.028) & 0.964 (0.024) \\
    naive & 2.501 (0.053) & 0.500 (0.022) & 2.500 (0.027) & 0.499 (0.011) \\
    CC    & 2.879 (0.128) & 0.773 (0.111) & 2.898 (0.059) & 0.777 (0.057)  \\
    MI    & 2.832 (0.139) & 0.729 (0.108) & 2.849 (0.071) & 0.730 (0.058) \\
    WQ-EE & 2.974 (0.069) & 0.899 (0.073) & 2.994 (0.029) & 0.949 (0.036) \\
    \hline
    \end{tabular}%
\begin{tablenotes}
    \small
    \item Note: Value, the value of the estimated regime; Opt\%, the percentage of patients correctly classified to their true optimal treatments in both stages; The numbers outside parentheses represent the sample means, while the numbers inside parentheses represent the sample standard errors.
\end{tablenotes}
  \end{threeparttable}
  \label{simutable1}%
\end{table}%

We further examined the coverage rates of the constructed confidence intervals. Following the recommendations and numerical studies of  \cite{Shao_Wang_2016}, we constructed the confidence intervals with the bootstrap method due to the complicated form of the asymptotic covariance matrix. For stage-2  parameters, non-regularity is not an issue, allowing the use of the ordinary $n$-out-of-$n$ bootstrap method to construct confidence intervals. For stage-1 parameters, we employed the $m$-out-of-$n$ bootstrap method described in Web Appendix G to construct confidence intervals.  

Because the kernel regression for estimating missingness probabilities in the WQ-EE method slows down the computation, the adaptive selection of $\alpha$ for determining the bootstrap resample size $m$ in each simulated dataset is computationally prohibitive. To reduce the computation burden in our simulations, we set $B_1 = B_2=200$ in the double bootstrap method and set $\alpha$ at the fixed value that was determined by calculating the average of the $\alpha$ values adaptively selected from 10 simulated datasets, each with a sample size 500. The parameter of interest in the double bootstrap procedure was set as $\psi_{10}$. 

With the determined $\alpha$, we conducted 500 bootstrap resampling for each of the 1000 simulated datasets to calculate the $n$-out-of-$n$ and $m$-out-of-$n$ percentile bootstrap confidence intervals for the stage-2 and stage-1 blip function parameters, respectively. And the coverage rate of the bootstrap confidence intervals was calculated by the percentage of the confidence intervals that covered the true value of corresponding parameter.

Web Table \ref{WQ-EE_CI_coverage} shows the results for the coverage rates of the bootstrap confidence intervals when employing the WQ-EE method in Simulation 1. We see that these coverage rates were close to the nominal level $95\%$.


\begin{table}[htbp]
  \centering
  \caption{Coverage rates of  the 95\% bootstrap  confidence intervals when employing the WQ-EE method in Simulation 1}
  \begin{threeparttable}
    \begin{tabular}{lcccccc}
    \hline
    &$m$ & $\psi_{10}$ & $\psi_{11}$ & $\psi_{20}$ & $\psi_{2A}$ & $\psi_{22}$ \\
    \hline
     Sample size 500 & 473 & 0.947 & 0.937 & 0.951 & 0.941 & 0.946 \\
     Sample size 2000 & 1956 & 0.947 & 0.945 & 0.953 & 0.950 & 0.945 \\
    \hline
    \end{tabular}
\begin{tablenotes}
    \small
    \item Note: $m$, the average  resample size in the $m$-out-of-$n$ bootstrap method in 1000 simulated datasets, rounded to integer.   For stage-1 parameters ($\psi_{10}, \psi_{11}$), the 95\% confidence intervals were constructed by $m$-out-of-$n$ bootstrap percentiles. For stage-2 parameters ($\psi_{20}, \psi_{2A}, \psi_{22}$), the 95\% confidence intervals were constructed by standard ($n$-out-of-$n$) bootstrap percentiles. 
\end{tablenotes}
  \end{threeparttable}
  \label{WQ-EE_CI_coverage}%
\end{table}%

\subsection{Simulation 2: comparative performance of the WQ-EE method when Assumption 5 is violated}

To investigate the robustness of the proposed WQ-EE method when Assumption 5 was violated, we kept all other data generating mechanisms unchanged, while altering the true model for $R_{pse,1}$ to $P(R_{pse,1}=1) = \left\{1+\exp(\gamma_z X_{1,1} + 0.5X_{1,2}-2A_1-Y_{pse,1})\right\}^{-1}$. When $\gamma_z=0$, the data generating mechanism was the same as that in Simulation 1 and Assumption 5 was satisfied. However, when $\gamma_z \neq 0$, $X_{1,1}$ was associated  with $R_{pse,1}$ conditional on $(X_{1,2},A_1,Y_{pse,1})$. Thus there were no nonresponse instrumental variables. We set $\gamma_z \in \{-0.4,-0.2,0.2,0.4\}$ such that $X_{1,1}$ was weakly associated with $P(R_{pse,1}=1)$ conditional  on $X_{1,2}$ and $Y_{pse,1}$.  We generated 1000 Monte Carlo samples with sample size $n=500$. 

Since the future-independent missingness assumption still held in Simulation 2, the estimators for $(\psi_{20},\psi_{2A},\psi_{22})$  in the WQ-EE method were still consistent. We report the performance of the estimators for $(\psi_{10},\psi_{11})$ and the value and correct classification rate of the estimated optimal DTRs. The results were calculated by averaging over  1000 simulated data sets and are shown in Web Table~\ref{simu_IV_robust}. We see that the proposed WQ-EE method was robust when $X_{11}$ was weakly related to $R_{pse,1}$. Despite the declining performance of the corresponding optimal DTR as the absolute value of $\gamma_z$ increased, the WQ-EE method consistently outperformed the other three methods with higher mean final outcomes and correct classification rates.

\begin{table}[htbp] 
\centering
\begin{threeparttable}
  \caption{Performance of the WQ-EE method without valid nonresponse instrumental variable in comparison with alternative methods in Simulation 2}\label{simu_IV_robust}
    \begin{tabular}{lcccccccc}
    \hline
           & \multicolumn{2}{c}{$\psi_{10}$}     & \multicolumn{2}{c}{$\psi_{11}$}  & \multicolumn{2}{c}{Value} &  \multicolumn{2}{c}{Opt\%} \\
          \cmidrule(lr){2-3}\cmidrule(lr){4-5}\cmidrule(lr){6-7}\cmidrule(lr){8-9}
          Method  & Bias & Std & Bias & Std & Mean & Std  & Mean & Std \\
          \hline
          \multicolumn{9}{c}{$\gamma_z = -0.4$} \\
    naive  & -1.005 & 0.168 &  \textemdash  &\textemdash  &2.498 &0.054 &0.499 &0.022 \\
    CC    & 0.068 & 0.188 & -0.447 & 0.277 & 2.881 & 0.131 & 0.773 & 0.113  \\
    MI    & 0.166 & 0.169 & -0.588 & 0.242 & 2.838 & 0.144 & 0.738 & 0.116 \\
    WQ-EE & -0.042 & 0.237 & 0.113 & 0.318 & 2.971 & 0.078 & 0.896 & 0.080 \\
              \multicolumn{9}{c}{$\gamma_z = -0.2$} \\
    naive & -1.008 & 0.166 &  \textemdash &\textemdash&2.498 &0.054 & 0.500 & 0.023 \\
    CC    & 0.062 & 0.184 & -0.443 & 0.266 & 2.885 & 0.126 & 0.781 & 0.113  \\
    MI    & 0.161 & 0.165 & -0.591 & 0.231 & 2.838 & 0.142 & 0.733 & 0.113 \\
    WQ-EE & -0.043 & 0.243 & 0.090 & 0.315 & 2.972 & 0.081 & 0.898 & 0.076 \\
              \multicolumn{9}{c}{$\gamma_z = 0.2$} \\
    naive  & -1.000 & 0.155 &  \textemdash &\textemdash& 2.502 & 0.052 & 0.499 & 0.022 \\
    CC    & 0.073 & 0.185 & -0.482 & 0.253 & 2.871 & 0.125 & 0.766 & 0.110  \\
    MI    & 0.174 & 0.164 & -0.627 & 0.227 & 2.816 & 0.144 & 0.718 & 0.113 \\
    WQ-EE & -0.040 & 0.239 & 0.031 & 0.310 & 2.974 & 0.070 & 0.899 & 0.072 \\
              \multicolumn{9}{c}{$\gamma_z = 0.4$} \\
    naive  & -1.007 &0.151  & \textemdash &\textemdash& 2.501 & 0.053 & 0.499 & 0.023 \\
    CC    & 0.088 & 0.185 & -0.495 & 0.274 & 2.862 & 0.132 & 0.755 & 0.114  \\
    MI    & 0.183 & 0.163 & -0.639 & 0.238 & 2.808 & 0.146 & 0.705 & 0.114 \\
    WQ-EE & -0.020 & 0.233 & 0.053 & 0.321 & 2.968 & 0.074 & 0.895 & 0.078 \\
    \hline
    \end{tabular}%
   \begin{tablenotes}
       \small
       \item Note: Value, the value of the estimated regime; Opt\%, the percentage of patients correctly classified to their true optimal treatments in both stages; Bias, the bias of the sample mean; Std, the sample standard error; Mean, the sample mean.
   \end{tablenotes}
\end{threeparttable}
\end{table}

We also considered scenarios where there were interactions between $y_{pse,t}$ and $\bm{u}_t$ in the true missingness probability model. The true model for $R_{pse,1}$ was modified as $P(R_{pse,1}=1) = \left[1+\exp\{0.5X_{1,2}-2A_1-Y_{pse,1}+\gamma_{uy}(X_{1,2}-1)Y_{pse,1}\}\right]^{-1}$, while the data generating mechanism for other covariates was the same as that in Simulation 1. The specific form for the interaction, $(X_{1,2}-1)Y_{pse,1}$, was chosen to ensure that the missing data proportions were largely similar between Simulations 1 and 2. 
We set $\gamma_{uy} \in \{-0.2,-0.1,0.1,0.2\}$ by  allowing weak interactions between $X_{1,2}$ and $Y_{pse,1}$.
We again generated  1000 simulated data sets with sample size $n=500$. The results of the compared methods are shown in Web Table~\ref{simu_inter_robust}. 
The WQ-EE method exhibited superior performance compared to the other three methods despite the introduction of interaction terms involving $Y_{pse,1}$, showcasing its robustness in the presence of mild model misspecification.

\begin{table}[htbp]
  \centering
    \caption{Performance of the WQ-EE method when interaction terms between $y_{pse,t}$ and $\bm{u}_t$ were incorrectly omitted, in comparison with alternative methods  in Simulation 2} \label{simu_inter_robust}
\begin{threeparttable}
    \begin{tabular}{lcccccccc}
    \hline
           & \multicolumn{2}{c}{$\psi_{10}$}     & \multicolumn{2}{c}{$\psi_{11}$}  & \multicolumn{2}{c}{Value} &  \multicolumn{2}{c}{Opt\%} \\
          \cmidrule(lr){2-3}\cmidrule(lr){4-5}\cmidrule(lr){6-7}\cmidrule(lr){8-9}
          Method  & Bias & Std & Bias & Std & Mean & Std  & Mean & Std \\
          \hline
          \multicolumn{9}{c}{$\gamma_{uy} = -0.2$} \\
naive & -1.011 & 0.160  & \textemdash &\textemdash& 2.500 & 0.053 & 0.500 & 0.022 \\
CC & 0.124 & 0.184 & -0.514 & 0.269 & 2.863 & 0.139 & 0.760 & 0.115 \\
MI & 0.215 & 0.162 & -0.656 & 0.233 & 2.808 & 0.152 & 0.712 & 0.115 \\
WQ-EE & -0.067 & 0.239 & 0.075 & 0.324 & 2.972 & 0.072 & 0.898 & 0.073 \\
\multicolumn{9}{c}{$\gamma_{uy} = -0.1$} \\
naive & -1.002 & 0.153  & \textemdash &\textemdash& 2.500 & 0.053 & 0.499 & 0.022\\
CC & 0.084 & 0.182 & -0.474 & 0.270 & 2.871 & 0.132 & 0.764 & 0.116 \\
MI & 0.185 & 0.159 & -0.624 & 0.230 & 2.820 & 0.141 & 0.715 & 0.115 \\
WQ-EE & -0.062 & 0.228 & 0.065 & 0.318 & 2.971 & 0.072 & 0.897 & 0.075 \\
\multicolumn{9}{c}{$\gamma_{uy} = 0.1$} \\
naive & -1.002 & 0.157  & \textemdash &\textemdash& 2.499 & 0.053 & 0.500 & 0.022\\
CC & 0.046 & 0.182 & -0.441 & 0.258 & 2.886 & 0.119 & 0.780 & 0.105 \\
MI & 0.147 & 0.163 & -0.595 & 0.231 & 2.832 & 0.143 & 0.734 & 0.107 \\
WQ-EE & -0.021 & 0.233 & 0.022 & 0.307 & 2.974 & 0.074 & 0.898 & 0.077 \\
\multicolumn{9}{c}{$\gamma_{uy} = 0.2$} \\
naive & -1.010 & 0.160  & \textemdash &\textemdash& 2.497 & 0.051 & 0.499 & 0.023 \\
CC & 0.046 & 0.188 & -0.444 & 0.263 & 2.883 & 0.124 & 0.772 & 0.112 \\
MI & 0.146 & 0.168 & -0.590 & 0.234 & 2.835 & 0.140 & 0.732 & 0.111 \\
WQ-EE & 0.023 & 0.242 & -0.011 & 0.308 & 2.972 & 0.079 & 0.895 & 0.082 \\
    \hline
    \end{tabular}%
\begin{tablenotes}
    \small
    \item Note: Value, the value of the estimated regime; Opt\%, the percentage of patients correctly classified to their true optimal treatments in both stages; Bias, the bias of the sample mean; Std, the sample standard error; Mean, the sample mean.
\end{tablenotes}
\end{threeparttable}
\end{table}%


\subsection{Simulation 3: The performance of the WQ-SA method and the proposed simulation-based approach to calibrating the magnitude of the sensitivity parameters}

To evaluate the performance of the WQ-SA method and the proposed simulation-based approach to calibrating the magnitude of the sensitivity parameters, we simulated  data from a 2-stage scenario with two covariates at each stage. At stage $t = 1, 2$, there existed two covariates, $X_{t,1}$ and $X_{t,2}$. Among these covariates, only $X_{1,2}$ and $X_{2,2}$ were partially observed, with their corresponding missingness indicators denoted as $R_1$ and $R_2$, respectively. The data-generation mechanism was as follows:

\begin{align*}
    & \left( \begin{matrix}X_{1,1}\\X_{2,1}\end{matrix}\right)  \sim N\left(\left( \begin{matrix}0\\0\end{matrix}\right), \left(  \begin{matrix}1 & 0.5\\0.5 & 1 \end{matrix}\right) \right) ;\\
    & X_{1,2} \sim \text{Uniform}(0,2); \\
    &R_1  \sim \text{Bernoulli} (\{1+\exp(-1-X_{1,2})\}^{-1});\\
    & A_1  \sim 2\times \text{Bernoulli}(\text{expit}(1-X_{1,1}-X_{1,2}+R_1)) - 1;\\
    & Y_1  =  A_1(\psi_{10}'+2+\psi_{11}'X_{1,2})+\beta_{10}'+\beta_{11}' X_{1,1} + \beta_{12}' X_{1,2} + \epsilon_1' ; \epsilon_1' \sim N(0,3);\\
    & X_{2,2} \sim \text{Uniform}(0,2); \\
    &R_2 \sim  \text{Bernoulli} ([1+\exp\{2.5-\exp(2X_{1,1})-0.5X_{1,2}+A_1+Y_{1}+X_{2,1}+0.5X_{2,2}\}]^{-1}) ; \\
    & A_2\sim 2\times \text{Bernoulli} (\text{expit}(-1+X_{1,1}-2A_1+X_{2,1}) ) - 1;\\
    & Y_2 = A_2(\psi_{20}'+\psi_{2A}'A_1+\psi_{22}'X_{2,2}) + \beta_{2A}'A_1 + \beta_{22}'X_{2,1}+ \beta_{23}'X_{2,2} + \epsilon_2'; \epsilon_2' \sim N(0,1). 
\end{align*}
The true values of the parameters $\bm{\theta}_1' = (\psi_{10}',\psi_{11}',\beta_{10}',\beta_{11}',\beta_{12}')$ and $\bm{\theta}_2' = (\psi_{20}',\psi_{2A}',\beta_{22}',\beta_{2A}',\beta_{22}',\beta_{23}')$ were $(-1,1,-2,-0.5,-1)$ and $(1,-1,1,-1,1,-0.5)$, respectively. The final outcome was set to be $Y=Y_1+Y_2$. 


The true optimal treatment rule at stage 2 was $2\mathbb{I}(1-a_1+x_{2,2}\geq 0)-1$. Since $a_1\in \{-1,1\}$ and $x_{2,2} \geq 0$, the optimal choice of $A_2$ was always 1. With this optimal treatment at stage 2, $Y_{pse,1} = 1-2A_1+Y_1+X_{2,1}+0.5X_{2,2}$. Then, 
the optimal treatment rule at stage 1 was $2\mathbb{I}(\psi_{10}+\psi_{11}x_{1,2}\geq 0)-1$. Since the missingness indicators of $X_{1,2}$ and $X_{2,2}$ were directly related to their own values, they were nonignorably missing. The future-independent missingness assumption held because $R_1$ only depended on $X_{1,2}$, and $R_2$ only depended on $(X_{1,1},X_{1,2},A_1,Y_1,X_{2,1},X_{2,2})$. Because $Y_{pse,1} = 1-2A_1+Y_1+X_{2,1}+0.5X_{2,2}$, $R_{pse,1} = R_2$, $P(R_{pse,1} =1) = \{ 1.5-\exp(2X_{1,1})-0.5X_{1,2}+3A_1+Y_{pse,1} \}^{-1}$, $Y_{pse,1}$ was nonignorably missing and there was no nonresponse instrumental variable. And the true value of the sensitivity parameter was  $\gamma_1'=1$ in this scenario.

We first evaluated the performance of the simulation-based approach to calibrating the magnitude of the sensitivity parameter. 

As discussed in Web Appendix A, we assumed the sign of the sensitivity parameter $\gamma_1'$ was known and chose $[0,7]$ as an initial range for $\gamma_1'$. To zoom in to a more plausible range, we employed the simulation-based approach proposed in Web Appendix A. Specifically, we first generated 1000 data sets with sample size $n=500$ and $2000$, respectively. For each data set, to evaluate the candidate values of $\gamma_1'$, we implemented the steps described in the end of  Web Appendix A. Particularly, in Step (2), $\pr(y_{pse,1}\mid x_{1,1},x_{1,2},a_1,R_1=1,R_2=1)$ was obtained by combing the kernel regression estimates of $E(y_{pse,1}\mid x_{1,1},x_{1,2},a_1,R_1=1,R_2=1)$ and its residual density estimates, which were obtained by kernel density estimation method with the R function \texttt{density}. And $E\{\exp(\gamma_1'y_{pse,1}) \mid x_{1,1},x_{1,2},a_1,R_1=1,R_2=1 \}$ was also estimated by kernel regression methods. In Step (3), based on the estimated $\pr(y_{pse,1}\mid x_{1,1},x_{1,2},a_1,R_1=0,R_2=1)$, we imputed $Y_{pse,1}^{\gamma_1',imp}$ for 1000 Monte Carlo replication times. And in Step (4), we estimated $\pr(y_{pse,1}^{\gamma_1',*}\mid x_{1,1},x_{1,2},a_1,R_1=1)$ with the correctly specified Q-function for the conditional mean and R function \texttt{density} for the residuals. Further, we generated the replications of observed pseudo-outcomes and compared them with the observed pseudo-outcome estimates using the Wilcoxon ranked sum test in Steps (5) and (6), respectively.
Finally, we calculated the medians of the Wilcoxon ranked sum test p-values from the 1000 Monte Carlo replicates for each simulated data set, and we reported the summary statistics of these  p-value medians from the 1000 simulated data sets and the  probability to be included in the  range of plausible values determined by the threshold of 0.05 in Web Table \ref{simu_sen_spec}.

\begin{table}[htbp]
    \centering
\begin{threeparttable}
    \caption{Performance of the proposed simulation-based approach to determining a plausible range of the sensitivity parameter 
 in Simulation 3}      \label{simu_sen_spec}
    \begin{tabular}{lcccccccc}
    \hline
    $\gamma_t'$ & 0 & 1 & 2 & 3 & 4 & 5 & 6  & 7 \\
    \hline
                  \multicolumn{9}{c}{Sample size = 500} \\
    minimum & 0.211 & 0.216 & 0.049 & 0.006 & 0.002 & 0.000 & 0.000 & 0.000\\
    median &0.591 & 0.615 & 0.452& 0.273& 0.201& 0.143& 0.101& 0.073\\
    mean & 0.578 &0.601 &0.440& 0.297 & 0.232 & 0.185 & 0.150 & 0.122\\
    maximum & 0.754 & 0.756 & 0.744 & 0.762 & 0.695 & 0.665 & 0.671 & 0.662\\
    inclusion probability  & 1.000 &  1.000 &  0.997 &  0.966 &  0.895 &  0.814 &  0.722 &  0.605\\
                   \multicolumn{9}{c}{Sample size = 2000} \\
    minimum & 0.105 & 0.338 & 0.001 & 0.000 & 0.000 & 0.000 & 0.000& 0.000 \\
    median & 0.377 & 0.624 & 0.045 & 0.010 & 0.003 & 0.001 & 0.000 &0.000\\
    mean & 0.381 & 0.618 & 0.061 & 0.015 & 0.008 & 0.005 & 0.003 & 0.002\\
    maximum & 0.681 & 0.757 & 0.342 & 0.225 & 0.089 & 0.108 & 0.094 & 0.072\\
    inclusion probability  & 1.000 & 1.000 & 0.448 & 0.055 & 0.026 & 0.013 & 0.012 & 0.007\\
    \hline
    \end{tabular}
    \begin{tablenotes}
        \small
        \item Note: minimum, minimum of p-value medians in the 1000 data sets; median, median of p-value medians in the 1000 data sets;   
        mean, mean of p-value medians in the 1000 data sets;
        maximum, maximum of p-value medians in the 1000 data sets;
        inclusion probability, the probability to be included in the selected range based on threshold 0.05.
    \end{tablenotes}
\end{threeparttable}
\end{table}




 We see that for both sample sizes, the summary statistics corresponding to the true value $\gamma_1'=1$ were largest among candidate values of $\gamma_1'$, and they decreased as the specified sensitivity parameter moving away from its true value. In analogy to the conventional significance level $0.05$ for hypothesis testing based on p-values, we considered a threshold of the p-value medians 0.05 to determine a plausible range  of $\gamma_1'$ for each simulated data set, where a value of $\gamma_1'$  was included in the determined range if its corresponding p-value median was greater than 0.05.  The true value for $\gamma_1'$ was $100\%$ included in the determined ranges when applying this threshold to the 1000 data sets. This indicated that applying the 0.05 threshold for the p-value medians can ensure that the determined plausible ranges  would cover the true value of $\gamma_1'$.

With larger sample sizes, the summary statistics of the p-value medians for the true value of $\gamma_1'$ was stable while those for other values of $\gamma_1'$ decreased. With a sample size of 2000, $\gamma_1'=3$ was included in the determined plausible range for sensitivity parameter in 54 out of 1000 data sets, and larger values of $\gamma_1'$ were selected even  less frequently.

We compared the performance of the WQ-SA method with $\gamma_1' \in [0,7]$ with other methods.  First we assessed their performance in estimating blip function parameters and optimal DTRs. Similar to Simulations 1 and 2, the future-independent missingness assumption still held in Simulation 3, and we reported the performance of the estimators for $(\psi_{10}',\psi_{11}')$ and the values and correct classification rates of the estimated optimal DTRs. The results were calculated by averaging over 1000 simulated datasets and are shown in Web Table \ref{WQ-SA_0506}.

\begin{table}[htbp] 
\centering
\begin{threeparttable}
  \caption{Performance of the WQ-SA method in Simulation 3}\label{WQ-SA_0506}
    \begin{tabular}{lcccccccc}
    \hline
           & \multicolumn{2}{c}{$\psi_{10}'$}     & \multicolumn{2}{c}{$\psi_{11}'$}  & \multicolumn{2}{c}{Value} &  \multicolumn{2}{c}{Opt\%} \\
          \cmidrule(lr){2-3}\cmidrule(lr){4-5}\cmidrule(lr){6-7}\cmidrule(lr){8-9}
          Method  & Bias & Std & Bias & Std & Mean & Std  & Mean & Std \\
          \hline
          \multicolumn{9}{c}{Sample size = 500} \\
All & 0.016 & 0.234 & -0.011 & 0.158 & -1.018 & 0.045 & 0.923 & 0.055\\
naive & 1.171 & 0.162 &  \textemdash & \textemdash & -1.498 & 0.035& 0.503& 0.023 \\
CC & -0.319 & 0.298 & -0.078 & 0.209 & -1.131 & 0.121 & 0.767 & 0.114 \\
MI & -0.269 & 0.264 & -0.178 & 0.186 & -1.184 & 0.138 & 0.721 & 0.113\\
WQ-SA ($\gamma_1'=0$) & -0.037 & 0.213 & -0.395 & 0.298 & -1.134 & 0.124 & 0.760 & 0.111\\
WQ-SA ($\gamma_1'=1$) & -0.056 & 0.241 &  0.034 & 0.335 & -1.033 & 0.071 & 0.893 & 0.084\\
WQ-SA ($\gamma_1'=2$) & -0.049 & 0.251 &  0.083 & 0.347 & -1.034 & 0.071 & 0.891 & 0.084\\
WQ-SA ($\gamma_1'=3$) & -0.056 & 0.252 &  0.062 & 0.339 & -1.036 & 0.070 & 0.889 & 0.086\\
WQ-SA ($\gamma_1'=4$) & -0.072 & 0.256 &  0.058 & 0.344 & -1.037 & 0.077 & 0.887 & 0.088\\
WQ-SA ($\gamma_1'=5$) & -0.070 & 0.264 &  0.025 & 0.349 & -1.041 & 0.077 & 0.883 & 0.091\\
WQ-SA ($\gamma_1'=6$) & -0.053 & 0.263 & -0.015 & 0.361 & -1.042 & 0.079 & 0.880 & 0.093\\
WQ-SA ($\gamma_1'=7$) & -0.060 & 0.271 & -0.023 & 0.359 & -1.044 & 0.080 & 0.879 & 0.095\\
\multicolumn{9}{c}{Sample size = 2000} \\
All & 0.005 & 0.118 & -0.002 & 0.080 & -1.003 & 0.020 & 0.965 & 0.025 \\
naive & 1.173 & 0.077 & \textemdash & \textemdash & -1.500 & 0.014 & 0.500 & 0.011 \\
CC & -0.324 & 0.153 & -0.075 & 0.106 & -1.101 & 0.053 & 0.774 & 0.057 \\
MI & -0.296 & 0.134 & -0.152 & 0.096 & -1.148 & 0.065 & 0.731 & 0.056\\
WQ-SA ($\gamma_1'=0$) & -0.036 & 0.102 & -0.421 & 0.143 & -1.120 & 0.056 & 0.756 & 0.055\\
WQ-SA ($\gamma_1'=1$) & -0.026 & 0.140 &  0.035 & 0.182 & -1.007 & 0.022 & 0.948 & 0.037\\
WQ-SA ($\gamma_1'=2$) & -0.019 & 0.132 & -0.036 & 0.177 & -1.008 & 0.023 & 0.943 & 0.043\\
WQ-SA ($\gamma_1'=3$) & -0.021 & 0.158 & -0.019 & 0.205 & -1.009 & 0.024 & 0.939 & 0.043\\
WQ-SA ($\gamma_1'=4$) & -0.020 & 0.163 & -0.066 & 0.200 & -1.012 & 0.026 & 0.931 & 0.051\\
WQ-SA ($\gamma_1'=5$) & -0.033 & 0.165 & -0.098 & 0.206 & -1.019 & 0.029 & 0.915 & 0.056\\
WQ-SA ($\gamma_1'=6$) & -0.032 & 0.164 & -0.137 & 0.203 & -1.027 & 0.035 & 0.897 & 0.062\\
WQ-SA ($\gamma_1'=7$) & -0.051 & 0.168 & -0.139 & 0.209 & -1.031 & 0.037 & 0.887 & 0.063\\
    \hline
    \end{tabular}%
   \begin{tablenotes}
       \small
       \item Note: WQ-SA ($\cdot$), the WQ-SA method based on corresponding sensitivity parameters; Value, the value of the estimated regime; Opt\%, the percentage of patients correctly classified to their true optimal treatments in both stages; Bias, the bias of the sample mean; Std, the sample standard error; Mean, the sample mean.
   \end{tablenotes}
\end{threeparttable}
\end{table}

We see that when the specified sensitivity parameter was equal to its true value, the WQ-SA method had the best performance. Notably,  when $\gamma_t'=0$ was specified, it was equivalent to assuming ignorable missingness, that is,  the missingness of $Y_{pse,1}$ was conditionally independent with its own values. Except for this specification, the WQ-SA method exhibited relative robustness to the sensitivity parameter specification and consistently outperformed the naive, CC, and MI methods with higher mean final outcomes and correct classification rates under the estimated optimal DTRs.  

Finally, we examined the coverage rates of the confidence intervals for blip function parameters when employing the WQ-SA method with the correctly specified sensitivity parameter value $\gamma_t'=1$. 
Similar to Simulation 1, the confidence intervals for stage-2 parameters were constructed by ordinary $n$-out-of-$n$ bootstrap method, while those for stage-1 parameters were constructed by $m$-out-of-$n$ bootstrap method described in Web Appendix G. And the bootstrap resample sizes were selected in the  same way as in Simulation 1. 
Web Table \ref{WQ-SA_CI_coverage} shows the coverage rates of the bootstrap confidence intervals, which were close to the nominal level $95\%$.   


\begin{table}[htbp]
  \centering
  \caption{Coverage rates of  the bootstrap 95\% confidence intervals when employing the WQ-SA method in Simulation 3}
  \begin{threeparttable}
    \begin{tabular}{lcccccc}
    \hline
     & $m$& $\psi_{10}$ & $\psi_{11}$ & $\psi_{20}$ & $\psi_{2A}$ & $\psi_{22}$ \\
    \hline
     Sample size 500 &413 & 0.961 & 0.958 & 0.950 & 0.960 & 0.961 \\
     Sample size 2000 &1864 & 0.946 & 0.944 & 0.947 & 0.948 & 0.939 \\
    \hline
    \end{tabular}
\begin{tablenotes}
    \small
    \item Note: $m$, the average resample size in the $m$-out-of-$n$ bootstrap method in 1000 simulated datasets, rounded to integer. For stage-1 parameters ($\psi_{10}, \psi_{11}$), the 95\% confidence intervals were constructed by $m$-out-of-$n$ bootstrap percentiles. For stage-2 parameters ($\psi_{20}, \psi_{2A}, \psi_{22}$), the 95\% confidence intervals were constructed by standard ($n$-out-of-$n$) bootstrap percentiles. 
\end{tablenotes}
  \end{threeparttable}
  \label{WQ-SA_CI_coverage}%
\end{table}%

In summary, Simulation 3 demonstrated that the proposed simulation-based approach could identify a plausible range for the sensitivity parameter. Additionally, it showed that the WQ-SA method was robust and outperformed the naive, CC, and MI methods within a wide range of the sensitivity parameter. Furthermore, it confirmed that the 95\% confidence intervals constructed by the bootstrap percentiles performed well when all the models were correctly specified for the WQ-SA method.

\subsection{Simulation 4: the impact of an increased number of total stages with nonignorable missing covariates}

To investigate the influence of the nonignorable missing covariates when the number of total stages increases, we designed a simulation study with three stages of intervention. To facilitate comparisons with the 2-stage scenario in Simulation 1, we used the same data generation mechanism for stage 1 and stage 2 as that in Simulation 1, and generated data for stage 3 as follows:  
\begin{align*}
    &X_{3,1} \sim N(0,1); X_{3,2} \sim \text{Uniform}(0,2); \\
    &R_3  \sim \text{Bernoulli} (\{1+\exp(-Y_1-Y_2-0.5X_{3,1}-|X_{3,2}-1|)\}^{-1});\\
    &A_3 \sim 2\times \text{Bernoulli}(\text{expit}(-1+X_{3,1}+X_{3,2})) - 1;\\
    &Y_3 = \beta_{30}+A_3(\psi_{30}+\psi_{3,1}X_{3,2})+\beta_{31}X_{31} +\epsilon_3; \epsilon_3 \sim N(0,1).
\end{align*}
The true values of the parameters $\bm{\theta}_3 = (\psi_{30},\psi_{31},\beta_{30},\beta_{31})$ were $(-1,1,-0.5,0.5)$. $R_3$ was the missingness indicator for the partially observed variable $X_{3,2}$, while $X_{3,1}$ was fully observed. Similar to Simulation 1, we generated 1000 data sets with sample sizes 500 and 2000.  

The final outcome was assumed to be $Y=Y_1+Y_2+Y_3$. The true optimal treatment rule at stage 3 was $2\mathbb{I}(\psi_{30}+\psi_{3,1}X_{3,2} \geq 0)-1$. With the optimal treatment at stage 3, the pseudo-outcome at stage 2 was $Y_{pse,2}=-0.5+|-1+X_{3,2}|+0.5X_{3,1}+Y_1+Y_2$. Because the data generation mechanism at stages 1 and 2 was unchanged from Simulation 1 and $E(Y_{pse,2}\mid \bm{H}_2,A_2,Y_2) = Y_1+Y_2$, the Q-functions, the optimal treatment rules at stages 1 and 2, and $(Y_{pse,1},R_{pse,1})$ were the same as those in Simulation 1. Since $R_3$ only depended on $(X_{3,1},X_{3,2})$, the future-independent missingness assumption held. Among the partially observed variables, only $X_{3,2}$ was employed in the calculation of $Y_{pse,2}$. Therefore, $R_{pse,2} = R_3$. Combining the form of the pseudo-outcomes and the missingness mechanism, we have $P(R_{pse,2}=1\mid \bm{H}_2,A_2,Y_{pse,2}) = \{ 1+\exp(-Y_{pse,2})\}^{-1}$. Since $(X_{2,1},X_{2,2})$ were associated with $Y_{pse,2}$ and independent of $R_{pse,2}$ given $Y_{pse,2}$, we employed $(X_{2,1},X_{2,2})$ as nonresponse instrumental variables when estimating the missingness model for $R_{pse,2}$.

\begin{table}[htbp]
  \centering
  \caption{Performance of estimated DTRs in Simulation 4}
\begin{threeparttable}
    \begin{tabular}{lcccc}
    \hline
          & \multicolumn{2}{c}{n=500}     & \multicolumn{2}{c}{n=2000} \\
          \cmidrule(lr){2-3}\cmidrule(lr){4-5}
          & Value & Opt\% & Value & Opt\% \\
          \hline
All & 2.983 (0.064) & 0.904 (0.059) & 2.995 (0.031) & 0.950 (0.028) \\
naive & 2.012 (0.064) & 0.255 (0.019) & 2.006 (0.031) & 0.252 (0.009) \\
CC & 2.650 (0.174) & 0.576 (0.114) & 2.622 (0.122) & 0.560 (0.070) \\
MI & 2.348 (0.192) & 0.450 (0.046) & 2.466 (0.103) & 0.479 (0.020) \\
WQ-EE & 2.947 (0.085) & 0.837 (0.097) & 2.983 (0.034) & 0.910 (0.049) \\
    \hline
    \end{tabular}%
\begin{tablenotes}
    \small
    \item Note: Value, the value of the estimated regime; Opt\%, the percentage of patients correctly classified to their true optimal treatments in all the stages;  The numbers outside parentheses represent the sample means, while the numbers inside parentheses represent the sample standard errors.
\end{tablenotes}
  \end{threeparttable}
  \label{simu_3stage_table}%
\end{table}%

Web Table \ref{simu_3stage_table} presents the performance of the methods in comparison in the 3-stage scenario. Compared to Web Table \ref{simutable1} in the 2-stage scenario, the means of the values under the estimated DTRs for all methods decreased. For the All and the WQ-EE method, this decrease was small and it became smaller as the sample size increased. It can be explained by the finite-sample bias accumulated in the backward-induction procedure. In contrast, for the naive, CC, and MI methods, the deterioration of performance was more pronounced, indicating that,  if it is not handled properly, the consequence of the nonignorable missing covariates problem would be more severe when the number of total stages increases. To better understand the reasons behind the performance deterioration, we examined the correct classification rates at different stages. Since we have presented the correct classification rates at stages 1 and 2 in Simulation 1, and the true Q-functions and optimal treatments rules at stages 1 and 2 in Simulation 1 and Simulation 4 were the same, we examined the correct classification rates at stage 3 and at stages 1 and 2, respectively.

\begin{table}[htbp]
  \centering
  \caption{Correct classification rates in Simulation 4}
  \begin{threeparttable}
    \begin{tabular}{lcccc}
    \hline
          & \multicolumn{2}{c}{n=500}     & \multicolumn{2}{c}{n=2000} \\
          \cmidrule(lr){2-3}\cmidrule(lr){4-5}
          & Stages 1\&2 & Stage 3 & Stages 1\&2 & Stage 3 \\
          \hline
All & 0.919  (0.060) & 0.980  (0.017) & 0.958  (0.027) & 0.990  (0.008) \\
naive & 0.501  (0.021) & 0.506  (0.021) & 0.501  (0.011) & 0.504  (0.010) \\
CC & 0.594  (0.117) & 0.958  (0.035) & 0.568  (0.067) & 0.981  (0.016) \\
MI & 0.485  (0.040) & 0.931  (0.058) & 0.506  (0.016) & 0.936  (0.026) \\
WQ-EE &0.873  (0.098) & 0.958  (0.035) & 0.928  (0.054) & 0.981  (0.016) \\
    \hline
    \end{tabular}%
\begin{tablenotes}
    \small
    \item Note: Stages 1\&2, the correct classification rate at stages 1 and 2; Stage 3, the correct classification rate at stage 3; The numbers outside parentheses represent the sample means, while the numbers inside parentheses represent the sample standard errors.
\end{tablenotes}
  \end{threeparttable}
  \label{simu_3stage_table2}%
\end{table}%

We used the results in  Web Table \ref{simu_3stage_table2} to investigate why    the increased number of total stages  led to  lower overall correct classification rate if nonignorable missing covariate were  not handled properly. For the naive method, its correct classification rate at stages 1 and 2 in Simulation 4 was almost the same as that in Simulation 1 (Web Table~\ref{simutable1}), and the decrease in its overall correct classification rate in Web Table~\ref{simu_3stage_table} was mainly explained by its poor average correct classification rate  at stage 3, which was near $0.5$. In contrast,  for the CC and the MI method, the average correct classification rates at stage 3 was higher than $0.93$, but their average correct classification rates at stages 1 and 2 dropped a lot compared to those in Simulation 1  (Web Table \ref{simutable1}). And this decrease was due to the bias from nonignorable missing pseudo-outcomes accumulated over more stages in the backward induction procedure.     



In summary, an increased number of total stages would generally lead to lower overall correct classification rates of the estimated optimal DTRs when covariates had nonignorable missing values. If the nonignorable missingness was properly addressed, the decrease in the overall correct classification rates would be small and could be mitigated by increasing the sample size. However, if the nonignorable missingness was not properly handled, the decrease in overall correct classification rates could be substantial due to the bias accumulated in the backward-induction procedure.

\section{Web Appendix I: Additional details  of  the MIMIC-III data and the estimation procedure}

\begin{figure}[htbp]
    \centering
    \includegraphics[scale=0.75]{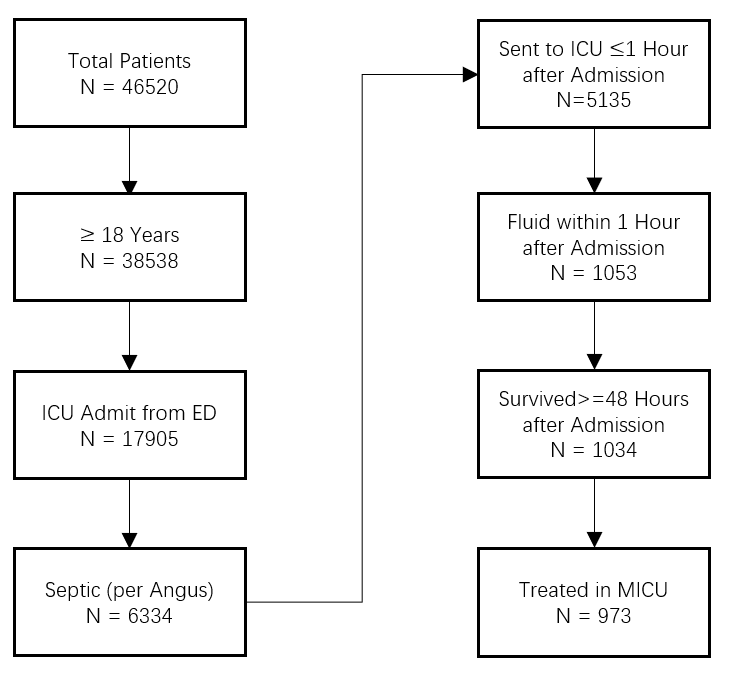}
    \caption{Eligibility criteria and patient population for the selected cohort from the MIMIC-III database.}
    \label{patient_select}
\end{figure}

The proposed methods were applied to the electronic medical records and administrative data from the MIMIC-III database \citep{Johnson_et_al_2016}. MIMIC-III is a retrospectively collected and freely available database accessible through PhysioNet \citep{Goldberger_et_al_2000}. The MIMIC-III database contains de-identified and anonymized health information from over 45,000 patients who were treated in the Intensive Care Units (ICUs) of Beth Israel Deaconess Medical Center in Boston, Massachusetts. We utilized similar patient selection criteria as in \cite{Speth_et_al_2022}. The detailed patient selection criteria are presented in Web Figure~\ref{patient_select}.

As depicted in Web Figure \ref{patient_select}, This study included adult patients (age $\geq$ 18) diagnosed with acute emergent sepsis who were transferred from the Emergency Department (ED) and admitted to the medical ICU. Patients must be sent to the medical ICU within 1 hour after hospital admission and had documented fluid intake within the first hour after hospital admission. Furthermore, following \cite{Speth_et_al_2022}, to ensure a relatively homogeneous patient cohort, we only included patients who survived for a minimum of 48 hours after ICU admission. 

Since there were very few patients with hemodynamic variables recorded precisely at 180 minutes after ICU admission, we calculated the mean of the hemodynamic variables recorded within a half-hour window around the third hour. That is, we used the data collected between 2.5 and 3.5 hours after ICU admission as surrogates for the values of the hemodynamic variables three hours after ICU admission. If a hemodynamic variable had no record during this specific time interval, we considered the corresponding record three hours after admission to be missing. Under this criterion, the missing data proportion of hemodynamic variables is provided in Web Table~\ref{missingtable}.

\begin{table}[htbp]
    \centering
    \caption{The missing proportion of hemodynamic variables 3 hours after ICU admission in the MIMIC-III data}
    \begin{tabular}{lc}
    \hline
        covariates &  missing proportion\\
        \hline 
        heartrate at 3 hours & 7.50 \% \\
        mean blood pressure at 3 hours & 10.2 \%  \\
        SpO2 at 3 hours  & 10.4 \% \\
        respiratory rate at 3 hours & 10.2 \% \\
        temperature at 3 hours & 19.8 \% \\
        urine output in 0-3 hours & 0.00 \% \\
    \hline
    \end{tabular}
    \label{missingtable}
\end{table}


Following \cite{xu2015regularized}, we performed cross-validation and $m$-out-of-$n$ bootstrap to assess the performance of the estimated optimal DTRs. Specifically, to estimate the expected out-of-sample improvement of SOFA score at 24 hours post-admission, we randomly split the MIMIC-III data for 1000 times. For each split, we used $4/5$ of the data (rounded to the nearest integer) as the training set and the remaining data as the testing set. 
The semiparametric missingness probability model and the Q-functions were re-estimated based on the training set of that split. The expected outcome in the testing set, under the estimated DTR, was then computed using the estimated missingness probability models and Q-functions. It should be noted that the optimal DTR may differ across different splits. Nevertheless, the average of the expected optimal outcomes in the testing sets from repeated splitting can be interpreted as the expected SOFA score at 24 hours after ICU admission that would have been obtained with out-of-sample data under the estimated regimes.

For the inference of the improvement of SOFA score at 24 hours post-admission, we further calculated the bootstrap standard deviation and the 95\% confidence interval  
using the $m$-out-of-$n$ bootstrap method based on the original data (i.e., without cross-validation). We used the double bootstrap method described in Web Appendix G to determine the bootstrap resample size $m$, with the number of bootstrap replications set as $B_1=B_2=200$. Specifically, when selecting $m$ and conducting $m$-out-of-$n$ bootstrap for the MI method, we followed the suggestion from \cite{Shen_Hubbard_Linn_2023} and used 25 imputations.  

\bibliographystyle{biom} 
\bibliography{DTR_MNAR_bib}